\documentclass[useAMS,usenatbib]{mn2e}
\usepackage{amsbsy,amssymb,amsmath,url,bm,epsfig,graphicx,fixltx2e}
\newcommand{\comment}{ }

\newcommand{\bfQ}{{\mathbfss{Q}}}

\newcommand{\bfA}{{\mathbfss{A}}}
\newcommand{\bfW}{{\mathbfss{W}}}

\newcommand{\rmi}{{\rm i}}

\newcommand{\bfk}{{\mathbf{k}}}

\newcommand{\bfs}{{\mathbf{s}}}

\newcommand{\bfC}{{\mathbfss{C}}}

\newcommand{\bfF}{{\mathbfss{F}}}
\newcommand{\bfD}{{\mathbfss{D}}}

\newcommand{\nhat}{{\bmath{\widehat{n}}}}
\newcommand{\Mpc}{{\rm Mpc}}

\begin{document}

\title{On using angular cross-correlations to determine source redshift
distributions}
\author[M. McQuinn and M. White]{Matthew McQuinn$^{1,2}$ and
Martin White$^{1,2}$\\ \\
$^{1}$ Department of Astronomy, University of California, Berkeley, CA 94720\\
$^{2}$ Department of Physics, University of California, Berkeley, CA 94720}

\pubyear{2013} \volume{000} \pagerange{1}

\maketitle\label{firstpage}

\begin{abstract}
We investigate how well the redshift distribution of a population of extragalactic objects can be reconstructed using angular cross-correlations with a sample whose redshifts are known.  We derive the minimum variance quadratic estimator, which has simple analytic representations in very applicable limits and is significantly more sensitive than earlier proposed estimation procedures.  This estimator is straightforward to apply to observations, it robustly finds the likelihood maximum, and it conveniently selects angular scales at which fluctuations are well approximated as independent between redshift bins and at which linear theory applies.  We find that the linear bias times number of objects in a redshift bin generally can be constrained with cross-correlations to fractional error $\approx \sqrt{10^2 \, N_{\rm bin}/{\cal N}}$, where ${\cal N}$ is the total number of spectra per $dz$ and $N_{\rm bin}$ is the number of redshift bins spanned by the bulk of the unknown population.  The error is often independent of the sky area and sampling fraction.  Furthermore, we find that sub-percent measurements of the angular source density per unit redshift, $dN/dz$, are in principle possible, although cosmic magnification needs to be accounted for at fractional errors of $\lesssim 10$ per cent.  We discuss how the sensitivity to $dN/dz$ changes as a function of photometric and spectroscopic depth and how to optimize the survey strategy to constrain $dN/dz$.  We also quantify how well cross-correlations of photometric redshift bins can be used to self-calibrate a photometric redshift sample.  Simple formulae that can be quickly applied to gauge the utility of cross correlating different samples are given.  
\end{abstract}


\section{Introduction}

In many spectral bands, the redshift distribution of a source
population is difficult to determine (e.g., the radio, microwave,
infrared, and X-ray).  Even in the optical, where photometric
techniques are widely applied to estimate source redshifts, these
techniques work better for certain galaxy types than for others.
However, extragalactic objects that are close together on the sky are
also likely to be close in redshift.  Thus, angular cross-correlations
between populations with poorly known redshifts and those with better
known redshifts can be used to improve the determination of the former's
redshift distribution.
Such reconstruction has a wide range of applications, from ascertaining
the redshift distribution of diffuse backgrounds to calibrating photometric
redshifts for the next generation of large-scale structure surveys.

Several previous studies have attempted to measure a population's redshift
distribution, $dN/dz$, by using its constituents' proximity on the sky to
sources with known redshifts, i.e., by computing angular cross
correlation statistics between the two populations
\citep{seldner79, phillipps87, ho08, erben09}.
Similar techniques have been used to search for contamination in
photometrically selected redshift slices or to bound the
median redshift of a sample
\citep{padmanabhan07,erben09,benjamin10, benjamin12}.
Different $dN/dz$ cross-correlation estimators have also been studied theoretically
\citep{phillipps85,newman08,matthews10,schulz10,matthews12}.
However, it is unknown how close any of these estimators are to being optimal.
It is also unclear which survey specifications (depth, area, sampling fraction,
etc.) are best for reconstructing the redshift distribution of an unknown
population.

This paper attempts to answer these questions.  We write down the
optimal $dN/dz$ estimator and show that in very applicable limits,
intuitive formulae describe how well the redshifts of a given source
population can be constrained from a population whose redshift
distribution is better known.  In the limit of a
dense spectroscopic survey, we show that the fractional error in the number of galaxies in the unknown population that fall in spectroscopic redshift bin $z$ can be estimated to the precision
\begin{equation}
  \frac{\delta N(z)}{N(z)} \sim   0.1 \left(\frac{\beta(z)}{0.1} \,  \frac{f_{\rm sky}}{10^{-3}}  \right)^{-1/2} \left( \frac{\ell_0}{10^3} \right)^{-1},
\end{equation}
where $f_{\rm sky}$ is the sky coverage of the survey, $\ell_0$ is the
multipole at which shot noise becomes equal to intrinsic clustering
in either sample, and $\beta(z)$ is
the fraction of the unknown auto-power (at multipoles less than $\ell_0$) that arises from redshift bin $z$.  
  However, the result is even simpler in the limit of a sparse spectroscopic
sample, having fewer than a thousand objects per sq.~deg.~per $\Delta z$: 
\begin{equation}
  \frac{\delta N(z)}{N(z)} \sim \left(\frac{{\cal N}^{(s)}}{10^3}\right)^{-1/2}   \left(\frac{\beta(z)}{0.1} \right)^{-1/2}, 
\end{equation}
where ${\cal N}^{(s)}$ is the total number of spectra per unit redshift.
In this `rare spectroscopic sample' limit, the fractional error on $N(z)$
depends on the total number of spectra but \emph{not} separately on the
density of spectra, the sky area, or the fraction of objects with spectra.

Angular cross-correlations to determine redshifts have applications beyond estimating $dN/dz$.  For example, they could
be used to measure the redshifts of unresolved cosmic infrared background
anisotropies (as was done in \citealt{kashlinsky07}) or to isolate
foregrounds in cosmic microwave background (CMB) and high-redshift
$21\,$cm maps.
Angular cross-correlations can additionally be used to reconstruct three-dimensional correlations from angular clustering measurements
\citep{seljak98, padmanabhan07}.
Furthermore, such cross-correlations are able to calibrate photometric
redshift errors even when the spectroscopic population is not intrinsically
identical to the unknown population.
Applications that are not in the vein of precision cosmology likely need no
better than a $10$ per cent fractional constraint on $dN/dz$.  However, percent-level or even better calibration of photometric redshifts is required to prevent
redshift errors from being the limiting factor for cosmological parameter
estimates with the next generation of weak lensing surveys \citep{huterer06,schneider06, bernstein10, zhang10, cunha12}.\footnote{While
photometric redshifts are object-specific, in practice weak lensing studies
will likely use the statistical distribution from photometric redshifts
owing to catastrophic errors \citep{cunha,mandelbaum08}.
In contrast, cross-correlations are not able to measure the redshifts of
individual objects, but they are another way to measure this statistical
distribution.}

There are a wide range of surveys to which cross-correlation techniques could
be applied.  Recent spectroscopic surveys have gone wide over hundreds
\citep{GAMA} or thousands of square degrees
\citep{eisenstein01,TwodFGRS,WiggleZ,ahn12}
or deep over $\sim 1~$sq.~deg. patches \citep{VVDS,newman12}.
Some are complete to a magnitude limit, whereas others more sparsely sample
the sources \citep{lawrence99, eisenstein01,AGES}.
The large spectroscopic data sets that should be available in the next decade include:\footnote{\url{http://www.sdss.org}, \url{http://www.gama-survey.org}, \url{http://deep.ps.uci.edu}, \url{http://cesam.oamp.fr/vvdsproject/}, \url{http://archive.eso.org/archive/adp/zCOSMOS/VIMOS_spectroscopy_v1.0/}}  
\begin{itemize}
\item the Baryon Oscillation Spectroscopic Survey (BOSS) galaxy sample, covering
$10,000\,{\rm deg}^2$ with $1.5$ million redshifts of massive galaxies
extending to $z\simeq 0.7$ \citep{dawson13}, and {\comment the WiggleZ survey with $240,000$ redshifts over $0.2 < z< 1$ \citep{WiggleZ},}
\item the Sloan Digital Sky Survey (SDSS)+BOSS quasar sample,
covering $10,000\,{\rm deg}^2$ with $2\times 10^5$ redshifts
\citep{schneider10,Shen11,ahn12},
\item the Galaxy and Mass Assembly (GAMA) survey,
covering $310\,$deg$^{2}$ with redshifts for $3.4\times 10^5$ galaxies
to a $z$-band magnitude limit of $19.8$ \citep{GAMA},
\item DEEP2 \citep{newman12},
the VIMOS-Very Large Telescope Deep Survey \citep[VDSS;][]{VVDS},
the $z$-Cosmology Evolution Survey \citep[zCOSMOS;][]{ZCOSMOS}
and, while not technically spectroscopic, COMBO-17; \citep{COMBO17},
each with $\sim 10^4- 10^5$ redshifts in $\sim 1\,$deg$^2$ fields.
\item the HETDEX survey gathering $10^6$ Ly$\alpha$ emitting galaxies over $200\,$deg$^2$ at $1.8 < z < 3.8$ \citep{HETDEX},
\item 21cm emission line surveys over wide fields with e.g.,
the Australian Square Kilometer Array Pathfinder \citep[ASKAP;][]{ASKAP}, which aims for $\sim 10^6$ galaxies to $z\lesssim 0.43$ \citep{duffy12}.
\end{itemize}
The proposed projects eBOSS and BigBOSS would increase the number of spectroscopically
identified galaxies and quasars by an order of magnitude over the existing
SDSS + BOSS samples \citep{BIGBOSS}.\footnote{\url{http://www.sdss3.org/future/eboss.php},
\url{http://bigboss.lbl.gov}}  Ultimately the Square Kilometer Array (projected for 2020) aims to capture a billion galaxies over half of the sky \citep{rawlings04}.

\begin{figure}
\begin{center}
\epsfig{file=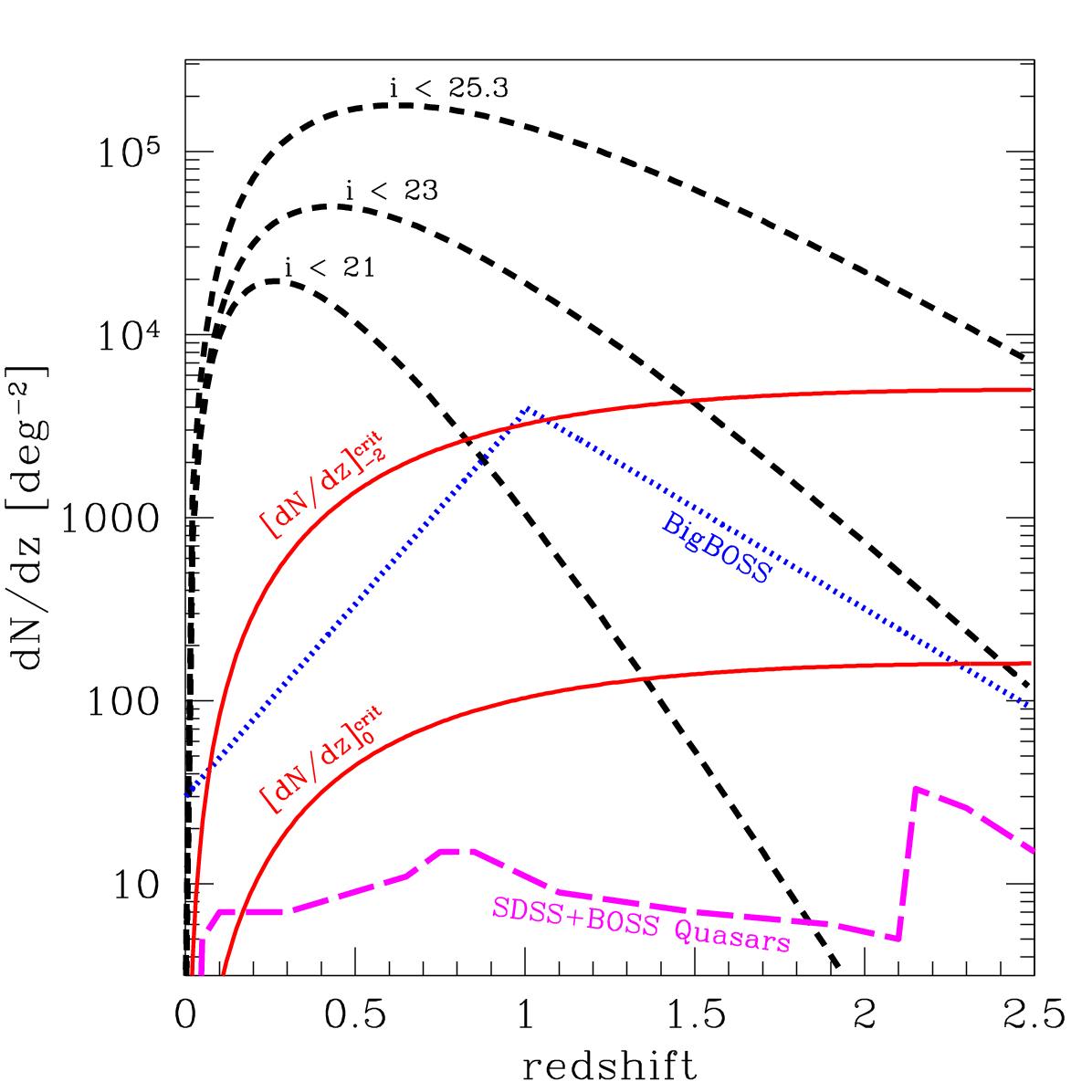, width=8.3cm}
\end{center}
\caption{Shown are the $dN/dz$ of different galaxy populations.  The dashed curves are for surveys complete to i-band magnitude limits of $21,\; 23,$ and $25.3$, calculated via Eq.~(\ref{eqn:pziband}).  Also shown are the density of SDSS+BOSS spectroscopic quasars and estimates for the future combined density of luminous red galaxies, emission line galaxies, and quasars with BigBOSS.  The solid curves represent the critical densities for whether a sample is in the rare galaxy limit (Section \ref{ss:rare}).
\label{fig:different_populations}}
\end{figure}

In addition, we are entering a new age of optical photometric surveys, with
the Kilo Degree Survey (KIDS; $1,500\,$deg$^2$ reaching an i-band magnitude limit of $\rmi=23$),
the Dark Energy Survey (DES; $5,000\,$deg$^2$ to $\rmi=25$) and
the HyperSuprimeCam Project (HSC; $2,000\,$deg$^2$ to $\rmi=26.2$)
all currently gathering data.
These surveys\footnote{\url{ http://kids.strw.leidenuniv.nl/}, \url{http://www.darkenergysurvey.org}, \url{http://www.naoj.org/Projects/HSC/HSCProject.html}, \url{http://www.lsst.org/lsst/}, \url{http://sci.esa.int/euclid}.}
will be followed in the next decade by Large Synoptic Sky Telescope (LSST), which aims to constrain the cosmological model
using a ``gold sample'' of galaxies with $\rmi<25.3$ over half of the sky, and
Euclid, which will provide high-resolution images of galaxies out to $z\sim 2$
over $15,000\,$deg$^2$.
While we do not model in detail any particular survey, we use the above to
guide our discussion.

Fig.~\ref{fig:different_populations} shows characteristic number densities with redshift
for some of the aforementioned spectroscopic surveys as well as for complete
surveys to the specified i-band limiting magnitude.  For these and ensuing
calculations, we have parametrized the galaxy redshift probability
distribution for an i-band magnitude limited sample as
\begin{eqnarray}
 p(z|\; \rmi) &=& \frac{1}{2 \, z_0} \left(\frac{z}{z_0} \right)^2
    \exp \left[ - \frac{z}{z_0} \right], \label{eqn:pziband} \\
  z_0	&=& 0.0417 \; \rmi - 0.74, \nonumber
\end{eqnarray}
with a total angular number density of $1.7\times 10^{5+0.31(i-25)}\,$deg$^{-2}$
(\citealt[``calibrated'' over the range $20.5<\rmi<25.5$, although the deepest data can only constrain $\rmi<23$ and the behavior above this threshold is inferred from mocks from semi-analytic galaxy-formation models applied to the Millennium simulation]{LSSTbook, coil04, hoekstra06};
see also \citealt{efstathiou91,brainerd96,benjamin10,hildebrandt12}).

Cross-correlation techniques can also be applied to maps in the X-ray such as those made with the X-ray Multi-Mirror Mission (XMM-Newton), in the ultraviolet such as with the Galaxy Evolution Explorer (GALEX), and in the infrared such as with the Wide field Infrared Survey Explorer (WISE) and the Herschel Space Observatory, the microwave such as with Atacama Cosmology Telescope (ACT) and the South Pole Telescope (SPT), and the radio such as with ASKAP.\footnote{\url{http://xmm.esac.esa.int}, \url{http://www.galex.caltech.edu}, \url{http://wise.ssl.berkeley.edu}, \url{sci.esa.int/herschel/}, \url{http://www.princeton.edu/act/}, \url{http://pole.uchicago.edu}, \url{http://www.atnf.csiro.au/projects/mira/}}  In many of these surveys, their angular resolution or depth makes redshift identification using overlapping optical surveys difficult.  Cross-correlations offer an independent means to gauge redshifts. 

This paper is organized as follows.
Section \ref{sec:formalism} sets up the formalism
used in this paper and applies it to an idealized $dN/dz$ problem
for illustration.
Section \ref{sec:approx} provides intuition into the mechanics of the
optimal estimator and discusses what scales contain the bulk of the information, setting the ground for the relevant examples discussed in Section \ref{sec:examples}.
Section \ref{sec:configspace} generalizes our Fourier space results to
configuration space and compares our estimator to the more familiar \citet{newman08}
estimator.
Section \ref{sec:bias} quantifies the estimator biases that result from
common simplifying approximations.
Penultimately, Section \ref{sec:photoz} shows how the results of the previous
sections apply to photometric redshift calibrations.  
  Finally, Section \ref{sec:mocks} demonstrates our estimator on mock surveys
and is followed by our conclusions.
We defer some technical details and derivations to a series of appendices,
which are referenced in the text.
The numerical calculations in this study take a flat
$\Lambda$CDM cosmological model with $\Omega_m=0.27$,
$\Omega_\Lambda=0.73$, $h=0.71$, $\sigma_8= 0.82$, $n_s=0.96$, and
$\Omega_b = 0.046$, consistent with recent measurements
\citep{larson11}.  We treat the background cosmology as perfectly known in all calculations.  Roman indices $\{i, \; j, \;k\}$ run from $1$ to some
maximum integer whilst Greek indices start from $0$, and repeated indices that do not appear in the same quantity are summed.  Table~\ref{tab:defs} provides definitions of some commonly appearing symbols.

\begin{table*}
\begin{center}
\begin{tabular}{l l}
 symbol ~~~ & description \\ \hline
 $\alpha_i^{(x)}$ & the faint-end power-law index of the cumulative source number counts of population $x$\\
  $\bfA(\ell, m)$ &  covariance matrix of $p(\ell)$ with $\mathbf{s}(\ell)$ with index $0$ referring to $p$ \\
      $b^{(x)}_i$ & linear bias of population $x$ in redshift bin $i$\\
 $\beta_i(\ell)$ & fraction of the total angular power contributed by redshift bin $i$ (Eq.~\ref{eqn:betadef})\\
  $C_{ij}(\ell)$ & matter density angular cross power spectrum between redshift bins $i$ and $j$ \\
 $\chi$ & the conformal distance; $d\chi = c\,(1+z) \,dt$\\
 $\delta^{(x)}(\ell, m)$ & overdensity in population $x$\\
  $\delta^{\rm K}_{ij}$ & Kronecker delta\\
 $D(z)$ & growth factor such that $D(0) = 1$; $D_i \equiv D(z_i)$\\
 $dN^{(x)}_i/dz$ & equal to $N^{(x)}_i/\Delta z_i$, where the subscript $i$ is dropped if redshift-independent\\
 ${\rm i}^{(x)}$ & i-band limiting magnitude of sample $x$ (assumed complete unless otherwise specified)\\
 $\bfF$ & Fisher matrix; generally $[\bfF^{-1}]_{ii}$ gives error in $z$-bin $i$ \\
  $\bfF^{\rm S}$ & Fisher matrix in Schur-Limber limit (Section~\ref{sec:schurlimber})\\
  $n$ & local power-law index of the density power spectrum such that $P(k) \sim k^{n}$\\
 $N_{\rm bin}$ &  number of redshift bins used in analysis \\
$N^{(x)}_i$ & average sky density in population $x$ in redshift bin $i$; $N^{(x)} = \sum_{i=1}^{N_{\rm bin}} N^{(x)}_i$  \\
 ${\cal N}^{(s)}_i$ & total number of spectroscopic galaxies per unit redshift in redshift bin $i$\\
   $\ell_0$ & multipole where shot noise is equal to cosmic variance\\
     $\ell_{\rm NL}$ & multipole at which linear theory errors at a factor of $2$ (Eq.~\ref{eqn:knl})\\
     $\ell_{PkX}$ & multiple where the logarithmic slope of $P(k)$ has $n=X$ \\
 $p(\ell, m)$ & multipole moment of photometric population \\
$P(k)$ & the $z=0$ linear-theory matter overdensity power spectrum \\
 $\mathbf{s}(\ell, m)$ & vector of multipole moments of spectroscopic $z$-bins
($s_i$ is component in redshift bin $i$)\\
 $S(\ell)$ & the `Schur parameter' (Eq.~\ref{eqn:schur});  $S \geq 1$, with equality holding in the rare limit\\
  $w_i^{(xy)}$ & stochastic component of the cross power between samples $x$ and $y$ in bin $i$; $w^{(x)} \equiv w^{(xx)}$\\
  $w_{ps_i}(\theta)$ & the angular cross correlation function between $p$ and $s_i$ \\
  $W_i(\chi)$ & the window function for redshift bin $i$; typically assumed to be a top hat
\end{tabular}
\end{center}
\caption{Definitions of commonly appearing symbols. The arguments are often dropped in the text, and hats on any symbol indicate an estimated value.}
\label{tab:defs}
\end{table*}

\section{Basic Formalism}
\label{sec:formalism}

We begin by introducing our notation and physical model, before deriving
the most general form for our $dN/dz$ estimator and applying it to idealized,
illustrative examples.  Useful limits of our expressions are
taken in Section \ref{sec:approx}, where we also build intuition for the
mechanics of the estimator.

\subsection{Model and notation}

Initially we will discuss galaxy clustering in the spherical harmonic
basis as our covariance matrix is maximally sparse in this space.
We shall write expressions as if the galaxy samples cover
the full sky, but often finite sky coverage can be included by simply multiplying
by the sky covering fraction ($f_{\rm sky}$).
Section \ref{sec:configestimator} generalizes our estimation methods to
configuration space, while Section \ref{ss:finitesky} discusses the
generalization to finite sky coverage.

We denote the multipole moments of a
`photometric' population of objects with unknown redshifts and a
`spectroscopic' sample in which the redshifts are perfectly known as
\begin{eqnarray}
  p(\ell, m) &=& N^{(p)} \,\delta^{(p)}(\ell, m)
  = \sum_{i=1}^{N_{\rm bin}} N^{(p)}_i \,\delta^{(p)}_i(\ell, m) \label{eqn:p},\\
  s_i(\ell, m) &=& N^{(s)} _i \,\delta^{(s)}_i(\ell, m),
\label{eqn:si}
\end{eqnarray}
respectively.
Here, $1 \leq i \leq N_{\rm bin}$ labels the redshift bin spanning the
range $z_{i-1} - z_i$, where the $z_i$ are ordered in increasing redshift,
and $\delta^{(x)} \equiv x/\langle x \rangle - 1$ is the overdensity in
population $x$, where $x$ denotes an angular source density field with
$\langle x \rangle = N^{(x)}$, the mean density per unit area.
Our calculations are more general than the case of a spectroscopic and
photometric galaxy sample:  the photometric sample can be thought of as any sample
for which the redshifts are unknown and the spectroscopic as
one for which they are known to precision $\Delta z/2$.
Our ultimate aim is to use a survey's estimates for the left-hand-side of
Eqs.~(\ref{eqn:p}) and (\ref{eqn:si}), $\widehat{p}(\ell, m)$ and
$\widehat{s_i}(\ell,m)$, to estimate the $N^{(p)}_i$.

Our discussion will be couched in terms of constraining the
$N_i^{(p)}$ for which the $\Delta z_i$ need to be chosen to be sufficiently narrow in order that there are not significant gradients in $dN^{(p)}/dz$ across the bin.  However, in many cases, particularly when the sensitivity to cross correlations is marginal, a smoother parametrization of $dN^{(p)}/dz$ may be desirable.  Our error estimates can be easily translated into the errors on other parameterizations of $dN^{(p)}/dz$ (like its mean and
variance or the empirically motivated parameterization of a power-law times an exponential;
see Appendix \ref{ap:other_bases} for more details).    

We model the $s_i(\ell, m)$ as Gaussian random variables with auto power
spectrum 
\begin{equation}
\langle s_i s_j \rangle(\ell) =  N^{(s)}_i \, N^{(s)}_j\,  b^{(s)}_i \, b^{(s)}_j \, C_{ij}(\ell) + w_i^{(s)} \delta^{\rm K}_{ij} ,
\label{eqn:sisj}
 \end{equation}
where we have dropped the $m$ dependence as different modes are orthogonal by statistical isotropy but have the same auto-power.  We denote by $C_{ij}$ the cross power between the matter overdensity in the $i$ and $j$ slices, and by $b^{(x)}_i$ the linear bias of population $x$ in redshift bin $i$.  The expression for the shot noise piece $w_{i}^{(s)}$ in the halo model results from
taking the large-scale limit of the one-halo term 
\citep[see e.g.][for a review]{cooray02}:
\begin{equation}
  w_{i}^{(xy)}= \int_{\chi_{i-1}}^{\chi_i} d\chi \int dm_h\, n_h(m_h)\;
    \langle {n}^{(x)}_g {n}^{(y)}_g | m_h \rangle, 
\label{eqn:w}
\end{equation}
where $n_h(m_h)$ is the halo mass function and
$\langle {n}^{(x)}_g {n}^{(y)}_g | m_h \rangle$
is the number of galaxies of type $x$ in a halo of mass $m_h$ times that in
type $y$ and averaged over all haloes at fixed mass.\footnote{The normalization of the stochastic component can potentially be reduced for dense samples by differently weighting sources \citep{seljak09, hamaus10} instead of the galaxy number weighting used here.}
This large-scale limit is a good approximation at the angular scales we consider.  We will also adopt the simplifying notation $w_{i}^{(x)} \equiv w_{ i}^{(xx)}$.  We note that a measurement of the $N^{(p)}_i$ is not limited by sample variance, and it can be perfectly measured in the limit that the stochastic component is zero.

The cross power spectrum of $s_i(\ell)$ and $p(\ell)$ is
\begin{equation}
\langle p \, s_i \rangle (\ell) =
    N^{(s)}_{i} \,b^{(s)}_{i} \sum_{j=1}^{N_{\rm bin}} N^{(p)}_{j}\, b^{(p)}_{j}
    \, C_{i j}(\ell) + w^{(ps)}_{i}.
\label{eqn:psj}
\end{equation}
Finally,\footnote{The total linear bias of the photometric sample is $b^{(p)} = \sum_{i=1}^{N_{\rm bin}} N^{(p)}_{i} b^{(p)}_{i}/N^{(p)}$.
} 
\begin{equation}
 \langle p^2 \rangle (\ell) =  \sum_{i=1}^{N_{\rm bin}} \sum_{j=1}^{N_{\rm bin}}  \left[ N^{(p)}_{i} b^{(p)}_{i} \, N^{(p)}_{j} b^{(p)}_{j}   \, C_{i j}(\ell) + w^{(p)}_{i} \, \delta^{\rm K}_{ij} \right].
\label{eqn:pp}
\end{equation}
We will add to Eqs.~(\ref{eqn:sisj}), (\ref{eqn:psj}) and (\ref{eqn:pp})
the generally smaller terms that owe to cosmic magnification later.

While our formalism is completely general, subsequent calculations (and the
figures we present) assume
\begin{equation}
  b_i^{(x)} = D(z_i)^{-1},
\label{eqn:bias_red_ind}
\end{equation}
where $D(z)$ is the linear growth factor normalized so that $D(0)=1$,
and we will interchangeably use $\chi$ and $z$ for its argument.
This choice leads to redshift-independent clustering, appropriate for several
cosmological populations, especially if they are rare objects.
In many instances this assumption will be benign, and our results can be simply rescaled  by fixing $N^{(x)}_{i} b^{(x)}_{i}$.  
We also assume
\begin{eqnarray}
  w^{(x)}_{i} &=&  \left( \frac{1+3\,f_{\rm sat}^{(x)}}{1+f_{\rm sat}^{(x)}}
                   \right) \, N^{(x)}_{i},\\
 w^{(ps)}_{i} &=& f_{\rm over}\;{\rm min}[w^{(s)}_{i},w^{(p)}_{i}],
\label{eqn:wx}
\end{eqnarray}
for the stochastic component of the power.
We take the `overlap fraction' to be $f_{\rm over} =1$ unless stated otherwise
(which means that the rarest ${\rm min}[N^{(s)}_{i}, N^{(p)}_{i}]$
sources are the same in both samples).
In addition, we take a satellite fraction of $f_{\rm sat}^{(x)}=0$.
Increasing $f_{\rm sat}^{(x)}$ to 25 per cent -- the largest fraction found for
the relevant galaxies in \citet[see their figs. 8 \& 12]{wetzel10} -- does
not change our results appreciably.\footnote{In the case of $f_{\rm over}=1$
and equal numbers in both the $s$ and $p$ samples, 
both populations trace the same large-scale cosmological plus stochastic perturbations and the
$N^{(p)}_{i}$ can be perfectly estimated.} 
 
The cross power in the matter overdensity is
\begin{eqnarray}
  C_{ij}(\ell) &=& \int_0^\infty \frac{2\,k^2 dk}{\pi}
  \ \alpha_\ell(k,z_i) \alpha_\ell(k,z_j)
  \, P(k),  \label{Cij}\\
  \alpha_\ell(k, z_i) &=& \int_0^\infty d\chi\;D(\chi)\;W_i(\chi)
  \;j_\ell(k \chi), \label{alphaij}
\end{eqnarray}
where, in our top hat $N^{(p)}_{i}$ bias, $W_i=\Delta \chi_i^{-1}$ for redshifts that fall in the range
$z_{i-1} - z_i$ and zero otherwise.  (For a discussion of how to evaluate
$j_\ell$ and these highly-oscillatory integrals over $j_\ell$ numerically see
Appendix \ref{ap:jayell}.)
While not required, we have assumed linear theory such that $P(k)$ is
the $z=0$ linear-theory matter overdensity power spectrum.
Eq.~(\ref{Cij}) ignores redshift space distortions (RSDs).
RSDs contribute a small fraction to the angular fluctuations on relevant
angular scales, {\comment with a larger impact on the fluctuations in the spectroscopic sample compared to the photometric} (Appendix~\ref{ap:rsd}).  

We note that linear scales can only be used to reconstruct
the \emph{product} of the large-scale bias, $b_i^{(p)}$, and the number density,
$N_i^{(p)}$, at any redshift
\citep{newman08,bernstein10,schulz10} as they always appear in combination.  
This product is sometimes the desired quantity (e.g., when cleaning a map
of diffuse backgrounds), but for many applications it is $N_i^{(p)}$
itself that is desired.
We discuss methods for breaking this degeneracy in Section \ref{sec:bias_number}.  {\it We will often write our constraints as on $N_i^{(p)}$ for notational simplicity, but please note that the constraints we quote are always on the combination $b^{(p)}_i N_i^{(p)}$}.

{\comment
Recently, \citet{menard13} advocated using nonlinear scales ($<1\,$proper~Mpc)
to constrain the $N_i^{(p)}$.
In fact, most of the constraint from the \citet{menard13} method appears to
derive from $<300\,$proper~kpc \citep{schmidt13}, scales that are likely to
reside within halos.
While small-scale measurements have the advantage that they can be applied to
data sets even if there are significant calibration problems \citep{menard13},
on nonlinear scales it is less clear how to map cross-correlation amplitude
to the redshift distribution of a population.  This is especially true on
intra-halo scales, as the correlations depend on how the two samples inhabit
the same halos\footnote{If there is significant evolution in the overlap of
the samples with redshift (or the size of halos), this method will lead to
artificial trends in the $N_i^{(p)}$ inferences.
There also may be pathological cases where two populations do not significantly
overlap (such as in the early and late type galaxies models considered in
\citealt{rossAJ09}), which would greatly impact small scale measurements
while having minimal impact on large scales.}.
We shall not use nonlinear scales for our estimator.
}

\subsection{Estimator} \label{sec:estimator}

To simplify notation, we define the combined covariance matrix of the photometric survey and the redshift slices of the spectroscopic survey: 
\begin{equation}
\bfA(\ell,m) \equiv \left \langle \left( \begin{array}{c}   \widehat{p}(\ell,m)^*\\  \widehat{\bmath{s}}(\ell,m)^* \end{array} \right) \,  \left(\widehat{p}(\ell,m) ~~\widehat{\bmath{s}}(\ell,m) \right) \right \rangle, 
\end{equation}
where $\widehat{\bmath{s}}^T = (\widehat{s}_1, \cdots, \widehat{s}_n)$ and {\comment note that $\bfA = \langle \widehat{\bfA} \rangle$}.  
The argument $(\ell,m)$ will typically be dropped in subsequent expressions.
The minimum variance estimator for $N^{(p)}_{i}$ that maximizes the likelihood function if it is Gaussian in this parameter near the maximum (as is likely if many modes are included in the estimate) is 
\begin{eqnarray}
\widehat{N}^{(p)}_{i} &=& [\widehat{N}^{(p)}_{i}]_{\rm last} +  \frac{1}{2} \, [{\bfF}^{-1}]_{ij}  \sum_{\ell, m}  \bigg[\left( \begin{array}{cc}  \widehat{p}  & \widehat{\bmath{s}}\end{array} \right) \bfQ_{j}   \left( \begin{array}{c}  \widehat{p} \\  \widehat{\bmath{s}}\end{array} \right) \nonumber \\
&-&  {\rm Tr}[\bfA^{-1} \bfA_{,j}] \bigg],\label{eqn:MQest}\\
{\bfQ}_{j}  &\equiv &  \sum_{\ell, m}  \bfA^{-1} \bfA_{,j} \bfA^{-1}, \label{eqn:Q}
\end{eqnarray}
(e.g. \citealt{bond98, tegmark98, dodelson}),
where all repeated indices are summed and subscript `$,i$' indicates
a derivative with respect to the $i^{\rm th}$ parameter, which for most of our discussion is the parameter $N_i^{(p)}$.  The parameter $[\widehat{N}^{(p)}_{i}]_{\rm last}$ is initially a guess and,
for subsequent iterations, the previous estimate.  {\comment In addition, the $[\widehat{N}^{(p)}_{i}]_{\rm last}$ appear in the $\bfA$ in the next iteration.  Despite this we do not include hats on the $\bfA$ (a slight notational
inconsistency).}
One can also trivially recast the estimated quantity in Eq.~(\ref{eqn:MQest}) to be $b_i^{(p)} N_i^{(p)}$ rather than $N_i^{(p)}$, since $b_i^{(p)} N_i^{(p)}$ is what is truly constrained.  Appendix \ref{ap:estwithprior} derives Eqs.~(\ref{eqn:MQest}) and (\ref{eqn:Q})
and shows how they generalize to the case with priors on the ${N}^{(p)}_{i}$.

In the limit that many modes are included in the estimate (which is appropriate; Appendix~\ref{ap:fullestimator}), 
\begin{equation}
F_{ij} = \frac{1}{2} \sum_{\ell, m}{\rm Tr} \left[\bfA^{-1} \, \bfA_{,i}  \,\bfA^{-1} \bfA_{, j} \right],
\label{eqn:fish1}
\end{equation}
and $\bfF$ is the Fisher matrix.  The estimator in this limit is the minimum variance quadratic estimator, and the variance of this estimator is $[\bfF^{-1}]_{ii}$ (e.g., \citealt{tegmark97}).  We will use Eq.~(\ref{eqn:fish1}) in our subsequent calculations.   

\citet{schulz10} and \citet{matthews12} considered a maximum likelihood estimator approach to constrain the ${N}^{(p)}_{i}$, at least for their most general expressions.  This approach should yield similar estimates to ours as the Fisher matrix, which sets our variance, saturates the Rao-Cramer bound (and so is optimal).  In fact, quadratic estimators are prone to find local extrema and so a
Markov Chain Monte Carlo approach to find the maximum likelihood may yield more robust estimates (e.g., \citealt{christensen01}).  However, the linearity of our estimator reduces the severity of this problem, and we show in Section \ref{sec:mocks} that it robustly finds the true minimum even when the initial guess for the $N^{(p)}_{i}$ is off by orders of magnitude. 
 
It is worth noting two subtleties in our approach:
First, we do not consider estimators for the $N^{(p)}_{i}$ that simultaneously
estimate the $w^{(ps)}_{i}$, although this would be a small generalization
of Eq.~(\ref{eqn:MQest}).
Instead, we assume that the $w^{(ps)}_{i}$ can be measured independently from the $N^{(p)}_{i}$, which should hold because of the much different scaling
of the cosmological and stochastic components in the $\langle p \, s_i \rangle$.
Larger $\ell$ can also be utilized for the $w^{(ps)}_{i}$ estimate
than are useful for constraining the $N^{(p)}_{i}$.
Secondly, our expressions do not consider the case in which the true
value for $N^{(s)}_i$ differs from the measured number density owing to
large-scale modes on the scale of the survey.
Such an error will be most important in narrow fields.
One can take this effect into account by using the measured number in a
prior on the field to field fluctuations and then marginalizing over the $N^{(s)}_i$ (Appendix \ref{ap:estwithprior}). 

\subsection{Idealized application}

\begin{figure}
\epsfig{file=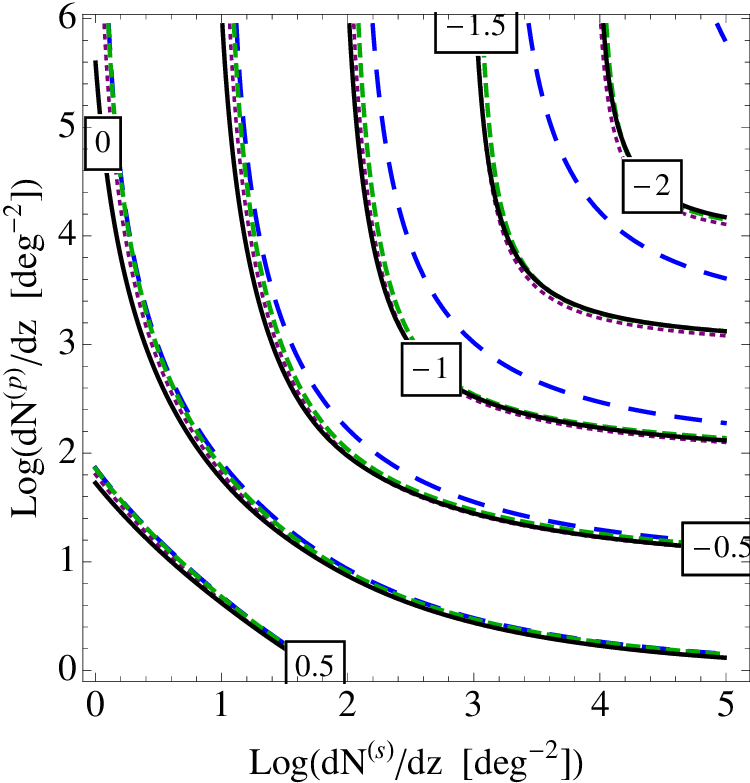, width=8.0cm}
\caption{{\comment The fractional error on the photometric number density for different spectroscopic and photometric samples.  The contours} represent $\log_{10}$ of the fractional error on $N_{i}^{(p)}$ with $i = N_{\rm bin}/2$.  They consider an idealized survey in which the $N_i^{(x)}$ are equal and span $z= 0-1$ with $10$ redshift bins of the same width, covering 1 per cent of the sky ($400\,$deg$^2$).  Contours are labelled for the solid curves, and the corresponding contour for the other curves is the adjacent curve at higher number densities. The calculations assume our fiducial parameters except $f_{\rm over} = 0$.  (For $f_{\rm over} = 1$, the curves buckle outwards when the number densities become equal.)  The black solid curves are the sensitivity of the optimal estimator.  The purple dotted curves show the approximation that sets to zero terms in $\bfF$ in which the derivatives hit $A_{00}$.   The short dashed green is the diagonal approximation to the remaining Fisher matrix, a limit that also works excellently.  The long dashed blue is the error on the estimator in the Schur-Limber limit (Section \ref{sec:schurlimber} and Eq.~\ref{eqn:fishC00large}).  
\label{fig:Covplots}}
\end{figure}

\begin{figure*}
\epsfig{file=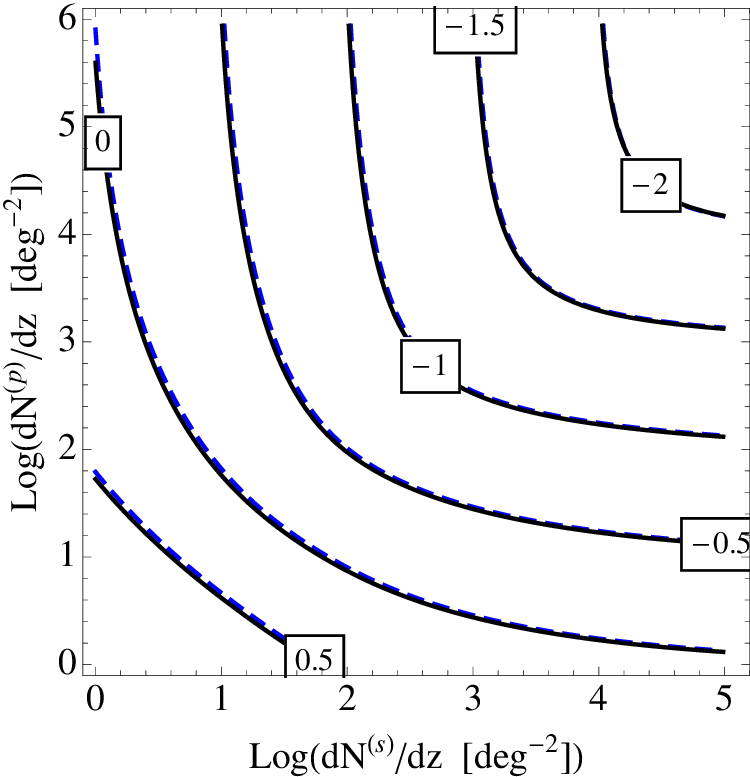, width=7cm} \epsfig{file=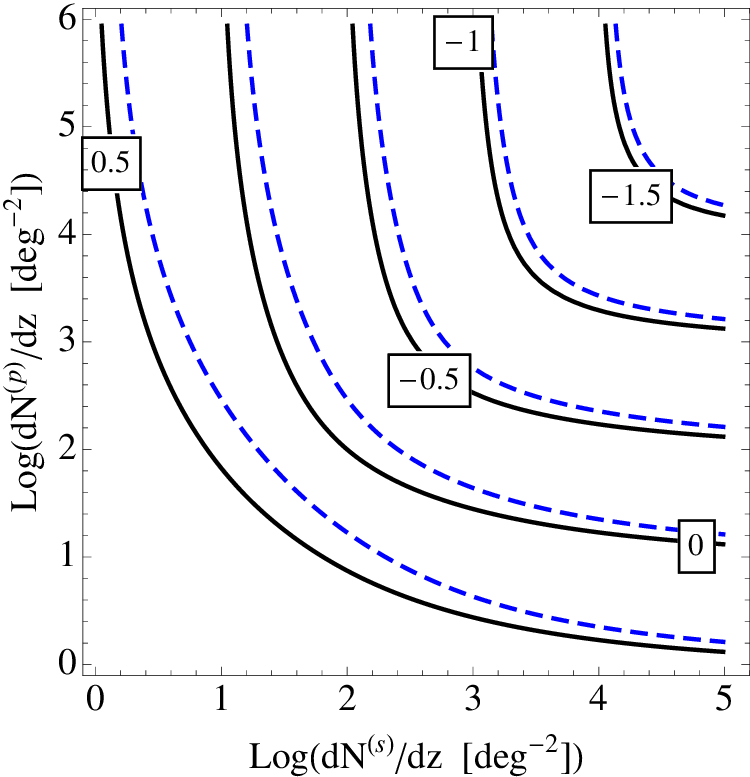, width=7cm}
\caption{Contours showing $\log_{10}$ of the fractional error in $b_{i}^{(p)}N_{i}^{(p)}$, where $i = N_{\rm bin}/2$ in the Limber approximation (black solid curves) and the full calculation without this approximation (blue dashed curves; which for the same fractional error fall immediately upwards of the solid curves).  The contours are calculated for a survey that spans $z= 0-1$ with $10$ (left panel) and $100$ (right panel) redshift bins of equal width over 1 per cent of the sky.  Roughly, the errors are $\sqrt{10}$ larger in the right panel than in the left panel.  This figure illustrates that the Limber approximation works well for the $\Delta z = 0.1$ case, but is starting to break down at $\Delta z = 0.01$.  While making the Limber approximation leads to errors in the uncertainty estimate, we find in Section~\ref{sec:bias} that the bias on $N^{(p)}_i$ is always quite small.  \label{fig:limber_impact}}
\end{figure*}

Eq.~(\ref{eqn:fish1}) allows us to estimate
the sensitivity of a hypothetical survey. The solid curves in
Fig.~\ref{fig:Covplots} show these estimates for an idealized case in which the
$N_i^{(x)}$ are equal, have redshift-independent clustering (see Eq.~\ref{eqn:bias_red_ind}), and span the redshift range $0-1$ with $10$
redshift bins.
The curves represent contours of constant sensitivity on the parameter
 $b_{i}^{(p)} N_{i}^{(p)}$ where $i = N_{\rm bin}/2$ (i.e., the fractional error on the bias times the angular number density of photometric objects in the fifth redshift bin)
as a function of the $dN^{(p)}/dz$ and $dN^{(s)}/dz$ used in the cross
correlations.  The labels on the black solid curves are $\log_{10}$ of the fractional error.
The solid curves in the right panel of Fig.~\ref{fig:limber_impact} are the same except
assuming a survey in which $z=0-1$ is spanned with $100$ redshift bins,
which approximately results in $\sqrt{10}$ larger errors.
The other contours in both figures show different approximations that are
developed in Section \ref{sec:approx}.
All of the curves are computed for a fractional sky coverage of
$f_{\rm sky} = 0.01$, but the errors scale as $f_{\rm sky}^{-1/2}$ for
surveys with areas $\gg 1\,$deg$^2$ (Section \ref{ss:finitesky}). 

While the contours in Figs.~\ref{fig:Covplots} and
\ref{fig:limber_impact} are for the simplistic case of constant
$dN^{(p)}/dz$ and $dN^{(s)}/dz$, they illustrate a few of our results.
First, the sensitivity to the $N^{(p)}_i$ saturates once either the
photometric or spectroscopic $dN/dz$ becomes larger than the other.
 Secondly, the contours show that percent-level constraints for $\Delta z = 0.1$ are possible for number densities of $dN^{(s)}/dz \sim dN^{(p)}/dz \sim 10^3\,$deg$^{-2}$
if $10$ per cent of the sky is utilized.

We find that the calculations in Figs.~\ref{fig:Covplots} and
\ref{fig:limber_impact} can be crudely applied beyond the assumption of constant
$dN^{(p)}/dz$, of constant $dN^{(s)}/dz$, or of the redshift at which they were
computed.  For example, if these calculations are used to estimate the sensitivity of the
LSST gold sample, which will have $dN^{(p)}/dz \sim 10^5\,$deg$^{-2}$ over
a quarter of the sky \citep{LSSTbook}, one finds that percent-level
determinations of the $N^{(p)}_i$ are possible in $\Delta z \sim 0.1$ bins
with spectroscopic follow up of $dN^{(s)}/dz\sim 10^3\,$deg$^{-2}$
(comparable to the sky density of BigBOSS emission line galaxies).
This estimate is consistent with the conclusions of more detailed calculations
in Section \ref{sec:examples}.
Also, the LEGACY plus the ongoing BOSS quasar samples on SDSS provide a
spectroscopic number density of $dN^{(s)}/dz\sim 10\,$deg$^{-2}$ out to
$z\approx 2.7$ over $\sim 10^4\,$deg$^{-2}$
(with double this number density at $z\sim2.3$; \citealt{ahn12}).
Fig.~\ref{fig:Covplots} suggests that cross-correlations with denser
photometric surveys should provide $\sim 10$ per cent errors on their
$N^{(p)}_i$ in $\Delta z = 0.1$ for $f_{\rm sky} \sim 0.1$, again consistent
with what we find later on.

We now turn to building intuition for the estimator presented in
Section \ref{sec:estimator}.

\section{Approximations and special cases}
\label{sec:approx}

In this section, we provide an understanding of the shape of the
contours in Figs.~\ref{fig:Covplots} and \ref{fig:limber_impact}, we discuss which scales contribute the $N_i^{(p)}$ estimate, and we provide intuitive formulae that can be quickly applied to gauge the utility of cross correlating different samples.  

\subsection{The Limber approximation}
\label{sec:limber}

If the theoretical power spectrum is smooth and our signal is coming
primarily from scales which are small compared to the width of each
redshift shell, then the Limber approximation applies
\citep{Limber53,Limber54} and our expressions simplify significantly.
The Limber approximation assumes that $P(\bfk_\perp,k_\parallel)$
varies slowly as a function of $k_\parallel$ compared to
$j_\ell(k_\parallel \chi)$ -- which should hold when
$\ell \gg \chi/\Delta \chi_i$.  Making use of the identity
\begin{equation}
  \int k^2 dk\ j_\ell (k \chi) j_\ell (k \chi') =\frac{\pi}{2 \chi^2} \delta^D(\chi -\chi'),
\end{equation}
where $\delta^D$ is the Dirac delta function, and the Limber approximation, $C_{ij}(\ell)$ -- Eq.~(\ref{Cij}) -- becomes diagonal \citep{kaiser92,whitehu00}
 \begin{eqnarray}
  C_{ij}(\ell) &=& \delta_{ij}^{\rm K} \, \,\int_0^\infty d\chi  \, D^2(\chi) \, W_i^2(\chi)
   \frac{P(\ell/\chi)}{\chi^2}, \\
  &\approx& \delta_{ij}^{\rm K} \,D^2(z_i)\, \frac{P(\ell/\chi)}{\chi^2 \,\Delta \chi_i},
\label{eqn:limberapprox}
\end{eqnarray}
{\comment where $\delta_{ij}^{\rm K}$ is the Kronecker delta.}
We discuss how the Limber limit is approached and compute the corrections owing
to RSDs in Appendix \ref{ap:rsd} (where we show that RSDs enter at ${\cal O}([\ell \,\Delta \chi/\chi]^{-2})$ in the photometric sample, which means they contribute negligibly on scales where the Limber approximation applies).

The majority of past studies
\citep{newman08, matthews10, schneider06}
have used the Limber approximation.  Fig.~\ref{fig:limber_impact} shows that this approximation provides
a good estimate for the variance of our $N^{(p)}_i$ estimator, with only a small error in the case of
$\Delta z = 0.1$ (left panel) and the error starting to become significant for $\Delta z = 0.01$ (right panel).
In both panels, compare the solid contours, which assume Limber, with the dashed contours,
which do not.  The Limber approximation is accurate because, as we will show,
much of the estimator's constraint derives from $\ell$ where it should hold. (The percent-level bias introduced by this approximation is quantified in
Section \ref{sec:bias}.)

The covariance matrix of the photometric and spectroscopic surveys simplifies
considerably in the Limber approximation, with only the $A_{0 \alpha}$ terms
and the diagonal components of $A_{i j}$ being nonzero, namely
\begin{eqnarray}
A_{00} &=& \sum_{i=1}^{N_{\rm bin}}
    \left(b^{(p)}_i N_i^{(p)}\right)^2 C_{ii} + w_i^{(p)}, \\
A_{0i} &=& b^{(p)}_i N_i^{(p)} b^{(s)}_i N_i^{(s)} C_{ii} + w_i^{(ps)}, \\
A_{ij} &=& \delta_{ij}^{\rm K} \;
 \left[ \left( b^{(s)}_i N_i^{(s)}\right)^2 C_{ii} + w_i^{(s)} \right],\\
 \left [A_{0i} \right]_{,i}  &=&  b^{(p)}_i  b^{(s)}_i N_i^{(s)} C_{ii}.
\end{eqnarray}
Furthermore, this $\bfA (\ell,m)$ can be inverted analytically, yielding
\begin{eqnarray}
  {[\bfA^{-1}]}_{00} &=& \frac{S}{A_{00}} ,\label{eqn:C00}\\
  {[\bfA^{-1}]}_{0i} &=& - \frac{S}{A_{00}} \, \frac{A_{0i}}{A_{ii}}
  ~~~~~~~~~~~= - \frac{S \, r_i^2}{A_{0i}}, \label{eqn:C0i}\\
  {[\bfA^{-1}]}_{ij} &=& \frac{\delta_{ij}^{\rm K}}{A_{ii}} + \frac{S}{A_{00}}
  \, \frac{A_{0i}A_{0j}}{A_{ii}A_{jj}}
  = \frac{\delta_{ij}^{\rm K}}{A_{ii}} +
  S \sqrt{\frac{r_i^2 r_j^2}{A_{ii}A_{jj}}} , \label{eqn:Cij}
\end{eqnarray}
with
\begin{equation}
  S = A_{00} \left(A_{00} - \sum_{i=1}^{N_{\rm bin}} \frac{A_{0 i}^2}{A_{ii}}\right)^{-1}
  = \left(1 - \sum_{i=1}^{N_{\rm bin}} r_i^2 \right)^{-1},
  \label{eqn:schur}
\end{equation}
where $r_i(\ell)  \equiv A_{0i}/(A_{00} \, A_{ii})^{1/2}$ is the cross
correlation coefficient between $p$ and $s_i$, and again we are using the
convention $i,\,j \in 1-N_{\rm bin}$.  
The above inverse can be derived using the Schur complement matrix identity
and the Woodbury formula (e.g., \citealt{matrixcookbook}). 

The `Schur parameter', $S$, is greater than or equal to unity and quantifies
the extent of correlation between the spectroscopic and photometric samples.
In the case of complete redshift overlap of the spectroscopic sample and in
the absence of shot-noise, $S\rightarrow\infty$ and the $N^{(p)}_i$
are perfectly constrained.
If the unknown sample is limited by shot-noise, or if the two samples
cover different redshift ranges, $S\rightarrow 1^{+}$.
The implication is that even a small amount of noise diminishes considerably
the constraining power of a mode.

In the analytic derivations that follow, we ignore derivatives that hit the $A_{00}$ in
Eqs.~(\ref{eqn:MQest}) and (\ref{eqn:fish1}), as this element provides only an integral-like
constraint on the $N^{(p)}_i$.
For all relevant limits, the approximation of ignoring the
$A_{00}$-derivatives is excellent:  Fig.~\ref{fig:Covplots} compares
the solid black error contours, which include the $A_{00}$-derivatives,
with the nearly-overlapping dotted purple contours, which do not.  With this additional simplification, the Limber-approximation Fisher matrix (Eq.~\ref{eqn:fish1}) is
\begin{eqnarray}
F_{ij} &\approx&  \sum_{\ell, \; m} \left( {[\bfA^{-1}]}_{ij} {[\bfA^{-1}]}_{00} +  {[\bfA^{-1}]}_{0i} {[\bfA^{-1}]}_{0j}  \right) [A_{0i}]_{,i}  [A_{0j}]_{,j}, \nonumber \\
&=&   \sum_{\ell, m} \frac{S}{A_{00}} \left(\frac{\delta^K_{ij}}{A_{ii}} + 2\,S \, \sqrt{\frac{r_i^2 r_j^2}{A_{ii} \, A_{jj}}} \right)  \, [A_{0i}]_{,i} \; [A_{0j}]_{,j}.
\label{eqn:limber_fish}\label{eqn:limber_fish1}
\end{eqnarray}
Furthermore, the minimum variance quadratic estimator becomes\footnote{We thank Andrew Johnson and Chris Blake for pointing out an error in an earlier version of this expression.  See also \citet{johnsoninprep}.}
\begin{eqnarray}
\widehat{N}^{(p)}_{k}  &=&   [\widehat{N}^{(p)}_{k}]_{\rm last} + [\bfF^{-1}]_{k i} \nonumber \\
& \times &  \sum_{\ell, m} \frac{S  [A_{0i}]_{,i}}{A_{00} A_{ii}} \, \Bigg \{ \left(\delta_{ij}^{\rm K} + 2 S \frac{A_{0i} \,A_{0j} }{A_{00} \, A_{jj}} \right)  \, \left( \widehat{p} \; \widehat{s}_j -A_{0j} \right) \nonumber \\
&& -   \frac{A_{0d}}{A_{dd}} \left(\delta_{ij}^{\rm K} +  {\frac{ S A_{0i} A_{0j}}{ A_{00} A_{jj}}} \right) \, \left(\widehat{s}_d \, \widehat{s}_j - A_{jj} \delta^{\rm K}_{dj} \right) \nonumber \\
   &&-  \frac{S \, A_{0i}}{A_{00}} \, \left( \widehat{p}\,^2 -A_{00} \right)   \Bigg\}, \label{eqn:minquad}
\end{eqnarray}
where repeated indices that do not appear in the same quantity are summed.  Note that since we have dropped the terms that include derivatives of $A_{00}$, the trace term in equation~(\ref{eqn:Q}) must be slightly altered to recover the unbiased estimator given by equation~(\ref{eqn:minquad}).  However, it is trivial to make the estimator unbiased by imposing that the estimator averages to $\widehat{N}^{(p)}_{k}$ when $\bfA =\hat \bfA$.  

\begin{figure}
\resizebox{8cm}{!}{\includegraphics{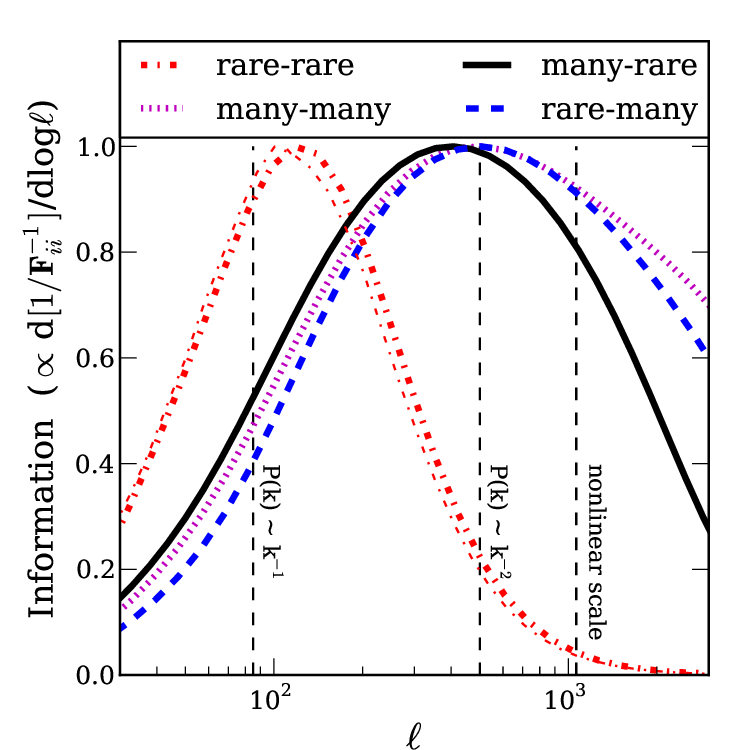}}
\caption{An illustration of the scales that contribute to the constraint on $N^{(p)}_i$ in different limits.  The areas under these curves, which are of $d[1/\bfF^{-1}_{ii}]/d\log \ell$, are proportional to the information that contributes to the estimate in the $i=6$ bin for a measurement in $10$ redshift bins with $\Delta z = 0.1$ and spanning $0 <z<1$.  For illustrative purposes, we have assumed constant $dN^{(p)}/dz$ and $dN^{(s)}/dz$.   The first adjective for each curve's label in the key describes the spectroscopic sample (rare=$10\,$deg$^{-2}$ and many=$10^{5}\,$deg$^{-2}$), and the second describes the photometric sample (rare=$100\,$deg$^{-2}$ and many=$10^{6}\,$deg$^{-2}$).  However, the curves are not significantly impacted at linear scales by the assumed densities as long as `many' equates to $\gtrsim 10^4\,$deg$^{-2}$ and `rare' to $\lesssim 10^3\,$deg$^{-2}$, with the exception being the many-many case.  In the text we describe why these limits select the scales that they do.  The vertical lines denote significant scales discussed in the text.  The thin red dot-dashed curve does not assume the Limber approximation whereas the corresponding thick curve assumes it.
\label{fig:information}}
\end{figure}

\begin{figure}
\epsfig{file=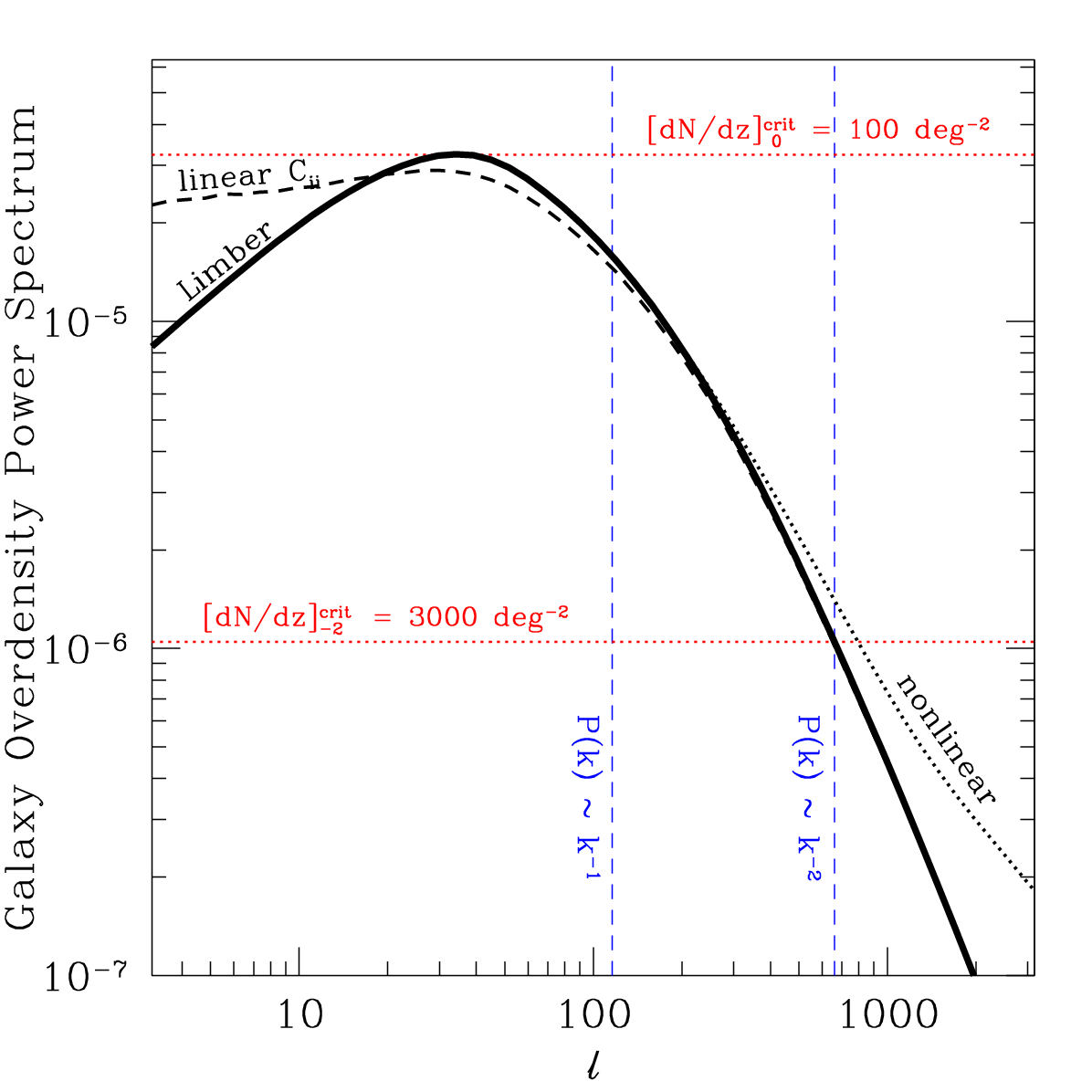, width=8.5cm}
\caption{{\comment The source clustering angular power spectrum under different approximations and for different source number densities.  Shown is the clustered component of the power,} $C_{ii}$, for $z_i=1$, $\Delta z_i = 0.1$, and our fiducial bias model.  The $C_{ii}$ are calculated under various approximations -- linear theory (dashed curve) and the Limber approximation (solid curves) -- and for the full \citet{peacockdodds} nonlinear power spectrum (dotted curve).  Also depicted are the stochastic component of the power for two characteristic number densities and  $f_{\rm sat} = 0$ (horizontal dashed lines).  The auto-power of spectroscopic bin $i$, $\langle s_i^2\rangle$, equals $C_{ii}$ plus the stochastic component.  The optimal quadratic estimator selects information that roughly falls in the range of the two vertical dotted lines (Section \ref{sec:approx}), between where $P(k)$ roughly scales as $k^{-1}$ and $k^{-2}$.  Conveniently, both linear theory and the Limber approximation apply around these scales.
\label{fig:Pkillustration}}
\end{figure}

\begin{figure}
\epsfig{file=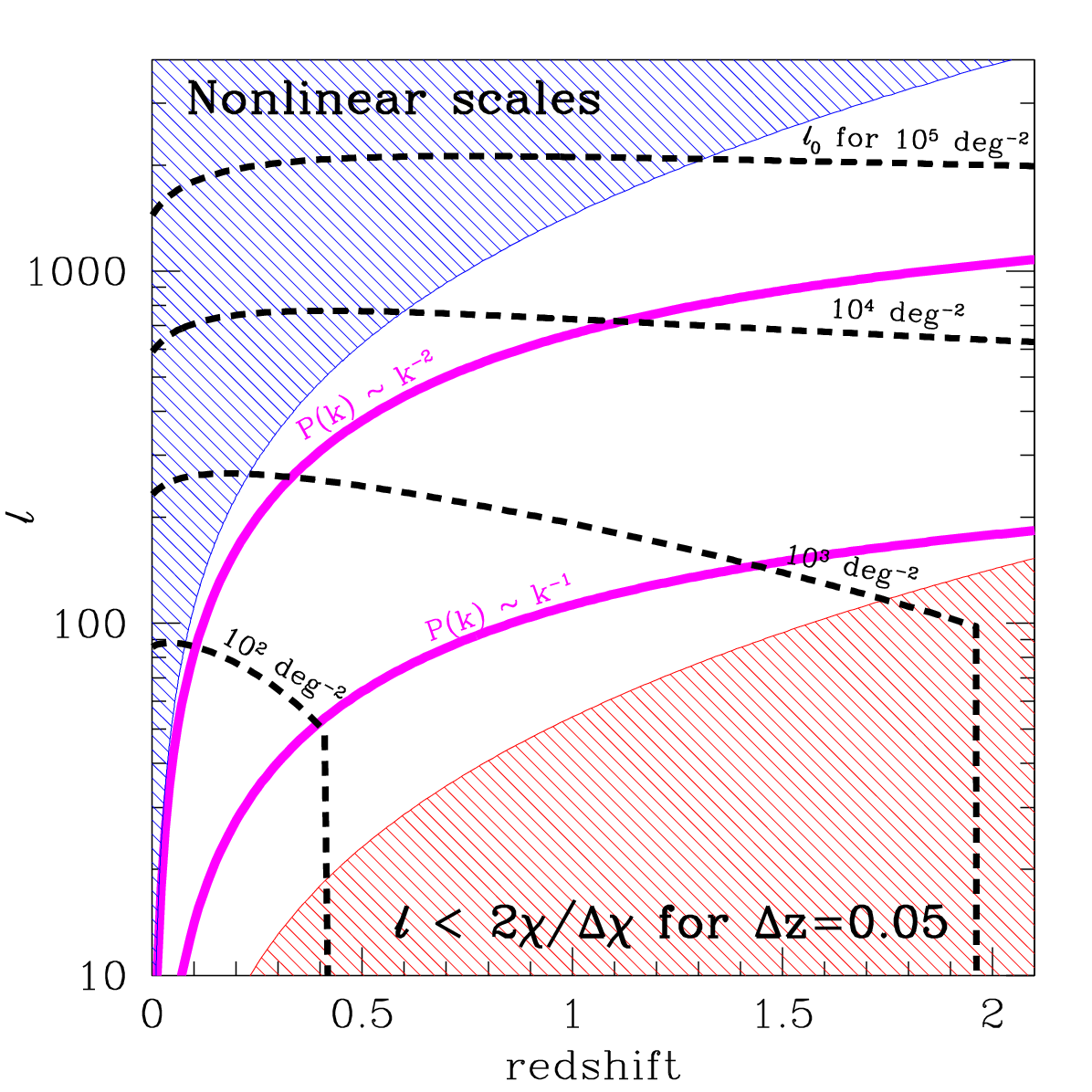, width=8.5cm}
\caption{Shown are characteristic $\ell$ values for cross-correlation analyses as a function of $z$.  The lower (red) shaded region delineates $\ell \leq 2 \chi(z)/ \Delta \chi$ for $\Delta z = 0.05$, approximately where the Limber approximation errors at $\sim 10\%$ (Appendix \ref{ap:limber}; Eq.~\ref{eqn:limberm2}).  The upper (blue) shaded region is where deviations from linear perturbation theory are a factor of $\geq 2$.  The other curves show characteristic scales at which $dN^{(p)}/dz$ estimates receive the bulk of their information.   The dashed curves show the multipole where the Poisson term is equal to the clustering term, which we denote as $\ell_0$, for surveys with number densities of $b^2 \, dN/dz = 10^2, \,10^3, \, 10^4$, and $10^5\,$deg$^{-2}$.   The magenta curves are the scales where the density power spectrum has power-law index $-2$ and $-1$, $\ell_{Pk-2}$ and $\ell_{Pk-1}$, respectively.  The optimal estimator applied to two rare samples (where rare is defined as having $\ell_0 \lesssim \ell_{Pk-1}$) utilizes modes with $\ell \sim \ell_{Pk-1}$ to constrain $dN^{(p)}/dz$.  However, rare and abundant samples use modes with $\ell \sim \ell_{Pk-2}$, whereas if both samples are abundant the estimate comes from $\ell \sim \ell_0$ in certain cases {\comment (unless windows are applied to e.g. downweight nonlinear scales).}  
\label{fig:lmax}}
\end{figure}

Figs.~\ref{fig:information}, \ref{fig:Pkillustration} and \ref{fig:lmax}
motivate why the approximations of Limber and linear theory are justified.
Fig.~\ref{fig:information} shows the scales that contribute to the estimator
for several different cases, plotting $d[1/\bfF^{-1}_{ii}]/d\log \ell$.
The areas under these curves are proportional to the information that
contributes to the estimate in the $i=6$ bin for a measurement in $10$
redshift bins with $\Delta z=0.1$ spanning $0<z<1$.
The first adjective for each curve's label in the key describes the
spectroscopic sample
(rare=$10\,$deg$^{-2}$ per $dz$ and many=$10^{5}\,$deg$^{-2}$ per $dz$)
and the second describes the photometric sample
(rare=$100\,$deg$^{-2}$ per $dz$ and many=$10^{6}\,$deg$^{-2}$ per $dz$),
where these number densities are assumed constant with redshift for
illustration.
This figure indicates that (at least for these extremities of the
parameter space) the bulk of the information derives from modes around
where the density power spectrum has power-law index $-2$ and $-1$,
$\ell_{Pk-2}$ and $\ell_{Pk-1}$, respectively.
As we shall discuss further, correlations between two rare samples
(where rare is defined as having $\ell_0\lesssim\ell_{Pk-1}$)
constrain $N^{(p)}_i$ primarily from multipoles with $\ell\sim\ell_{Pk-1}$.
Rare and abundant samples use multiples with $\ell\sim\ell_{Pk-2}$, which also holds in the case in which both samples are extremely abundant. 
It is also possible in less extreme examples (in which both samples are relatively abundant) for the
information to derive primarily from the scale $\ell_0$.

To orient the reader, Fig.~\ref{fig:Pkillustration} shows estimates for the
$C_{ii}$ at $z=1$ and for $\Delta z=0.1$ that use linear theory, the Limber
approximation, and the \citet{peacockdodds} nonlinear power spectrum.
The vertical lines show $\ell_{Pk-2}$ and $\ell_{Pk-1}$.
$\ell_0$ is the scale at which the (horizontal) stochastic power becomes
equal to the $C_{ii}$, i.e.~where the red dotted lines intersect the black solid curve.
We show the stochastic terms for two illustrative number densities.  In particular, the upper horizontal line in Fig.~\ref{fig:Pkillustration} is the lowest number density at which
$w^{(s)}_i > [b_i^{(s)} \, N^{(s)}_i]^2 \, C_{ii}$ is satisfied at \emph{all} $\ell$, which we denote as $\left[\frac{dN}{dz}\right]^{\rm crit}_{0}$, where
\begin{equation}
\left[\frac{dN}{dz}\right]^{\rm crit}_{0} \simeq {300} \, b^{-2}
\, \left(\frac{1+z}{2} \right)^{1.8} \quad{\rm deg}^{-2}.
\label{eqn:Ncrit0.01}
\end{equation}
Eq.~\ref{eqn:Ncrit0.01} uses the Limber approximation, takes
$f_{\rm sat}^{(s)} = 0$, and approximates the redshift dependence as a
power-law evaluated at $z=1$.  In addition, the lower horizontal line is the number density at which $w^{(s)}_i = [b_i^{(s)} \, N^{(s)}_i]^2 \, C_{ii}(\ell_{Pk-2})$, or
\begin{equation}
  \left[\frac{dN}{dz}\right]^{\rm crit}_{-2} \simeq {8000}  \, b^{-2}
  \, \left(\frac{1+z}{2} \right)^{1.8} \quad {\rm deg}^{-2}.
\label{eqn:Ncrit0.2}
\end{equation}
Both critical number densities are shown in Fig.~\ref{fig:Pkillustration} for our fiducial bias model.  We return to the significance of these numbers in future sections.

We often will approximate the scale at which linear theory no longer holds as
\begin{equation}
  k_{\rm NL} \simeq  0.25 \, (1+z)\ \ {\rm Mpc}^{-1},
\label{eqn:knl}
\end{equation}
which we find is close to the scale in which the \citet{peacockdodds}
nonlinear density power spectrum overshoots linear theory by a factor of $2$
for the redshifts of interest.
We define $\ell_{\rm NL} \equiv \chi \, k_{\rm NL}$, which is plotted in Figs.~\ref{fig:information} and \ref{fig:lmax} and throughout as the limit of
validity of our assumptions.
Fig. \ref{fig:lmax} shows that $\ell_0$ falls in the range in which both linear theory and the
Limber approximation more or less apply across all relevant redshifts and number densities.
Linear theory also applies for $\ell_{Pk-1}$ and (more approximately) $\ell_{Pk-2}$.  We note that
$\ell_{Pk-1}$ [$\ell_{Pk-2}$] corresponds to a transverse physical scale of $k\simeq 0.03\,\Mpc^{-1}$ [$k=0.2\,\Mpc^{-1}$] (Table \ref{tab:neff}).

\begin{table}
\begin{center}
\begin{tabular}{cc|cc|cc}
 $k$  & $n_{\rm eff}$ & $k$ & $n_{\rm eff}$ & $k$ & $n_{\rm eff}$ \\ \hline
 0.01 & 0.05 & 0.1 & -1.7 & 1 & -2.4\\
 0.02 & -0.7 & 0.2 & -2.0 & 2 & -2.5\\
 0.05 & -1.3 & 0.5 & -2.3 & 5 & -2.7
\end{tabular}
\end{center}
\caption{The instantaneous power-law slope of the $\Lambda$CDM linear
theory power spectrum as a function of
wavenumber, $k$, in Mpc$^{-1}$ ($n_{\rm eff} \equiv d\log P/d \log k$).  The values were computed using the \citet{eisenstein98} matter transfer function without baryon acoustic features and for the fiducial cosmological parameters.}
\label{tab:neff}
\end{table}

\subsection{The Schur-Limber limit}
\label{sec:schurlimber}

We now investigate the above Limber-approximation estimator in the limit
$S(\ell)\rightarrow 1^+$ and show that a small tweak to this limit
captures almost all of the information in the general case.  We refer
to the $S\rightarrow 1^+$ limit as the `Schur limit' henceforth.  In
this limit the information originates from modes where 
$\sum_i r_i^2 \ll 1$, either because of incomplete overlap of the spectroscopic
survey or because shot noise is important.  In many interesting cases
this limit at least marginally holds.  Importantly, both $\bfA$ and
$\bfF$ are diagonal in the Schur limit, viz
\begin{eqnarray}
 F_{ij}^S &\approx& \sum_{\ell, m} \frac{[A_{0i}]_{,i}^2}{A_{00} \,  A_{ii}  } \delta^{K}_{ij},\label{eqn:fishC00large}
\end{eqnarray}  
where the superscript $S$ denotes the Schur limit.  Furthermore, the estimator becomes
\begin{eqnarray}
  \widehat{N}^{(p)}_{i} &=&   [\widehat{N}^{(p)}_{i}]_{\rm last} +  \frac{1}{F_{i i}^S} \sum_{\ell, m} \, \frac{[A_{0i}]_{,i}}{A_{00} \, A_{ii}}  \left \{  \widehat{p} \; \widehat{s}_i  -  A_{0i}  \right \}, \label{est:diag}
\end{eqnarray}
such that the number density in each bin is
now estimated independently and is proportional to the cross-power,
$\widehat{p} \,\widehat{s}_i$, minus a constant.  The Schur-Limit approximation yields the long-dashed blue curves for the errors
on the ${N}^{(p)}_{i}$ shown in Fig.~\ref{fig:Covplots}.  These
trace the contours in the full calculation (compare with the
solid contours) at $dN/dz \lesssim 10^3~$deg$^{-2}$, but deviate if both samples have higher number densities, as is
expected.

Three notes in passing: (1) The structure of $F^S$ is 
reminiscent of the optimal weight in the \citet{FKP} definition
of the effective volume.  While our expression is in harmonic space, the structure 
has the form $\left[ \bar{n}P/(1+\bar{n}P)\right]^2$ just as in
\citet{FKP}.  This is not surprising as our estimator is asking a similar question to ``What is the significance that the cross power can be detected?'' {\comment (2)  It is simple to show that the Schur-Limber estimator has the same error as fitting the amplitude of the cross power as done in \citet{ho08} to constrain the redshift distribution of the NVSS catalogue.}  (3) The Schur-Limber estimator is exact in the limits where Limber holds and $S=1$, and does not require dropping certain derivative terms as was required to derive Eq. (\ref{eqn:minquad}).\\

To see how the Schur-Limber estimator works, we take the case in which a single $\ell, m$ mode contributes to the estimate such that
\begin{equation}
  \widehat{N}^{(p)} =  [\widehat{N}^{(p)}_i]_{\rm last}  +
\frac{\widehat{p}\,\widehat{s}_i-A_{0i}}{b^{(p)}_i\,N^{(s)}_i\,b^{(s)}_i\,C_{ii}}.
\label{eqn:singleell}
\end{equation}
If the true $N^{(p)}_i$ differs from the fiducial model,
$[N^{(p)}_i]_{\rm last}$, by $\delta N_i^{(p)}$, we have the relations
\begin{equation}
  \widehat{p}\,\widehat{s_i} = \left([N^{(p)}_i]_{\rm last}+
    \delta N^{(p)}_i\right)
    N^{(s)}_i b^{(s)}_i b^{(p)}_i C_{ii}^{\rm data} + w^{(ps)}_i,
\end{equation}
where $C_{ii}^{\rm data}$ is the actual density power in this harmonic, and 
\begin{equation}
  A_{0i} = [N^{(p)}_i]_{\rm last} \times N^{(s)}_i b^{(s)}_i b^{(p)}_i C_{ii}
  + w^{(ps)}_i.
\end{equation}
Plugging these into Eq.~(\ref{eqn:singleell}) yields
\begin{equation}
  \left\langle \widehat{N}^{(p)}_i \right\rangle
  = [N^{(p)}_i]_{\rm last} +  \delta N^{(p)}_i  = N^{(p)}_i,
\end{equation}
noting that $\langle C_{ii}^{\rm data}(\ell, m) \rangle = C_{ii}$.  
  Thus, the iteration converges in a single step, and
the estimate is unchanged with subsequent iterations.  The former is
no longer the case when multiple $\ell$ are used in the estimate, but
we show in Section \ref{sec:mocks} that the estimator still converges in just a few
iterations.\\

The structure of the formula for the Fisher matrix in this Schur limit (Eq.~\ref{eqn:fishC00large})
is also quite simple, and is most easily brought out by considering the
case where the underlying power spectrum is a power-law, $C_{ii}=c_i\ell^n$:
\begin{equation}
\bfF^{S}_{ij} =  [{N}^{(p)}_i]^{-2} \sum_{\ell, m} \frac{c^{(p)}_i c^{(s)}_i \ell^{2n}\ \delta^{K}_{ij}}{(c^{(p)} \, \ell^{n} + w^{(p)}) \, (c_i^{(s)} \, \ell^{n} + w^{(s)}_i)},
\label{eqn:fishC00largeapprox}
\end{equation}     
where we have written $c^{(x)}_i=[N^{(x)}_i b^{(x)}_i]^2\,c_i$ and
$c^{(p)} = \sum_i c^{(p)}_i$.
The CDM case can often locally be thought of a power-law where the spectrum has a power-law index 
which becomes increasingly negative towards smaller scales
(see Table \ref{tab:neff}).
{\comment Eq.~(\ref{eqn:fishC00largeapprox}) -- which we remind the reader is valid in the Schur-Limber limit -- provides intuition into the shape of
the contours in Fig.~\ref{fig:Covplots}.  In particular, we now focus on three sub-limits that bracket different regimes for the densities of galaxies being correlated.}

\subsection{Abundant galaxy limit}
\label{ss:abundant}

At $\ell$ where neither the photometric nor the spectroscopic survey is limited
by shot noise, all $\ell$ contribute equally and the argument in the sum in
Eq.~(\ref{eqn:fishC00largeapprox}) is roughly constant in $\ell$.
However, once shot noise becomes appreciable for either survey ($\ell >\ell_0$), the
argument in the sum scales as $\ell^n$.
At scales where $n<-2$, which becomes increasingly satisfied at smaller
scales with CDM spectra (see Table \ref{tab:neff}), this scaling cuts
off the sum as shells of increasing $\ell$ contribute progressively less
to $\bfF$.
If $n>-2$, this is not true, and there is information until scales
where both surveys are limited by shot noise (or $n$ has steepened).
This explanation is reflected by the contours in Fig.~\ref{fig:Covplots}.
For number densities where $\ell_0$ occurs at scales at which $n < -2$ ($dN/dz > 8000 \, b^{-2}\,$deg$^{-2}$),
information is gained all the way until $\ell \sim \ell_0$.
In this case, the contours are very boxy and Eq.~(\ref{eqn:fishC00large}) can be approximated as being clustering dominated at $\ell < \ell_0$ and being $0$ at $\ell > \ell_0$:
\begin{equation}
\frac{\delta N^{(p)}_i}{N^{(p)}_i} \equiv \frac{\sqrt{[\bfF^{S \,-1}]_{ii}}}{N^{(p)}_i} \sim \left( \langle \beta_{i} \rangle \,  f_{\rm sky}  \,[\ell_0^2 - \ell_{\rm min}^2] \right)^{-1/2},\label{eqn:lmax}
\end{equation}
where $\ell_{\rm min}$ is the minimum wavenumber used, and $\langle
\beta_i \rangle$ is the $\ell$-averaged fraction of the angular power
in the photometric sample that comes from $z$-bin $i$:
\begin{equation}
  \beta_i \equiv \frac{[N^{(p)}_i b^{(p)}_i]^2 \,C_{ii}(\ell, m)}{\sum_{j=1}^{N_{\rm bin}} [N^{(p)}_j b^{(p)}_j]^2 \, C_{jj}(\ell, m)}.
\label{eqn:betadef}
\end{equation}
For the simple case of slices of fixed number and distant observers (i.e., $\chi$ not changing appreciably across the sample), $\langle \beta \rangle \sim N_{\rm bin}^{-1}$.  The left panel in Fig.~\ref{fig:Constn} shows how the sensitivity is increased with increasing $dN^{(p)}/dz$, fixing the photometric population (here a survey complete to $\rmi = 23$) and the survey area. It shows that the prediction of Eq.~(\ref{eqn:betadef}) of a number density--independent error comes into full effect at
$dN^{(s)}/dz > 10^5\,$deg$^{-2}$, which is on par with the maximum
number densities that for medium-future experiments (see Fig.~\ref{fig:different_populations}).  Values of the Schur parameter greater than unity (Eq.~\ref{eqn:lmax} sets $S=1$) result in some number density dependence even at high $dN^{(s)}/dz$.\footnote{In fact, Eq.~(\ref{eqn:lmax}) should be regarded as an upper bound on the error since
we set $S=1$.
When $S$ is large (and here we take $w^{(s)}_i > w^{(p)}_i$ and
$w^{(s)}_i > w^{(sp)}_i$, although similar conclusions apply regardless),
$S\propto \sum_{i=1}^{N_{\rm bin}} C_{ii}/w^{(s)}_i$.
Including $S$ in the summation in Eq.~(\ref{eqn:fishC00largeapprox})
makes the kernel peak at $\ell_{Pk-2}$ for high number densities rather than $\ell_0$.  This results in the many-many case peaking at $\ell_{Pk-2}$ in Fig.~\ref{fig:information}.  However, the constraint on $dN^{(p)}/dz$ only improves by a factor of $\sim 2$
for physically realizable number densities when accounting for $S \neq 1$ (as can be gleaned by comparing the Schur estimator's error -- the long-dashed blue curve -- to the full estimator's error -- the solid black curve -- at high densities in Fig.~\ref{fig:Covplots}). 
}  Also, evaluating Eq.~(\ref{eqn:lmax}) for parameters that match the case given in the left panel of Fig.~\ref{fig:Constn} -- $\ell_0 = 2000$ (see Fig. \ref{fig:lmax}), $\beta = 0.1$, and $100~$deg$^2$ -- yields $\delta N/N = 0.03$, which is comparable to the values for the largest $dN^{(s)}/dz$ in this plot.

We have used linear theory in our computations, but scales with $\ell>\ell_{\rm NL}$ should not be used in our formalism.  Hence, a large enough patch of sky must be chosen to sample $\ell<\ell_{\rm NL}$ such that cross correlations are fruitful.  
Evaluating Eq.~(\ref{eqn:lmax}) with $\ell_0\rightarrow \ell_{\rm NL}\sim 10^3$
implies that a square degree is required for cross-correlations to provide an
${\cal O}(1)$ constraint on $dN^{(p)}/dz$ with our method.

\subsection{Rare spectroscopic sample}
\label{ss:rare}

\begin{figure*}
\epsfig{file=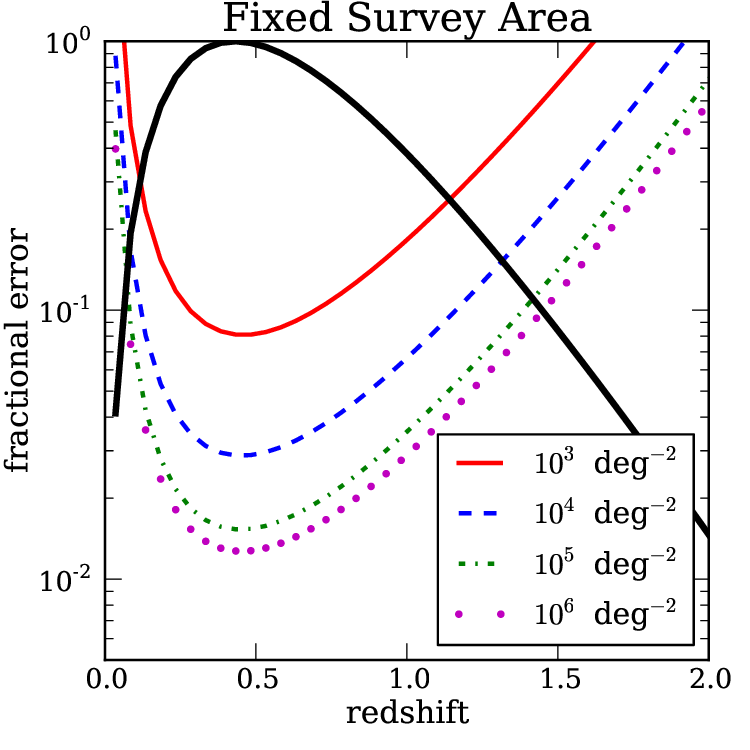, width=5.7cm}
\epsfig{file=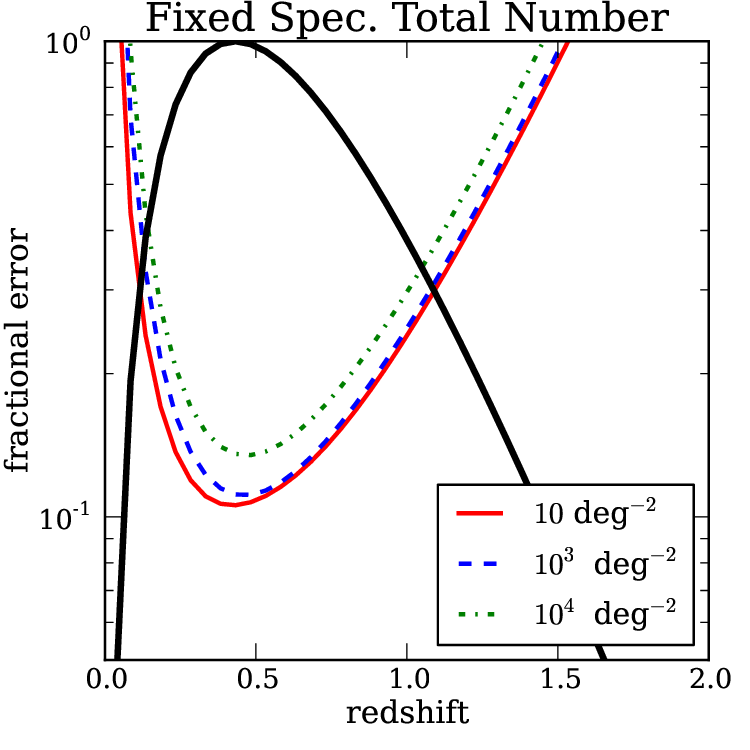, width=5.7cm}
\epsfig{file=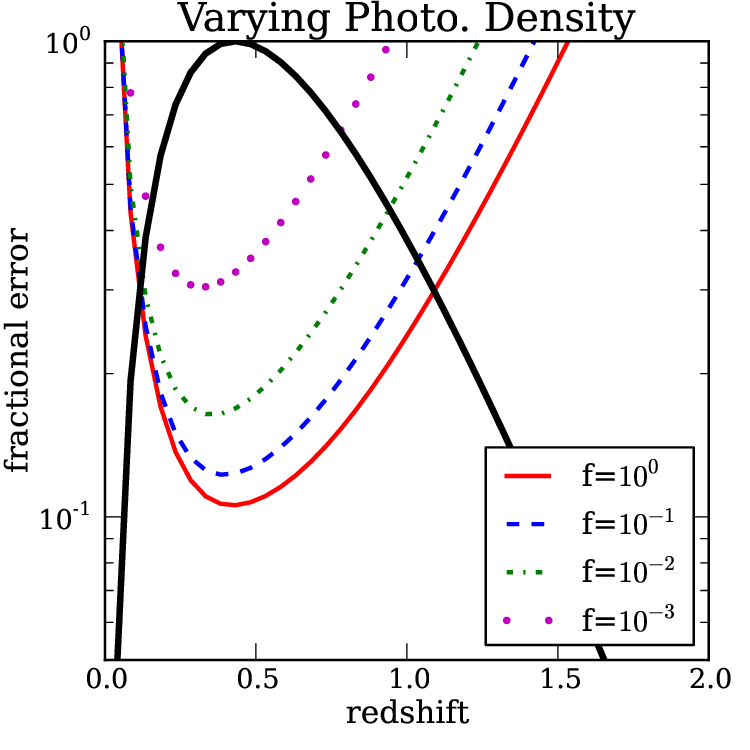, width=5.7cm}
\caption{Illustration of how the {\comment fractional constraints from cross correlations on the photometric sample's bias times number,} $b^{(p)}_i \, N^{(p)}_i$, depend on area, total number, and densities of the samples.  All panels take $\Delta z = 0.05$, a photometric sample down to a limiting
  magnitude of $\rmi^{(p)} =23$ (and $100$ per cent complete except in the right panel), and a spectroscopic sample for which $dN^{(s)}/dz$ is a constant out to $z=2$.  The $p(z|\rmi)$ of the photometric sample is given by the thick
  solid curve. Left panel:  The curves assume a $100\,$deg$^2$ survey and the specified $dN^{(s)}/dz$.  At lower $dN^{(s)}/dz$ the sensitivity improves as the square root of $dN^{(s)}/dz$, as anticipated in the rare-spectra limit, but at high densities the sensitivity does not depend on depth, as anticipated by our abundant limit.   Middle panel: The
  three curves show a spectroscopic sample with fixed total of $10^5$ galaxies and the specified sky densities.  
  {The similarity of the sensitivity between these much different densities demonstrates our analytic result that in the rare tracer limit the fractional error
  scales as the total number of spectroscopic galaxies.}  Right panel:  Varying the fraction, $f$, of photometric galaxies that are used with a spectroscopic sample with angular density $10\,$deg$^{-2}$, and $10^5$ spectroscopic galaxies.  In the limit in which both the photometric and spectroscopic samples are rare, the fractional sensitivity scales as $f^{-1/2}$.
\label{fig:Constn}}
\end{figure*}

Another relevant limit of the Schur-Limber estimator is when the spectroscopic
sample is sparse enough that it is dominated by shot noise.  In this limit, the Schur approximation ($S\approx1$) is always justified, and
our equations simplify further so that the Fisher
matrix becomes
\begin{equation}
F_{ij} = N^{(s)}_i \sum_{\ell, m}
  \frac{[b^{(p)}_i \, \ b^{(s)}_i\, C_{ii}]^2\ \delta_{ij}^K}
          {\sum_k (b^{(p)}_k N^{(p)}_k)^2 C_{kk} + w^{(p)}_k}
  \propto N^{(s)}_i f_{\rm sky}
  \label{eqn:rarefish}
\end{equation}
for $f^{(x)}_{\rm sat} = 0$.
Thus, in this limit the error on the $N^{(p)}_i$ scales as the
total number of spectra -- it does not depend on the
density of spectroscopic sources.  It turns out that in many relevant
cases cross-correlations will be in this regime (as discussed in
Section \ref{sec:examples}).  

What $dN^{(s)}/dz$ are required to be in the rare limit?
If $dN^{(s)}/dz < \left[\frac{dN}{dz}\right]^{\rm crit}_{0}$, or roughly a hundred per square degree (Eq.~\ref{eqn:Ncrit0.01}), the sparse tracer
limit certainly holds as the shot component always dominates.  However, for even much larger number densities, we find that the rare spectroscopic limit is a good approximation.  The Fisher information at each $\ell$ for a rare spectroscopic sample
(but an abundant photometric sample) keeps increasing until $\ell_{Pk-2}$
(as $dF^S_{ii}/d\log \ell \propto \ell^{n +2}$ so that the contribution to $F^S_{ii}$
decreases in bins of $\log\ell$ once $n<-2$).
  Thus, to be in the rare limit, it is less important that shot noise dominate
at all $\ell$ and more important that shot noise dominates by $\ell_{Pk-2}$.
Therefore, once $dN^{(s)}/dz < [{dN}/{dz}]^{\rm crit}_{-2}$ (see Eq.~\ref{eqn:Ncrit0.2})
the rare limit applies, and the constraint on the $N^{(p)}_i$ only depends
on the total number of galaxies.

The middle panel in Fig.~\ref{fig:Constn} tests this argument.  
  It plots the
constraints on $b^{(p)}_i \, N^{(p)}_i$ for a photometric sample down to a limiting magnitude of
$\rmi=23$, assuming $\Delta z= 0.05$.  The three curves each take a spectroscopic sample comprised of $10^5$
galaxies and differing $dN^{(s)}/dz$, where $dN^{(s)}/dz$ is taken to be
constant up to $z=2$ as specified in the figure key.
Thus, the three curves represent surveys with the same number of
spectroscopic galaxies.  The sensitivity changes negligibly with increasing number
until $10^4\,{\rm deg}^{-2}$ {\comment (or roughly $[{dN}/{dz}]^{\rm crit}_{-2}$), in agreement with the argument that the constraint depends only on the total number of spectroscopic galaxies at low densities.}

The middle panel in Fig.~\ref{fig:Constn}, combined with our argument that $\delta N^{(p)}_i \propto [{\cal N}^{(s)}_{i}]^{-1/2}$, where ${\cal N}^{(s)}_{i}$ is the total number of spectroscopic galaxies per unit redshift, suggests that a minimum of $\sim 10^3$ spectroscopic galaxies are
needed to have an order unity constraint on $b^{(p)}_i N^{(p)}_i$ (and somewhat fewer if the population is more
localized in redshift than in our example or if they
are more strongly clustered than in our fiducial model).
That $\sim 10^3$ spectroscopic galaxies are required is also apparent from evaluating
Eq.~(\ref{eqn:rarefish}) in the limits of an abundant photometric and rare spectroscopic survey, which yields
\begin{equation}
  \frac{\delta N^{(p)}_i}{N^{(p)}} \approx  \frac{0.6}{b_i^{(s)} D_i} \, \left(\frac{{\cal N}^{(s)}_{i}}{10^3} \, \frac{\langle \beta_i \rangle_C}{0.1} \right)^{-1/2}   \left(\frac{1 +z}{2}\right)^{-0.5},
  \label{eqn:rareNplimit1}
\end{equation}
where we have assumed bins of fixed $\Delta z$, $\langle \beta_i \rangle_C$ is defined analogously to $\langle \beta_i \rangle$ but weighted by $C_{ii}$, and the redshift factor owes to how lengths map to angles and redshift intervals with $z$ (which we evaluated at $z=1$, but this formula holds to $20$ per cent for $0.1 <z <3$).

\subsection{Rare-rare limit}

The final limit we consider is when the fluctuations in both samples are dominated by shot noise.  In this limit, $dF^S_{ii}/d\log \ell \propto \ell^{2 \, n +2}$ such that the contribution to $F^S_{ii}$ decreases in bins of $\log \ell$ once $n<-1$.  As with the abundant--rare limit previously considered, we can also evaluate Eq.~(\ref{eqn:fishC00large}) in the rare-rare limit, which yields
\begin{equation}
  \frac{\delta N^{(p)}_i}{N^{(p)}}\approx  \frac{1.7}{b_i^{(s)} b_i^{(p)} D_i^2} \left(\frac{{\cal N}^{(s)}_{i}}{10^3} \, \frac{dN^{(p)}_i/dz}{10^2\,{\rm deg^{-2}}} \, \frac{f_{i}}{0.1} \right)^{-1/2} \left(\frac{1 +z}{2}\right)^{0.4},
  \label{eqn:rareNplimit2}
\end{equation}
where $f_i$ is the fraction of the photometric galaxies in redshift bin $i$
(and equals the distant observer $\beta_i$ in the case of redshift independent
clustering).
This expression shows that at a minimum
\begin{equation}
 {\cal N}^{(s)}_{i}\times dN^{(p)}_i/dz \gtrsim 10^6\,{\rm ~~~deg^{-2}}
 \end{equation}
  is required for cross-correlations to be fruitful.
The right panel in Fig.~\ref{fig:Constn} shows the constraints on the $N^{(p)}_i$, again with the specifications  $\rmi^{(p)}=23$ and $10^5$ total spectroscopic galaxies, but taking $dN^{(s)}/dz =10\,$deg$^{-2}$ for all the curves and assuming that only a fraction, $f$, of photometric galaxies are used in the cross correlations.
When both the photometric and spectroscopic galaxies are in the rare limit,
Eq.~(\ref{eqn:rareNplimit2}) shows that the sensitivity scales as $f^{-1/2}$.
We note that the peak of $dN/dz$ for a survey complete to $\rmi =23$ equals
$5\times10^4\,$deg$^{-2}$, so the $f\lesssim 0.01$ curves should be in this limit, and we indeed find this scaling in this regime. This panel illustrates that cross-correlations can be used to constrain the redshift distribution of peculiar objects, comprising a part in $10^3$ of the photometric sample in the case shown, and not just of the full sample.

The derivations that led to Eq.~(\ref{eqn:rareNplimit2}) implicitly assumed that the bias of the spectroscopic sample is known from
auto-correlation function measurements.
However, in the limit of a rare spectroscopic sample, the auto correlations
can be much noisier than the cross correlations, calling into question this assumption.
We show in Appendix \ref{ap:estwithprior} that in this case the fractional variance on the  ${\cal N}^{(p)}_{i}$ is simply the fractional variance quoted
in this section added to the fractional variance in the bias measurement.

Because the two limits given by Eqs.~(\ref{eqn:rareNplimit1}) and
(\ref{eqn:rareNplimit2}) yield similar ${\delta N(z)}/{N(z)}$ at
the transition between the two regimes
(at $dN^{(p)}/dz\sim  0.1\,[dN^{(p)}/dz]^{\rm crit}_{-2}$),
the sensitivity of an arbitrary photometric survey can be estimated by
interpolating between them.

\subsection{Generalizing the Schur Limit}
\label{sec:generalizedschur}

We showed that in the Schur-Limber limit the Fisher matrix is
diagonal.  However, empirically we find that the inverse of the full Fisher matrix of the minimum variance quadratic estimator is quite diagonal and is well approximated by the inverse of $\sum_{\ell, m} S \, \bfF^S(\ell)$ (i.e., to ignoring the off diagonal elements in $\bfF$).
This is illustrated by the dashed green
contours in Fig.~\ref{fig:Covplots}, which show the variance calculated with this expression for $\bfF^{-1}$.

The approximation of ignoring off diagonals when computing the
estimator variance from $\bfF$ is equivalent to not marginalizing over
parameters other than $N^{(p)}_i$.  That $\bfF^{-1}$ is approximately
diagonal thus means that one does not have to simultaneously estimate each
of the $[\widehat{N}^{(p)}_{i}]$ and rather can estimate each
parameter independently for $[\widehat{N}^{(p)}_{i}]_{\rm last}$ near
the peak of the likelihood.

\section{Applications}
\label{sec:examples}

\begin{figure}
\epsfig{file=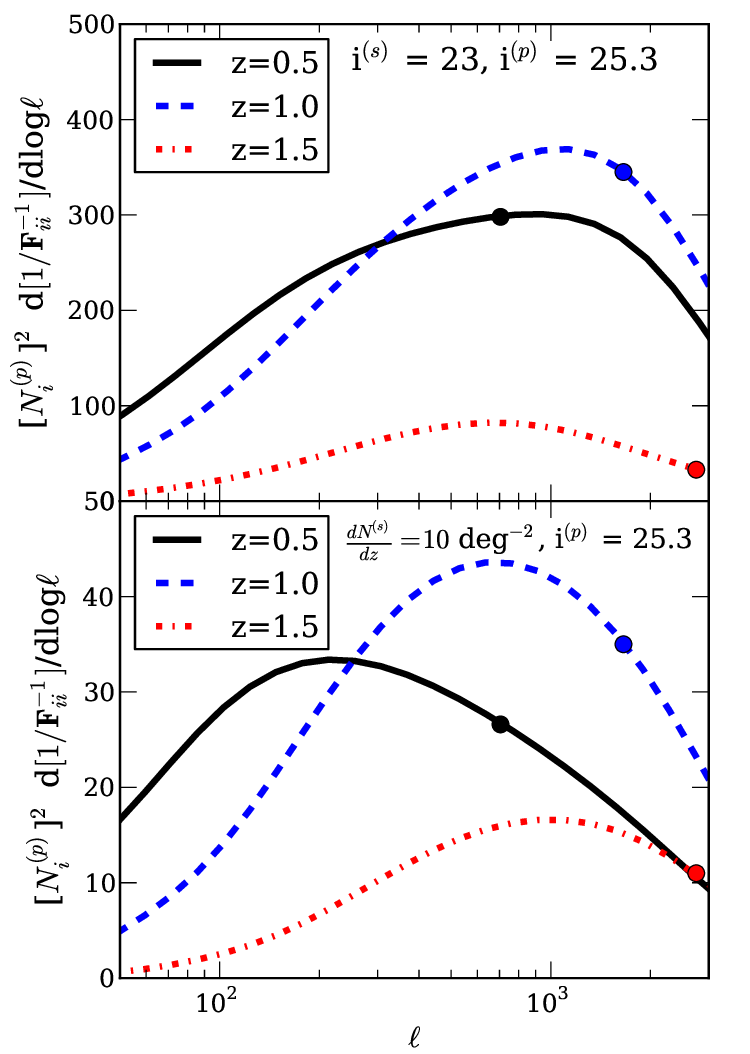, width=8cm}
\caption{Similar to Fig.~\ref{fig:information}, but for {\comment physically motivated galaxy samples}.  Plotted is the information as a function of $\ell$ {\comment that contributes to the $dN^{(p)}_i/dz$ estimate}, $d[1/\bfF^{-1}_{ii}]/d\log \ell$.  The variance in $N^{(p)}_i$ is the inverse of the area under these curves.  The filled circles show $\ell_{\rm NL} \equiv \chi \, k_{\rm NL} $ (c.f., Eq.~\ref{eqn:knl}).  The top (bottom) panel considers a $40\,$deg$^2$ ( $10^4\,$deg$^2$) survey and takes bins of $\Delta z=0.05$ spanning $0<z<2.5$.
\label{fig:info_example}}
\end{figure}

The previous section built intuition for the behavior of the estimator.  To bring out the appropriate limits we considered
simple $dN/dz$ distributions, such as constants.
This section considers more physically motivated parameterizations for the
extragalactic populations.
Fig.~\ref{fig:info_example} is analogous to Fig.~\ref{fig:information} but
quantifies the scales that contribute to the constraint on the $N^{(p)}_i$ for
\emph{realistic} source models, plotting $d[1/\bfF^{-1}_{ii}]/d\log \ell$.
In particular, Fig.~\ref{fig:info_example} considers the following models:
\begin{description}
\item[{\it top panel}:] $\rmi^{(s)} = 23$ over $40\,$deg$^2$, 
 and $\rmi^{(p)} = 25.3$ -- characteristic of the LSST gold sample,
\item[{\it bottom panel}:] $dN^{(s)}/dz = 10\,$deg$^{-2}$ over $10^4\,$deg$^2$ and $0<z<2.5$ -- characteristic of SDSS quasars -- , and again $\rmi^{(p)} = 25.3$.
\end{description}
In the model in the bottom panel, the kernel peaks near the scale $\ell_{Pk-2}$, which corresponds to $\ell=400$, $700$ and $900$ at $z=0.5$, $1$, and $1.5$.  This is as expected when at least one sample is abundant.
In the model in the top panel, the information has a broad peak that falls between $\ell_{Pk-2}$
and $\ell_0$, where $\ell_0=800$, $2000$, and $3000$ for the three redshifts
considered.
This is consistent with our arguments for the case of two abundant samples.
In both of the models considered in Fig.~\ref{fig:info_example}, the majority
of the information arises from linear scales (scales which fall leftward of
the filled dot on each curve, representing $\ell_{\rm NL}$ (z)).
We find similar conclusions apply for a range of models.

 \begin{figure}
 \begin{center}
{\epsfig{file=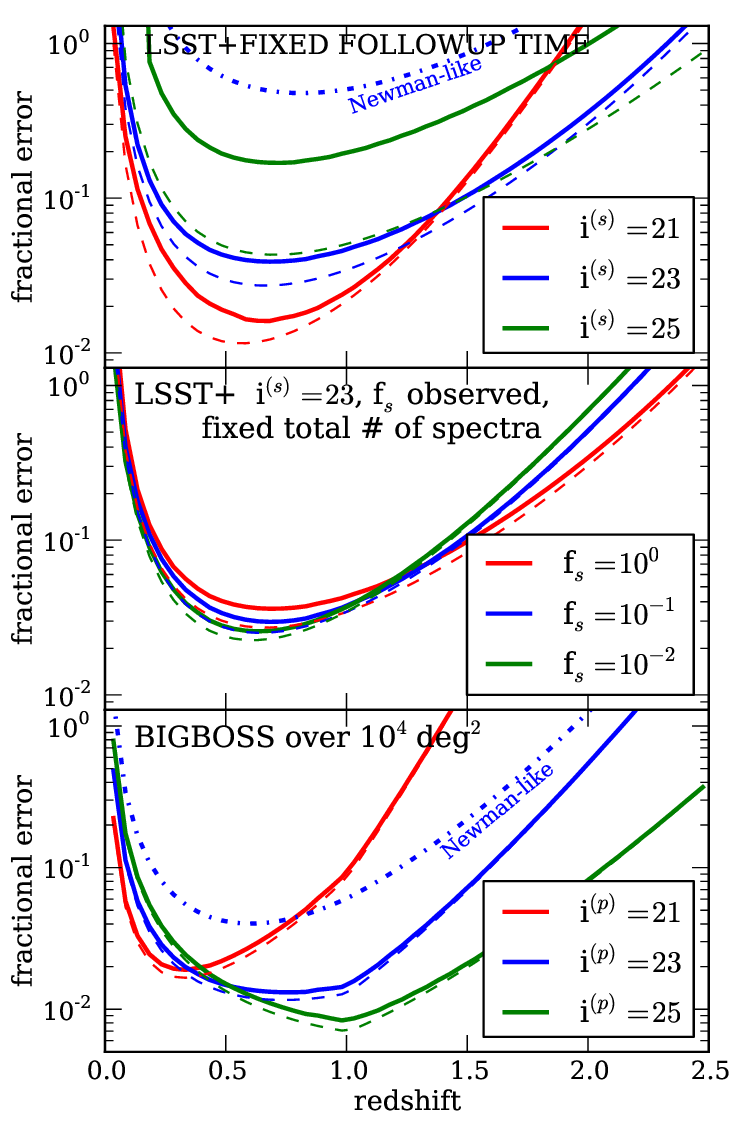, width=8.2cm}}
\end{center}
\caption{Shown are estimates for the fractional sensitivity to reconstruct the redshift distribution of the {\comment photometric sample's bias times number} in redshift bins of $\Delta z = 0.05$ and spanning $0<z<2.5$.  The top panel is for a photometric sample with the specifications of the LSST gold sample ($\rmi^{(p)} = 25.3$; see text) and for different  spectroscopic samples that could be obtained for the same total telescope time:  The spectroscopic followup covers $1\,$deg$^{2}$ to $\rmi^{(s)} = 25$, $40\,$deg$^{2}$ to $\rmi^{(s)} = 23$, or $1,600\,$deg$^{2}$ to $\rmi^{(s)}= 21$.  The middle panel is similar to the top panel but assumes that $f_s$ of galaxies to $\rmi^{(s)} = 23$ are observed over a region of  $40 \, f_s^{-1}~$deg$^{2}$.  The bottom panel is for a spectroscopic sample with the specifications of BigBOSS and the specified limiting photometric magnitudes.  This panel assumes that the surveys' overlap is $10^4\,$deg$^{2}$, but the quoted error scales as the square root of the survey area. 
In both panels, the dashed curves use all $\ell$ values, whereas the solid exclude information from $\ell > \ell_{\rm NL}$.  The dot-dashed curves (shown only for the $i=23$ case) in the top and bottom panels are the variance of the Newman-analogue estimator {\comment discussed in Section \ref{ss:comparison}} without any cutoff at nonlinear scales.
\label{fig:CovplotsIband}}
\end{figure}

Fig.~\ref{fig:CovplotsIband} investigates the tradeoffs of depth
versus area for attempts to constrain the $N^{(p)}_i$ in $50$ redshift
bins with $\Delta z = 0.05$ and spanning $0 < z<2.5$.  The top panel
shows {\comment the fractional error on $b^{(p)}_i \; N^{(p)}_i$} for a photometric sample with the specifications of the LSST gold
sample (which has $dN^{(p)}/dz > 10^4\,$deg$^{-2}$ over the entire redshift
range) and for three spectroscopic samples that could be obtained with
the same total time on a telescope. (More correctly, the limiting flux
squared divided by the survey area is held constant.)  We assume that
the spectroscopic followup covers $40\,$deg$^{2}$ at $\rmi^{(s)} = 23$.
Hence, it covers $1,600\,$deg$^{2}$ at $\rmi^{(s)}= 21$ and
$1.0\,$deg$^{2}$ at $\rmi^{(s)} = 25$.  This panel illustrates that
deeper is not necessarily better (compare only the solid curves for
the time being).  This conclusion arises because the spectroscopic
galaxies are more or less in the abundant limit (particularly near
their peak in  $dN^{(s)}/dz$) where the fractional error does not depend on depth and instead scales as
$f_{\rm sky}^{1/2}$.  However, the scaling $f_{\rm sky}^{1/2}$ -- a
factor of $6$ between the three cases considered in the top panel -- over predicts the differences between the curves in this panel.  This arises because these samples are only marginally in the rare $dN/dz$ regime where we find that this scaling holds (Fig.~\ref{fig:Constn}).  
 The $\rmi^{(s)} = 21$ sample is in the rare sample limit at the highest redshifts shown, and hence its errors blow up there.  By contrast, while the $\rmi^{(s)}=25$ sample is the least sensitive to $dN^{(p)}/dz$ at intermediate redshifts (owing to its small $f_{\rm sky}$), it is the most able to determine the distribution at the highest redshifts. 

The middle panel in Fig.~\ref{fig:CovplotsIband} is similar to the top panel but assumes that a random fraction, $f_s$, of all galaxies with $\rmi^{(s)} = 23$ are observed over a region of  $40\,f_s^{-1}\,$deg$^{2}$ such that the total number of galaxies is fixed.  This panel reinforces our result that the constraint on the $N^{(p)}_i$ depends primarily on the total number of spectroscopic galaxies and not their angular density, even though the case with $f_s = 1$ is in our abundant limit in which we no longer expect this scaling to hold exactly.  We still find that this result \emph{approximately} holds. 
  
The bottom panel in Fig.~\ref{fig:CovplotsIband} shows the case of a spectroscopic sample with the specifications of BigBOSS (whose $dN/dz$ is shown in Fig.~\ref{fig:different_populations}) and the specified limiting photometric magnitudes.\footnote{BigBOSS aims for a combined $dN/dz$ that we crudely parametrize as $30\times 10^{2.1 \, z}$deg$^2$ for $z< 1.0$ and $4000\times 10^{-1.1 \, (z-1)}$deg$^2$, to approximate what is quoted at \url{http://bigboss.lbl.gov}.}  This panel assumes that the surveys' overlap is $10^4\;$deg$^{2}$, but the error scales as the square root of the overlapping area.  Despite the lower number densities of galaxies in the BigBOSS case compared to those in the top panel, BigBOSS has a total number of galaxies that exceeds the other cases by more than an order of magnitude and, thus, is the most sensitive of all the cross-correlation examples considered in Fig.~\ref{fig:CovplotsIband}.  
We note that to reach the $10^{-2}$ sensitivity quoted here, BigBOSS would likely need to correct for magnification bias (which is discussed in Section \ref{sec:bias}).

Omitting nonlinear scales or introducing a redshift cutoff in the spectroscopic coverage has little impact on our results.
The dashed curves in Fig.~\ref{fig:CovplotsIband} include information from $\ell > \ell_{\rm NL}$, whereas the solid curves do not.  Excluding nonlinear modes in the analysis has only a modest impact on the estimator, except in the $\rmi^{(s)} = 25$ case in the top panel, where the constraint is reduced by a factor of $3$.  This case is most impacted because (1) its $\ell_0$ falls at the most nonlinear scales of the cases plotted and (2) the small $1\,$deg field assumed in this case has already limited the scales that can contribute.  Similar losses for each of the plotted cases also occur for a factor of $2$ smaller $\ell_{\rm NL}$.  In addition, we have assumed that the spectroscopic sample spans the entire redshift range of the photometric sample.  A cutoff in the coverage of a spectroscopic sample, as could occur if an emission line falls out of the spectroscopic band of a survey, has little impact on our results below that cutoff.  It has no impact to the extent that $S = 1$.   When the additional condition $dN^{(s)}/dz = 0$ was imposed for $z>1.5$, which forces $S$ to be small, we found no change to the $\rmi^{(s)} = 21$ case in the top panel of Fig.~\ref{fig:CovplotsIband}, but a factor of $2.5$ shift upward for $\rmi^{(s)} = 25$ in that panel.

The photometric sample can often be divided into magnitude bins or into photometric redshift bins. For magnitude cuts, extra sensitivity is often gained by dividing the primary photometric sample because galaxies in different magnitude bins are more likely to also be at different redshifts.  In particular, in the rare spectroscopic galaxy limit but where the photometric galaxies are more abundant than $[dN/dz]^{\rm crit}_{-2}$, the signal scales inversely with the redshift extent of the photometric sample and does not depend on the amplitude of $dN^{(p)}/dz$ (Eq.~\ref{eqn:rareNplimit1}).  Thus, the sensitivity is not improved by going deeper.  The redshift distribution of galaxies given by our parameterization for $P(z| \rmi)$ (Eq.~\ref{eqn:pziband}) has mean $3 \, z_0$ and variance $3 \, z_0^2$.  Because the variance of $P(z| \rmi)$ increases with depth, deeper surveys will be somewhat less sensitive at the peak of $P(z|\rmi)$ unless the sample is partitioned.\footnote{This statement holds as long as $dN^{(p)}/dz >[dN/dz]^{\rm crit}_{-2}$. This inequality is satisfied near the peak of $P(z|\rmi)$ down to the lowest magnitudes for which Eq.~(\ref{eqn:pziband}) is calibrated, $\rmi=20.5$ (see Fig.~\ref{fig:different_populations}).}  A partitioned sample can be easily accommodated in the quadratic estimator formalism.  In Section \ref{sec:photoz}, we discuss the gains from dividing by photometric redshift.

\section{Configuration space}
\label{sec:configspace}

The previous derivations were done in spherical harmonic space as this is the simplest basis for calculating the minimum variance estimator.  However, when dealing with actual data it can be more difficult to work with spherical harmonics as the survey window function enters nontrivially in convolution.  Hence many galaxy clustering analyses are done in configuration space.  In this section we show that the minimum variance estimator can be easily applied in this dual space (Section \ref{sec:configestimator}), we compare with previous configuration space $dN/dz$ estimators (Section \ref{ss:comparison}), and finally discuss the impact of finite sky coverage (Section \ref{ss:finitesky})

\subsection{Configuration space estimator}
\label{sec:configestimator}


The harmonic space quadratic estimator can be written in the form
\begin{equation}
  \sum_{\ell, m}  v_i({\ell}) \, \widehat{p}(\ell,m)^\star \widehat{s}_i (\ell,m),
\label{eqn:factor}
\end{equation}
for some $v_i(\ell)$, plus analogous terms proportional to the auto correlations.
Writing $\widehat{p} \, \widehat{s}_i(\ell, m) =  \int d\nhat  \;\widehat{p} \, \widehat{s}_i(\nhat) \, Y_\ell^m(\nhat)$,
Eq.~(\ref{eqn:factor}) becomes
\begin{equation}
\int d\nhat \, d\nhat'\ \widehat{p}(\nhat')\,
    v_i(\nhat\cdot\nhat')\, \widehat{s}_i(\nhat),
\label{eqn:nhat}
\end{equation}
where we have used the addition theorem for spherical harmonics
\citep{AbramowitzStegun72},
$P_\ell$ is the Legendre polynomial of order $\ell$, and
\begin{equation}
  v_i(x) = \sum_\ell \frac{2\ell+1}{4\pi} v_i(\ell) \, P_\ell(x).
  \label{eqn:vifullsky}
\end{equation}
If we define $\widehat{\omega}_{ps_i}(x) \equiv \langle \widehat{p}\;
\widehat{s}_i\rangle_x$, as the correlation function estimate where
$x=\nhat \cdot \nhat'$ and $\langle \ldots \rangle_x$ represents an
average over all separation angles $x$ in the survey, Eq.~(\ref{eqn:nhat})
can be re-expressed as
\begin{equation}
  8 \pi^2 \int dx\ v_i(x)\, \widehat{\omega}_{ps_i}(x).
\end{equation}

Thus, the configuration space estimator in the Schur-Limber limit is 
\begin{eqnarray}
[\widehat{N}^{(p)}_{i}] &=&
   [\widehat{N}^{(p)}_{i}]_{\rm last} +
  \frac{8\pi^2}{F_{ii}} \sum_{\alpha} \Delta x_\alpha\ v_i(x_\alpha)
  \nonumber \\
  &\times&  \; \bigg\{ \widehat{\omega}_{ps_i}(x_\alpha)
                             - {\omega}_{ps_i}(x_\alpha) \bigg\},
\label{eqn:configspaceest}
\end{eqnarray}
where $\alpha$ runs over the bins in (cosine of the) angle.
A similar configuration space estimator can be written for the full
minimum variance quadratic estimator (Eq.~\ref{eqn:MQest}).

For $\theta\ll 1\,$radian (the scales that we will show are of primary
interest), the result can be further simplified by making the flat sky
approximation.  Then, the Parseval identity,
$\int d^2\ell\ v_i^\star(\ell)\, \widehat{p} \,\widehat{s}(\mathbf{\ell})/(2\pi)^2
= \int d^2\theta\ v(\theta)\,\widehat{p} \,\widehat{s}(\theta)$,
can be directly applied to Eq.~(\ref{eqn:factor}) to yield
Eq.~(\ref{eqn:configspaceest}) with
$\Delta x_\alpha\rightarrow\theta_\alpha\, \Delta\theta_\alpha$ and
\begin{equation}
  v_i(\theta) =  \int_0^\infty \frac{\ell\,d\ell}{2\pi}
    \ v_i(\ell) \, J_0(\ell \,\theta) \, .
\label{eqn:vthetadef}
\end{equation}
The same expression can be derived from Eq.~(\ref{eqn:vifullsky})
by writing the small-angle limit of $P_\ell$ in terms of $J_0$
\citep{AbramowitzStegun72}.

We note that in the Schur-Limber limit 
\begin{equation}
v_i(\ell) = \frac{b^{(p)}_i b^{(s)}_i N_i^{(s)} C_{ii}}{ \left[
    \sum_i\left(b^{(p)}_i N_i^{(p)}\right)^2 C_{ii} + w_i^{(p)}
  \right] \left[ \left( b^{(s)}_i N_i^{(s)}\right)^2 C_{ii} +
    w_i^{(s)} \right]}.
\label{eqn:vell_SchurLimber}
\end{equation}

\begin{figure}
\begin{center}
\epsfig{file=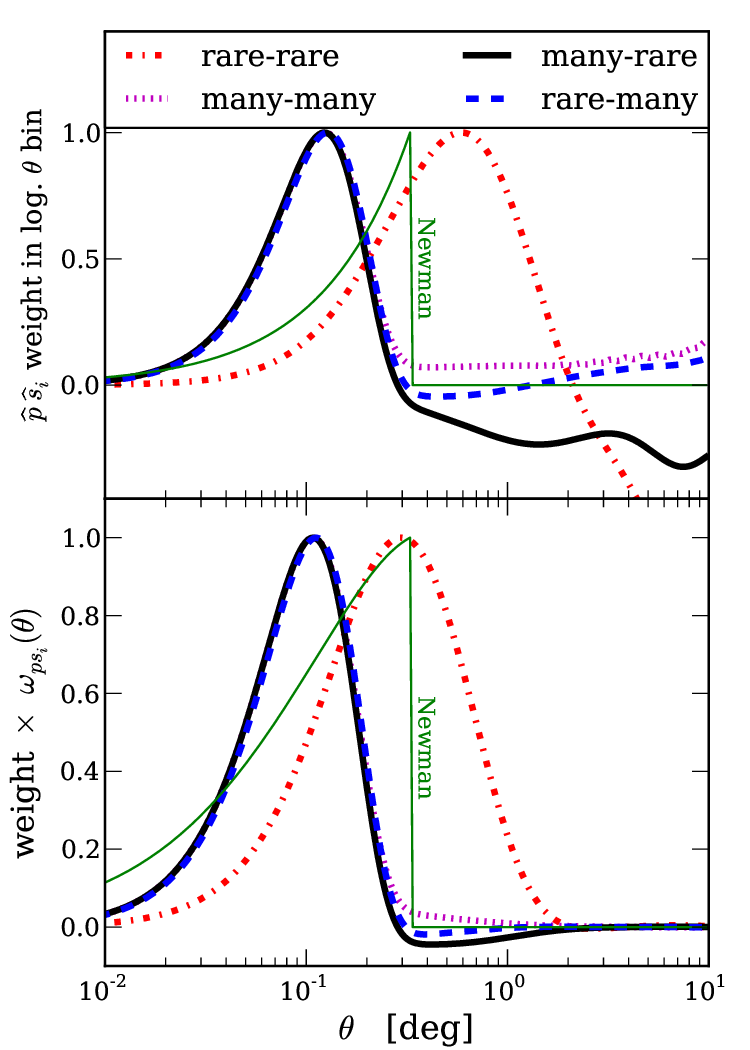, width=8cm}
\end{center}
\caption{{\comment The top panel shows $\theta \times v_i(\theta)$ for the illustrative cases considered in Fig.~\ref{fig:information}, again for the $i=6$ redshift bin.  The $\theta \times v_i(\theta)$ are the optimal estimator weights of the logarithmically binned cross-correlation function, $\omega_{ps_i}(\theta)$.}
  The bottom panel is $\theta \times v_i(\theta) \times \omega_{ps_i}$, which shows explicitly which angular scales the
  information derives.  The thin solid green curve in each panel is
  the weighting scheme used in our analogue of the \citet{newman08} estimator, with $r_{\rm max} = 10\,h^{-1}$Mpc.   All of the curves, aside from the Newman-analogue ones, have down weighted nonlinear modes by the factor $\exp[-\ell^2/\ell_{\rm NL}^2]$.  {\comment The curves in both panels are computed in the Limber and flat sky approximations.}\label{fig:weights}}
\end{figure}

The thick curves in the top panel in Fig.~\ref{fig:weights} show the flat sky
weighting kernel for the same example surveys as in Fig.~\ref{fig:information}, down weighting nonlinear modes by multiplying $v_i(\ell)$ by the factor $\exp[-\ell^2/\ell_{\rm NL}^2]$.
These calculations show that if any sample is in the abundant limit,
the window peaks at $\theta\sim 0.1\,$deg separations, whereas if both surveys are
in the rare limit the peak occurs at $\theta\sim 1\,$deg.
Both cases have non-negligible weight at super-degree scales. 

The bottom panel in Fig.~\ref{fig:weights} shows
$\theta \,v_i(\theta) \times \omega_{ps_i}$, which better represents
the $\theta$ that contribute to the final estimate.
Since measured correlations are weaker on large scales than small, the
$\theta>1\,$deg behavior of $v_i(\theta)$ is down-weighted and really only
sub-degree scales contribute significantly.

In practice, whether weights are applied during or after the
computation of the correlation function depends on the survey to which
cross-correlations are applied.  In the case where the survey's
contiguous area is much larger than the kernel of $v_i(x)$ ($\gg
0.1-1\,$deg), the exact details of the survey window are irrelevant.
The $\omega_{ps_i}(\theta)$ can be estimated with standard
techniques \citep[e.g.][]{landy93,hamilton93,bernstein94} and then
multiplied by the approximate $v_i$.
This is the regime most large-scale photometric and spectroscopic surveys,
such as SDSS, WiggleZ, BOSS, GAMA, DES, and LSST.
The second regime, where the survey area is comparable to or
smaller than the weighting kernel (e.g. with DEEP or HST fields) is
more complex.  Section \ref{ss:finitesky} discusses this case.

\subsection{Comparison to earlier work}
\label{ss:comparison}

Using cross-correlations to estimate redshift distributions has been
championed by \citet{newman08}.
The configuration space expression for the optimal quadratic estimator (c.f., Eq.~\ref{eqn:configspaceest}) allows us
to compare explicitly with the \citet{newman08} method.
Though the \citet{newman08} method is neither optimal nor
unbiased, it has some similarities to our estimator as we shall see.

The estimator in \citet{newman08} (and also \citealt{matthews10}) involves nonlinear, power-law fits to correlation functions over a
specified range of scales and with specified, diagonal
(i.e.~ignoring bin-to-bin correlations in $\theta$ and $z$) weights.
The estimator is thus a nonlinear functional of the measured two-point functions.
However since the power-law fit is used mainly to divide out trends and fit
for an amplitude, we can write an {\it analogous\/} estimator to
\citet{newman08} that contains essentially the same information.
Our analogue-estimator becomes very similar to that of \citet{newman08} for
power-law models.

Our analogue of the \citet{newman08} estimator is\footnote{While \citet{newman08} does not explicitly subtract a shot-noise term,
we have subtracted $w_i^{(ps)}$ so that the estimator is well-defined in
both configuration and harmonic space.} 
\begin{equation}
  \widehat{N}_i^{(p)} = \eta_i^{-1} \sum v_i^{\rm New} \,
  \left(\widehat{p} \,\widehat{s}_i - w_i^{(ps)}\right),
\label{eqn:Newmanest}
\end{equation}
where
\begin{equation}
 \eta_i = \left( \sum v_i^{\rm New}  b_i^{(p)} b_i^{(s)} \,
  N_i^{(s)} \,C_{ii}\right).
\end{equation}
This estimator returns $N_i^{(p)}$ if the Limber approximation holds
and the underlying power spectra and biases are correctly guessed.
When the sum in Eq.~(\ref{eqn:Newmanest}) is over configuration
space pixels (as in \citealt{newman08}), the weighting is
\begin{equation}
  v_i^{\rm New}(r) = \Bigg\{ \begin{array}{lc}   1 & r_{\rm min} <r < r_{\rm max}\\  0 & {\rm otherwise} \label{eqn:vnewman}   \end{array}
\end{equation}
where \citet{newman08} chooses $r_{\rm min} = 0$ and $r_{\rm max} = 10\,h^{-1}$Mpc.
Fig.~\ref{fig:weights} compares the weights of our optimal
estimator to that of our Newman-analogue estimator.
The thin green solid curve in the top panel is 
$\theta \,v^{\rm New}_i(\theta)$, and this curve in the bottom panel corresponds to
$\theta \,v^{\rm New}_i(\theta) \times \omega_{ps_i}(\theta)$.
The thick curves are the same quantity for the optimal estimator for the same four
extreme cases as considered earlier.
The Newman-analogue estimator uses similar scales to those selected by the
optimal estimator, especially in the rare-rare case.

While the weights for the optimal quadratic and
Newman-analogue estimators are superficially similar, it becomes apparent
that the estimators behave differently when examining the weights in
more detail.
The optimal estimator in the shot noise-limited regime has configuration-space
weights given by the density correlation function.
However, the Newman-analogue weights are simply a constant.
The structure of the Newman-analogue estimator is also much different in the
signal-dominated regime.
The optimal estimator has weight
$v_i(\theta) \propto \int \ell\,d\ell\ C_{ii}^{-1}J_0(\ell \theta)$,
in the Schur-Limber approximation, in contrast
to the constant configuration space weights in our Newman analogue estimator.

The variance of these estimators also differs.  The covariance of the minimum variance estimator is $\bfF^{-1}$, whereas the covariance of the Newman-analogue estimator (in the Limber approximation) is
\begin{eqnarray}
 {\rm cov}[\widehat{N}_i^{(p)}, \widehat{N}_j^{(p)}] &=&  \eta_i^{-1} \eta_j^{-1}  \sum_{\ell, m}    v_i^{\rm New}(\ell) \,  v_j^{\rm New}(\ell) \\
&\times &   \left[ A_{0i}(\ell)A_{0j}(\ell) + A_{00}(\ell) A_{ii}(\ell) \, \delta_{ij}^K \right], \nonumber
\end{eqnarray}
where the Fourier space (flat sky) Newman weights are the Hankle transform of Eq.~(\ref{eqn:vnewman}):
\begin{equation}
  v_i^{\rm New}(\ell)  = \frac{\chi_i}{\ell}
  \left( \frac{J_1(\ell \, r_{\rm max}/\chi_i)}{r_{\rm max}} -
           \frac{J_1(\ell \, r_{\rm min}/\chi_i)}{r_{\rm min}} \right).
\end{equation}
 The rapid oscillations at higher $\ell$ damp the contribution of these modes.  The dot-dashed curves in Fig.~\ref{fig:CovplotsIband}
(shown only for the $i=23$ case)
in the top and bottom panels are the variance of the Newman-analogue estimator
without any nonlinear cutoff in $\ell$.
The Newman-analogue estimator performs substantially worse than the optimal
estimator: a factor of $3-10$, with the factor of $10$ applying to the
abundant galaxy case (which is most similar to the cases investigated in
\citealt{newman08} and \citealt{matthews10}).

\subsection{Finite sky coverage}
\label{ss:finitesky}

Until now many of our expressions have implicitly assumed that the surveys cover the full sky, which is
unlikely to be the case in practice.  For surveys whose narrowest dimension
is much larger than the scales where our estimator peaks, the correction
for finite sky coverage is benign: we simply have a factor of
$f_{\rm sky}$ to correct the number of modes in our Fisher matrix
\citep[e.g.][]{Scott94,Jungman96,Tegmark96,Knox97},
as we have assumed in our prior example calculations.
The effects of finite sky coverage have been studied extensively in
the CMB \citep[e.g.][]{Hivon02,Hansen02,Efstathiou04} and
large-scale structure literature
\citep[e.g.][]{FKP, peacock91, park94, tegmark98}.


The case of a general survey window function can be complex,
but, if the width and height of the window are comparable, the effects
of windowing are easily understood.
Due to the convolution with the window function, $\ell$-modes which are separated
by less than $2\pi/\Theta$ (where $\Theta$ is the angular extent of the
window function and for simplicity we are working in the flat sky approximation) are almost completely correlated and, thus, contain largely
redundant information.  In contrast, for modes separated by much more than $2\pi/\Theta$,
the effects of the window function can be largely ignored.

Thus the effects of finite sky coverage can be taken into account by
replacing our sums over $\ell$ with sums over $L$ values which are integer
multiples of $2\pi/\Theta$ and defining the $C_L$ as bin-averages of the
$C_\ell$.
A simpler approximation, valid if the theoretical spectra are smooth, is
to simply integrate from $2\pi/\Theta$ to infinity rather than zero to
infinity in Eq.~(\ref{eqn:vthetadef}).  
  If in computing the
correlation function or power spectrum we estimate the mean density from
the survey itself, then the power is suppressed on large scales (often known
as the integral constraint; \citealt{peebles80}).
An approximation to this suppression is to multiply $C_\ell$ by
$\left|1-W(\ell)\right|^2$ where $W(\ell)$ is the window function normalized
so that $W\to 1$ as $\ell\to 0$.

\section{Bias of approximate estimators}
\label{sec:bias}

The minimum variance quadratic estimator under the approximation that off-diagonal terms in the Fisher matrix are zero is unbiased as long as the diagonal entries are appropriately calculated.  In addition, dropping derivative terms in the quadratic estimator is unbiased since each derivative explores separate dependences.  However, there are a few approximations that could incur bias: the Limber approximation, ignoring RSDs, including nonlinear scales, cosmic magnification, and assuming the incorrect cosmology.  We do not consider the latter because it should be reduced to the per cent--level with the coming generation of cosmological probes, but we consider the others.\footnote{If $dN/dz$ is being estimated as part of a program aimed at constraining the cosmology, e.g.~with gravitational lensing, the cosmology and $dN/dz$ will have to be simultaneously varied.}  We can compute the bias of these approximations by substituting the full $\langle ( \widehat{p} ~ \widehat{\mathbf{s}})^\dagger \times ( \widehat{p} ~ \widehat{\mathbf{s}}) \rangle$ that includes the ignored terms into the approximate estimator and evaluating both near the input $N^{(p)}_i$.  Using this formalism, we address these biases here.\\

\noindent {\it Limber approximation and RSDs:}\\
In the Limber approximation, which has been assumed by most previous investigations of $dN/dz$ estimation from cross-correlations, the diagonals are accurately estimated in the limit $\ell \, \Delta \chi \gg \chi$ (although, in practice this condition has to be just weakly satisfied).  Fig.~\ref{fig:lmax} suggests that most scales that contribute to our estimate are safely in the Limber regime for $\Delta z \sim 0.1$.  This will be less true for smaller $\Delta z$.  On angular scales favored by our estimator, at which the matter power spectrum is decreasing with increasing $k$, the Limber approximation results in an over-prediction of the $C_{ii}$.  Hence, our Schur-Limber estimator will result in an under-prediction.  However, setting to zero the $\langle p_i s_j\rangle$ for $i\neq j$ in the Limber approximation has the opposite effect.  We find that the former effect is larger such that Limber results in an under-prediction, with a fractional error of $-(2-3)\times10^{-3}$ for $\Delta z = 0.01$ and $0 <z<1$ for the cases where most of the information derives from $\ell_{Pk-2}$ (i.e., where one of the populations is abundant) and $-(0.3-1)\times10^{-2}$ for cases where most of the information derives from $\ell_{Pk-1}$.\footnote{We speculate that the surprising smallness of the biases in Limber results  because of a near cancellation of the two competing effects.}  For $\Delta z = 0.1$, the biases are of course significantly smaller than for $\Delta z = 0.01$.  Thus, the Limber approximation will likely result in a bias that is smaller than the estimator's variance even for applications with very large source populations.

The fact that the Limber approximation is as successful as it is suggests that redshift space distortions (RSDs) will also induce a small bias (as RSDs are negligible on scales at which the Limber approximation holds; Appendix~\ref{ap:rsd}).  However, for reasons discussed in Appendix \ref{ap:rsd}, including RSDs is difficult in our current formalism as it requires a basis switch from our choice of top hat redshift bins, which spuriously magnify the impact of RSDs.  Thus, we do not quantify the magnitude of their small bias on the estimator.  {\comment RSDs could be more important for calculating the $\langle s_i^2 \rangle$, terms that do not appear in the Schur-Limber estimator (Appendix~\ref{ap:rsd}).} \\

\noindent {\it Nonlinear scales and the one halo term:}

 Using scales that are nonlinear can bias the estimator.  The Schur-Limber estimator for $N^{(p)}_i$ is biased by nonlinear effects that occur at the redshift of the estimate, $z_i$, and (fortunately) not by nonlinearities at other redshifts.  This is not the case for the minimum variance quadratic estimator (a fact that we have ignored).  In our estimates in Section \ref{sec:examples} and Fig.~\ref{fig:CovplotsIband}, we masked nonlinear wavenumbers at $z_i$ that met the criterion $k > k_{\rm NL}(z_i)$ (defined in Eq.~\ref{eqn:knl}), and found that this operation does not have a large impact on the sensitivity, except for the densest samples that were considered.  This result owes to the broad range in $\ell$ that contributes the information, which generally peaks at  $\ell  < \ell_{\rm NL}$ (Fig.~\ref{fig:information}).  We find that if we reduce $k_{\rm NL}$ by an additional factor of $2$, which corresponds to a wavenumber where the nonlinear density power spectrum deviates from linear theory by just $10$ per cent, the constraints are additionally degraded by a similarly small factor. 
 
As long as they are modeled properly, nonlinearities that trace the density field do not necessarily bias a measurement of $N^{(p)}_i$ as the galaxies still trace the same large-scale density fluctuations.  A bias will arise if intra-halo correlations contribute at scales where they are not in the white noise regime (as we have assumed).  Fortunately, deviations from the large-scale limit generally occur at wavenumbers that are larger than $k_{\rm NL}$, especially if clusters and large, low-redshift groups are excluded from the cross-correlation analysis \citep[see plots in][]{cooray02}.  
\\ 

\noindent {\it Magnification bias:} 


\begin{figure}
\begin{center}
\epsfig{file=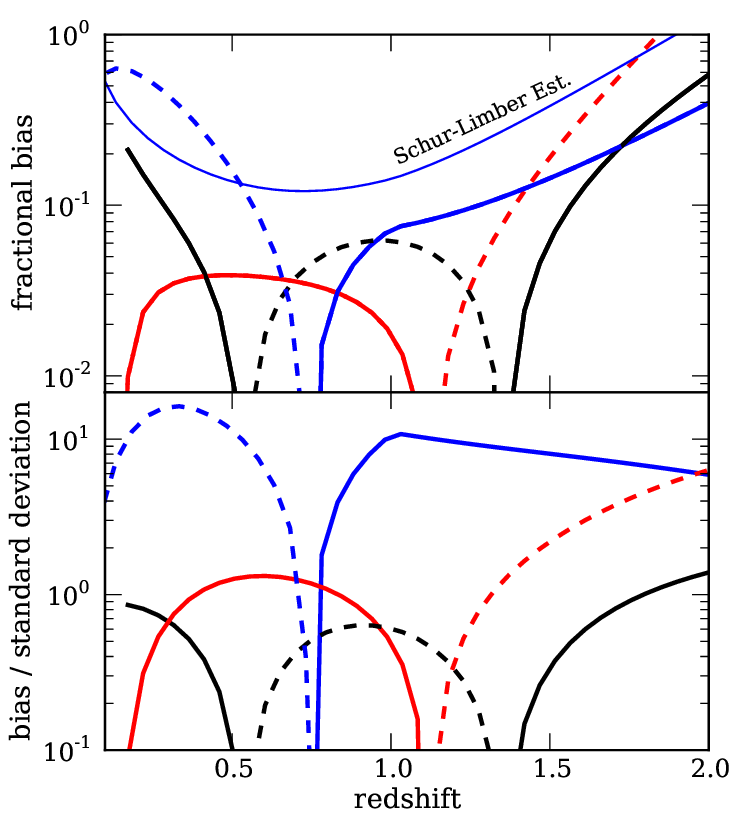, width=8cm}
\end{center}
\caption{The estimator bias arising from cosmic magnification for estimators that ignore this effect.  The curves assume that the photometric sample consists of all galaxies with limiting magnitude $\rmi^{(p)}=25.3$ and $0 < z< 2.5$ (the bias will be smaller in lower redshift samples) and different spectroscopic samples.  The thick blue curves are the full quadratic estimator for BigBOSS and an overlap area of $10^4\,$deg$^2$,  the black curves are for a spectroscopic sample with $dN^{(s)}/dz = 10\,$deg$^{-2}$ covering $0< z< 2.5$ and over $10^4\,$deg$^2$, and the red curves are  an example survey with $\rmi^{(s)} = 23$ and $40\,$deg$^2$.  The solid (dashed) curves indicate that the bias results in an overestimate (underestimate).  The top panel shows the bias relative to $N^{(p)}_i$, 
 and the bottom panel shows this relative to the fractional error.  All curves simplistically assume that the flux number counts of both populations have the \emph{rather steep} power-law index of $\alpha^{(x)} = -2$, to emphasize the effect.  The labelled thin blue curve in the top panel is the BigBOSS case with just the diagonal Schur-Limber estimator.
  \label{fig:mag_bias}}
\end{figure}

Magnification bias is the most significant of the biases we considered.
Cosmic magnification results in additional off-diagonal terms in $\bfC$
that were zero in the Limber approximation.
These terms are suppressed relative to the $j-j$, diagonal Limber term
(Eq.~\ref{eqn:limberapprox}) by the factor 
\begin{equation}
 R^{(x)}_{ij} \equiv -\frac{\alpha^{(x)}_i+1}{b^{(x)}_j}
  \left[\frac{(1+z_j) \; \chi_j  \; \Delta \chi_j}{2\times 10^7\,{\rm Mpc}^2}\right]
  \left(1- \frac{\chi_j}{\chi_i}\right)
\label{eqn:fracbias}
 \end{equation}
for $i > j$, where $\alpha^{(x)}_i$ is the power-law index of the cumulative {\comment in decreasing flux} source
number counts in bin $i$ above a certain flux threshold (see Appendix \ref{ap:magnification}).
Eq.~(\ref{eqn:fracbias}) ignores magnification-magnification correlations,
which are smaller except perhaps for surveys at $z\gg 1$ (e.g., \citealt{heavens11}).
 
For our simple Schur-Limber estimator, it is easy to compute the $N^{(p)}_i$ estimator bias, being 
\begin{equation}
 {\rm frac.~bias~from~mag.} = \sum_{k,\; k>i}  \frac{N^{(p)}_k}{N^{(p)}_i} R^{(p)}_{ki} +  \sum_{k, \;k<i}  \frac{N^{(p)}_k C_{kk}}{N^{(p)}_i C_{ii}} R^{(s)}_{ik},
\label{eqn:magbias}
\end{equation}
where $C_{ii}$ is defined in Eq.~(\ref{eqn:limberapprox}).  Thus, this estimator results in an overestimate when $-\alpha^{(x)} - 1 > 0$.
Evaluating this for our toy case of constant $dN/dz$ from $0< z< 1$,
one finds an $\approx -0.05\, (\alpha+1)$ per cent bias that is roughly constant with $z_i$.  In addition, Eq.~(\ref{eqn:magbias}) shows that if $N^{(p)}_i$ is well below the peak in $i$, this bias can be particularly severe.  

Fig.~\ref{fig:mag_bias} illustrates the importance of magnification
bias for a case in which the photometric sample consists of all
galaxies with $\rmi^{(p)}<25.3$  and different spectroscopic samples, all
covering $0 <z< 2.5$. (Lower redshift samples would be less biased by
magnification.)  For simplicity, we take $\alpha^{(x)}_i = -2$ for all
populations, which emphasizes the effect (being characteristic of the
bright end of quasar counts; fainter quasars have a slope $\alpha \sim -0.5$; \citealt{bartelmann01, scranton05}, and the faint-end slope for galaxies is $-(0.5-1)$; \citealt{bouwens12}).  The thick blue curves represent BigBOSS and $10^4\,$deg$^2$,  the black curves a survey with $dN^{(s)}/dz = 10\,$deg$^{-2}$ over $10^4\,$deg$^2$, and the red curves a survey with $\rmi^{(s)} = 23$ and $40\,$deg$^2$.  Solid (dashed) curves indicate  that the bias results in an overestimate (underestimate).  The top panel is the bias relative to $N^{(p)}_i$, and the bottom panel is this relative to the fractional error.   At $z< 1.5$, the bias is $\sim 1$ standard deviation for two of the cases.  However, for BigBOSS (which has fractional errors of $\sim 10^{-2}$), the bias is $10\,\sigma$ over many of the redshift bins of interest.   For all the cases, the biases are largest at $z< 0.5$ and $z>1.5$, redshifts at which there is a significant fall off in the photometric population.  The fact that these curves can become negative contrasts with the Schur-Limber estimator, which would always be biased high.  The thin blue curves are the Schur-Limber estimator for the case with BigBOSS.  We find that the bias of the Schur-Limber estimator (Eq.~\ref{eqn:magbias}) is typically larger than the bias of the full minimum variance quadratic estimator (that ignores magnification).

In all cases, magnification bias can be computed given an estimate for the $\alpha^{(x)}_i$ and removed.  The main issue is uncertainty in the $\alpha^{(x)}_i$.  It should be reasonably straightforward to remove the bias at redshifts greater than the peak in $dN^{(p)}/dz$ (where it is most severe) as the spectroscopic galaxies act as the sources and their $\alpha^{(s)}_i$ is easily measured. 
However, uncertainty in $\alpha^{(x)}_i$ could be the limiting factor
in $N^{(p)}_i$ constraints at redshifts where the photometric galaxies
act as the source, particularly in surveys that can place
percent-level errors on the $N^{(p)}_i$ and that extend to high
redshifts.  In such cases, the error will be approximately set by the
fractional bias of $N^{(p)}_i$ owing to magnification (what is plotted
in Fig.~\ref{fig:mag_bias}) times the fractional uncertainty in
$\alpha^{(x)}$.  Knowledge of $\alpha^{(x)}$ to $10\,|\alpha^{(x)}+1|$
per cent precision is required for this not to be the limiting factor
for the BigBOSS case considered above.  Since magnification only
depends on the sources' $N_i$ and not their $b_i$, the significant bias
of BigBOSS also suggests that it can use magnification to break this
degeneracy and separately estimate the $b^{(p)}_i$ to $10\,  |\alpha^{(x)}+1|$ per cent precision.   We revisit the impact of magnification in Section~\ref{sec:photoz}, showing that it is less onerous in the cases of (1) photo-$z$ calibration and (2) estimating the redshift distribution of diffuse backgrounds.

Analogous to magnification, intervening dust can also correlate background galaxies with foreground ones for surveys in the optical and bluer wavelengths  \citep{menard10}.  At linear scales, this effect will induce correlations that are a biased tracer of the projected density.  The magnitude of this effect with redshift could be determined with multi-band photometry using a population with uniform spectra, e.g. quasars, and this information would allow it to be corrected for in cross correlation studies again to the extent that the $\alpha^{(x)}_i$ are known.

\section{Calibrating photometric redshifts and cleaning correlated anisotropies from maps}
\label{sec:photoz}

Our previous results can be generalized to spectroscopically calibrate the $dN/dz$ of a photometric population that is partitioned by photometric redshift, an application which is relevant for large-scale clustering and weak lensing analyses on photometric populations.  When the catastrophic failure rate of the photometric redshift estimate is small, then it may be fruitful to self-calibrate by internal cross-correlations between different photometric redshift bins.  However, if the catastrophic failure rate is large, there can be degeneracies in the reconstruction from self calibrations, and it may be more robust to calibrate photometric redshifts with a spectroscopic sample.  In Section \ref{ss:specphotoz}, we discuss the latter, and Section \ref{ss:scalphotoz} discusses the former.  This section also addresses the more general problem of estimating the redshift distribution of a photometric sample in which other constraints exist for the sample's redshift distribution.  Finally, in Section \ref{ss:cleaning} we discuss how our results can be used to statistically clean diffuse background maps.

\subsection{Spectroscopic calibration}
\label{ss:specphotoz}

Consider binning the photometric sample by some property that we refer to as its ``photo-$z$'', and we denote the sample in photometric redshift bin `$m$' as `$pm$'.  One can think of $m$ as, for example, indexing a probability distribution of the sample's redshift as estimated from photometry.  The goal is to use cross-correlations with a spectroscopic sample to constrain this probability distribution.  The primary difference with the calculations in prior sections and this calculation is that the fluctuations from each photometric redshift bin are more likely localized in redshift than the full photometric sample.  (We defer discussion of internal correlations between different photo-$z$ bins to Section~\ref{ss:scalphotoz}.)

If this is the case, our approximate formulae for the sensitivities in different limits (Eqs.~\ref{eqn:rarefish}, \ref{eqn:rareNplimit1}, and \ref{eqn:rareNplimit2}) are altered so that
 $f_i \approx N^{(pm)}_i/N^{(pm)}_{\rm tot}$ and $\beta_i \approx [T^{(pm)}_i]^2/[T^{(pm)}_{\rm tot}]^2$, where 
 \begin{equation}
 T^{(pm)}_i \equiv D_i \, b^{(pm)}_i  N^{(pm)}_i,
\label{eqn:Tpm}
 \end{equation}
and $N^{(pm)}_i$ [$b^{(pm)}_i$] is the sky density [linear bias] of the photometric galaxies in redshift bin $m$ that are actually at redshift $i$.  Also, $N^{(pm)}_{\rm tot} \equiv \sum_i N^{(pm)}_i$, and $T^{(pm)}_{\rm tot} \equiv (\sum_i [T^{(pm)}_i]^2)^{1/2}$.  These relations for $f_i$ and $\beta_i$ are exact in the distant observer approximation. 
With these replacements, we can recast our formulae in the rare and abundant limits for the case of photo-$z$ calibration.

If the spectroscopic sample is in the rare limit, the potential constraint on the population in photo-$z$ bin $m$ that is actually in redshift bin $i$ follows from Eq.~(\ref{eqn:rareNplimit1}) and is
\begin{equation}
  \frac{\delta T^{(pm)}_i}{T^{(pm)}_{\rm tot}} \approx  \frac{0.06}{b_i^{(s)} D_i} \, \left(\frac{{\cal N}^{(s)}_{i}}{10^4}  \right)^{-1/2}   \left(\frac{1 +z}{2}\right)^{-0.5}.
\label{eqn:dNpz}
\end{equation}
Note that $\delta T^{(pm)}_i/ T^{(pm)}_{\rm tot}$ equals the outlier fraction for bin $i\neq m$ in the limit that $pm$ primarily falls in redshift bin $m$ and that the clustering is redshift independent.  For $pm$ to be in the dense galaxy limit (as Eq.~\ref{eqn:dNpz} assumes) requires that the redshift span of the photo-$z$ bin is sufficiently concentrated that $\sum_i [D_i \, b^{(pm)}_i]^2 C_{ii} > [N^{(pm)}_{\rm tot}]^{-1}$, which roughly should hold if $dN^{(pm)}/dz$ at the full width half-maximum is greater than $[dN^{(p)}/dz]^{\rm crit}_{-2}$.

In the contrasting case of a dense spectroscopic and photometric sample, it follows from Eq.~(\ref{eqn:lmax}) that
\begin{equation}
  \frac{\delta T^{(pm)}_i}{T^{(pm)}_{\rm tot}} \approx
  0.03 \left(\frac{f_{\rm sky}}{0.001} \right)^{-1/2}\,
  \left( \frac{\ell_0}{10^3}\right)^{-1}.
\label{eqn:dN2pz}
\end{equation}
Eqs.~(\ref{eqn:dNpz}) and (\ref{eqn:dN2pz}) demonstrate that cross-correlations can be used to constrain the fractional number (times bias) from $pm$ in bin $i$ at the part in a hundred level with $10^5-10^6$ spectra per unit redshift (for rare spectra) or $f_{\rm sky} =10^{-3}$ (for high spectral densities).

\begin{figure}
\begin{center}
{\epsfig{file=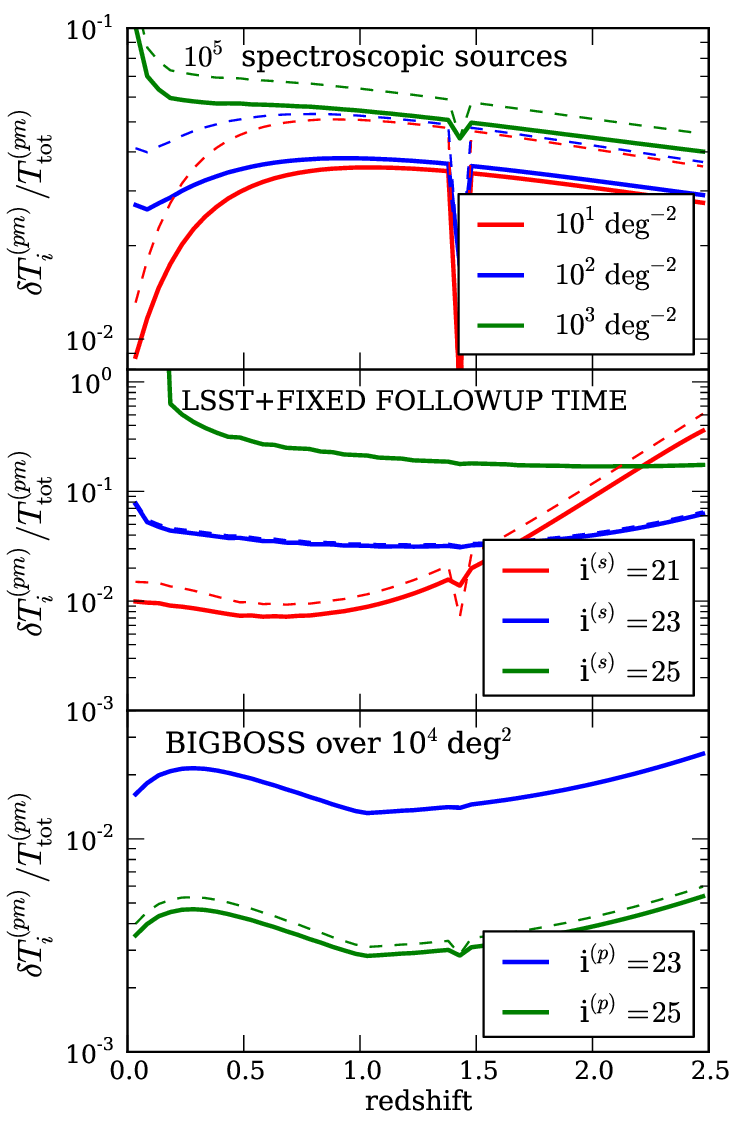, width=8.2cm}}
\end{center}
\caption{Estimates for how well the redshift distribution of sources (times their bias) in the photo-$z$ bin $pm$ can be reconstructed with cross-correlations. {\comment Shown is the error in redshift bin $i$ divided by the total number of galaxies in bin photo-$z$ $pm$ (i.e., $\delta T^{(pm)}_i/T^{(pm)}_{\rm tot}$, Eq.~\ref{eqn:Tpm}), assuming redshift bins of size $\Delta z = 0.05$.}  Our calculations assume that much of $pm$ resides in the $z_m=1.45$ bin, with ``outlier'' galaxies distributed uniformly in the range $0 <z < 2.5$, and that the number density at $z_m$ is that of a survey complete to $\rmi^{(p)}=25.3$ unless specified otherwise.  The solid curves take half of the galaxies in this photo-$z$ bin to reside outside of $z_m$, uniformly distributed so that $N^{(pm)}_i/N^{(pm)}_{\rm tot} = 10^{-2}$ for $i\neq m$.  The dashed curves are the same but for $N^{(pm)}_i/N^{(pm)}_{\rm tot} = 10^{-3}$ (so that most galaxies reside at $z_m$).  The top panel shows the constraints from different spectroscopic samples with the specified constant $dN^{(s)}/dz$ over $0 <z < 2.5$ and with $f_{\rm sky}$ adjusted so that there are $10^5$ total spectra.  The middle panel shows three different spectroscopic samples that could be obtained for the same total telescope time (with the same specifications as in the top panel in Fig.~\ref{fig:CovplotsIband}). The bottom panel is for a spectroscopic sample with the specifications of BigBOSS and the specified limiting photometric magnitudes.  All curves truncate the summation over $\ell$ at $\ell_{\rm NL}$.   \label{fig:photozcalib}}
\end{figure}

Fig.~\ref{fig:photozcalib} presents estimates for how well the
redshift distribution of a photo-$z$ bin can be reconstructed in bins
of size $\Delta z = 0.05$ with cross-correlations for the $z_m=1.45$
photo-$z$ bin, assuming that the ``outlier'' photo-$z$'s that are not
actually at the redshift $z_m$ are distributed uniformly in the range
$0 <z < 2.5$.  The solid curves assume that half of the galaxies in this
photo-$z$ bin reside outside of it, uniformly distributed so that $N^{(pm)}_i/N^{(pm)}_{\rm tot} = 10^{-2}$ for $i\neq m$.
The dashed curves are the same but for an outlier fraction of
$N^{(pm)}_i/N^{(pm)}_{\rm tot} = 10^{-3}$ so that only $5$ per cent of
galaxies reside outside the photo-$z$ bin $z_m$.  {\comment Despite these rather artificial outlier distributions, their comparison is useful for diagnosing how sensitive our results are to the details of the true outlier distribution.}

The top panel in Fig.~\ref{fig:photozcalib} shows the constraints from different spectroscopic samples with the
specified $dN^{(s)}/dz$, which is held constant over $0 <z < 2.5$ and
for fixed total number of spectra.  This panel shows that
Eq.~(\ref{eqn:dNpz}) is in qualitative agreement with these
estimates, noting that here ${\cal N}^{(s)} = 4\times 10^4$.  
  (We discuss the dip at $z_m=1.45$ below.)
Especially for the two lower number densities, the constraint depends weakly on the density of spectra as Eq.~(\ref{eqn:dNpz}) predicts.  {\comment The cases in this panel appear to depend modestly on the outlier fraction (compare the dashed and corresponding, and slightly more sensitive, solid curves).}   

The middle panel in Fig. \ref{fig:photozcalib} is for a photometric sample with the specifications of the LSST gold sample ($\rmi^{(p)} = 25.3$) and for different spectroscopic samples that could be obtained for the same total telescope time (with the same specifications as in the top panel in Fig.~\ref{fig:CovplotsIband}).  In this case, both the photometric and spectroscopic galaxies are at least marginally in the dense limit such that Eq.~(\ref{eqn:dN2pz}) applies, and the sensitivity scales roughly as $f_{\rm sky}^{1/2}$.  In the three cases plotted, $f_{\rm sky}$ equals $2.5\times10^{-4},~10^{-3},~$ and $4\times10^{-2}$.  The predictions in this panel depend weakly on the outlier fraction (compare the solid and dashed curves, which in two of the cases lie on top of each other).  The sensitivity of followup to $\rmi^{(s)}=21$ also falls off substantially with increasing redshift, which reflects that the spectroscopic galaxies are entering the rare regime.

The bottom panel shows the cases of a spectroscopic sample with the specifications of BigBOSS, the specified limiting photometric magnitudes, and where the surveys' overlap is $10^4\,$deg$^{2}$.  {\comment These cases depends negligibly on the outlier fraction.}  The BigBOSS sample is on the borderline of the rare limit (especially at the lowest and highest $z$) such that this panel is most difficult to relate to our predictions.   The rare-abundant limit given by Eq.~(\ref{eqn:dNpz}) appears to be most applicable for the case of $i^{(pm)}=25$ -- BigBOSS has ${\cal N}^{(s)}_i \sim 10^7$ at $z\sim 1$.  However, this limit does not appear to describe the error for the $i^{(pm)} = 23$ case as this case is considerably less sensitive:  $i^{(pm)} = 23$ is on the borderline of being in the rare limit with $dN^{(pm)}/dz = 5,000\;$deg$^{-2}$ at $z_m$.

Auto-correlations (which were dropped in the derivations that led to Eqs.~\ref{eqn:dNpz} and \ref{eqn:dN2pz}) add additional information. 
We find that auto correlation estimates do not improve the sensitivity for redshift bins that contain only a small fraction of $pm$ galaxies.  However, for the redshifts that contain the bulk of $pm$, they can improve the constraint on $\delta T^{(pm)}_i/T^{(pm)}_{\rm tot}$ by an order of magnitude.  This can be seen by focusing in on the dip at $z_m=1.45$ in Fig. \ref{fig:photozcalib}, which corresponds to the redshift that contains half or more of the galaxies.  Eqs.~(\ref{eqn:dNpz}) and (\ref{eqn:dN2pz}) do not predict a dip.  Especially with a rare spectroscopic sample as investigated in the top panel (where the cross-correlations can be quite noisy) and a low outlier fraction, much of the constraint on the number at $z_m$ owes to the large value of $\widehat{pm}^2$, which indicates many galaxies are concentrated in a narrow range in redshift. 

{\comment \citet{bernstein10} found that $0.0015$ error on the fractional number on `all outlying peaks' in the photo-$z$ distribution is required for uncertainty in the redshift distribution of the lenses to not to be the limiting
factor for the next generation of photometric weak lensing surveys.} 
Eqs.~(\ref{eqn:dNpz}) and (\ref{eqn:dN2pz}) [and Fig. \ref{fig:photozcalib}]
show that such an error in the true redshift
distribution of $pm$ would be difficult to achieve with spectroscopic
cross-correlations (even ignoring that the $b^{(p)}_i$ also need to be
constrained to ${\cal O}(10^{-3} \, f_c^{-1})$, where $f_c$ is the
contamination fraction).  The case of BigBOSS cross-correlations with a photometric sample complete to $\rmi^{(p)}=25$
over $10^4\,$deg$^2$ (green curves in bottom panel of Fig.~\ref{fig:photozcalib})
achieves the smallest error of the cases considered.  {\comment However, its error on $\delta T^{(pm)}_i/T^{(pm)}_{\rm tot}$ in redshift bin $i$ with $\Delta z= 0.05$ is still only $\sim 0.003$.  If, for example, an outlying peak in the photo-$z$ distribution spanned a redshift range of $0.2$, this would require four redshift bins and make the fractional error on the total number $\sim 0.006$.  While this does not appear sufficient to satisfy the \citet{bernstein10} requirement, it is possible that the calibration requirements are less severe owing to canceling effects \citep[who found than an $\sim 0.01$ outlier fraction may be tolerable]{cunha12}.  Quantitatively} answering the question of whether a BigBOSS-like survey is sufficient for futuristic weak lensing surveys requires an analysis of the bias on cosmological parameters induced by the pattern of uncertainties we find.

Thus far we have ignored prior information on the redshift distribution of the photo-$z$ subsample $pm$.  Often it is the case that we have prior information on the distribution of $N^{(pm)}_i$, e.g.~from
the photometric redshift PDF per galaxy \citep{lima08,freeman09,sheth10}.
In this case our formalism has only minor modifications.
Appendix \ref{ap:estwithprior} reviews how the quadratic estimator formalism generalizes to include prior information.  For a Gaussian prior on the $N_i$ (dropping $pm$ superscripts for simplicity),
the estimator with a prior becomes
\begin{eqnarray}
  \widehat{N}_{i} &=& [\widehat{N}_{i}]_{\rm last} +
   [{\bfF} +  {\bfF}_{\rm P}]_{ij}^{-1} \Bigg\{
   \sum_{\ell,\; m}  \bigg[\left( \begin{array}{cc}  \widehat{p}  &
   \widehat{\bfs}\end{array} \right) \bfQ_{j}
   \left( \begin{array}{c}  \widehat{p} \\  \widehat{\bfs}\end{array} \right)
   \nonumber \\
   &-&  {\rm Tr}[\bfA^{-1} \bfA_{,j}] \bigg] +
   [\bfF_{\rm P}]_{jk} \left({N}_{{\rm P}, k} -
   [\widehat{N}_{k}]_{\rm last} \right)  \Bigg\},
   \label{eqn:Nprior}
\end{eqnarray}
where ${\bfF}_{\rm P}$ and ${N}_{{\rm P}, i}$ are respectively the inverse covariance
matrix and mean of the prior.
The prior pulls the estimated quantity towards ${N}_{{\rm P}, k}$, and this pull
dominates if the prior is more peaked than the likelihood of the data. 

The final subtlety we address with regard to photo-$z$ calibration is cosmic magnification.  
Section~\ref{sec:bias} showed that cosmic magnification can be a significant bias if unaccounted for redshift estimation of the entire photometric sample.  Magnification may be less onerous for photo-$z$ calibration to the extent that the redshifts of the photo-$z$ samples are well localized {\comment because the locations of sources and lenses are more constrained.  However, it is also true that the $\alpha_{i}^{x}$ may be less constrained in fine photo-$z$ bins than less restricted populations.}  Appendix~\ref{ap:magnification_pz} addresses how magnification can be accounted for in the case of photo-$z$'s.  
 
\subsection{Self calibration of photometric sample}
\label{ss:scalphotoz}

Self-calibration of redshifts by cross correlating different photo-$z$ bins
within a photometric sample has the potential to achieve a tighter constraint on the $N^{(pm)}_i$ than calibration using correlations with spectroscopically identified galaxies, since spectroscopic samples are likely to be either sparser in number or distributed over  narrower fields than photometric ones.
Self-calibration of a photometric survey with cross-correlations has been investigated in
several studies \citep{huterer06,schneider06, benjamin10}.
Here we show that the maximum sensitivity to $dN^{(pm)}/dz$ that can be achieved with photometric self-calibrations is strikingly similar to the previously considered case of abundant spectroscopic and photometric samples.

For self-calibration to be successful, the redshift distribution of the photometric sample $pm$ needs to be much better known than in the case of calibration with spectroscopic cross-correlations.  This is because the redshift of $pn$ for all $n$ is the only knowledge one has to measure the redshift of $pm$:  If $pn$ is not centered around a single redshift, it is unclear how finite $\langle \widehat{pm} \, \widehat{pn} \rangle$ translates into the redshift distribution of sample $pm$.   To avoid this difficulty, we assume that most of sample $pm$ falls into redshift bin $z_m$.  This assumption is the best case scenario, and will allow us to put a lower bound on the constraint from self calibrations.\footnote{This assumption requires a highly artificial top hat photo-$z$ distribution at $z_m$ for consistency.  However, we expect that our result is more general than this choice.} Thus, the covariance matrix of the different photo-$z$ bins is
\begin{eqnarray}
  B_{mn} &\equiv& \langle p^{(m)} p^{(n)} \rangle
  = \sum_{ij} T^{(pm)}_i \, T^{(pn)}_j C_{ij} +
  w^{(pmpn)}_{ii} \delta^{\rm K}_{ij}, \nonumber \\
  & \approx & \sum_{i=m,n} T^{(pm)}_i \, T^{(pn)}_i C_{ii} + w^{(pmpn)}_{ii},
\end{eqnarray}
and we have assumed the same discretization in redshift to specify both the photometric and actual redshift bins.
In the second line, the sum is evaluated at only one value of $i$ if $m=n$
(i.e.~the auto-correlation).
The approximate equality in the last line follows from assuming that $C_{ij}$ is
diagonal (as holds in the Limber approximation), that
$T^{(pm)}_m \equiv  D_m b^{(pm)}_m N^{(pm)}_m\gg\sum_{i\neq m} T^{(pm)}_i$, and from
keeping terms that are $\mathcal{O}(T^{(pm)}_i/ T^{(pm)}_m)$ or larger. 
This is the limit in which the fraction of catastrophic photo-$z$'s is small
and where the covariance matrix $B_{mn}$ is diagonally dominated.
In this limit, and to lowest order in $\alpha_{m, i} \equiv T^{(pm)}_i/ T^{(pm)}_m$,
the Fisher matrix with respect to the $T^{(pm)}$ is
\begin{equation}
  \bfF_{T^{(pm)}_n T^{(pn)}_m} \approx \sum_{\ell, m} \left(\frac{T^{(pn)}_n C_{nn} \,T^{(pm)}_m C_{mm} }{B_{mm} \,B_{nn}} \right),
\label{eqn:F}
\end{equation}
where $B_{nn} \approx [T^{(pn)}_n]^2 C_{nn} + w^{(pnpn)}_{nn}$, and the matrix is zero between other combinations of parameters.  The quadratic estimator for $T^{(pm)}_n$ in this limit can also easily be written as it only involves correlations between the photometric samples $m$ and $n$.    Thus, in the diagonally dominated limit, the parameter $T^{(pm)}_n$ only correlates with $T^{(pn)}_m$, and there is a perfect degeneracy that must be broken by adding a prior (often catastrophic errors occur in one redshift direction) or going to higher order terms that are suppressed by another factor of $\alpha_{m, i}$. {\comment (Including cosmic shear would also break this degeneracy; \citealt{zhang10}.)}
  In the case of the prior that constraints $T^{(pn)}_m$ to be zero, many of our previous results hold as Eq.~(\ref{eqn:F}) is the same as Eq.~(\ref{eqn:fishC00large}) [and its subsequent incarnation in Eq. (\ref{eqn:lmax})] with the replacement $\beta(z) =1$ and a slightly different number dependence. (In fact, we do not need the additional approximation of $S=1$, as was made there.)  Thus, if $T^{(pn)}_n \gg 10^4 \ b^{-2} \, \Delta z$~deg$^{-2}$, so that the abundant limit holds,
\begin{eqnarray}
    \frac{\delta T^{(pm)}_n}{T^{(pm)}_m} &\approx& 10^{-3} \, f_{\rm sky}^{-1/2} \left( \frac{\ell_0}{10^3} \right)^{-1}.
\end{eqnarray}
Photometric self-calibration over a significant fraction of the sky is
capable of part in $10^3$ accuracy required by the next generation of
weak lensing surveys (e.g., \citealt{bernstein10}), but with the same
caveats as noted in the previous subsection that {\comment (1) this method does not break the degeneracy between number and linear bias, and (2) we have not calculated 
the bias on cosmological parameters as is necessary to truly quantify the potential of this method.}  In addition, this error only applies to the case of a single
catastrophic error direction.  If the latter does not hold, the
constraint is likely to be weakened by the factor $\sqrt{\alpha_{m,
i}}$.  

More generally, the full covariance matrix of the photo-$z$ bins,
$B_{mn}$, (plus overlapping spectroscopic populations) can be used as the covariance matrix in the minimum variance
quadratic estimator.  This self-calibration estimator is likely to be more sensitive
than the algorithm discussed in \citet{benjamin10}, the only self-calibration method that we are aware of, as that algorithm uses linear combinations of the $A_{\alpha \beta}$ that encapsulate a subset of the full covariance and does not weight scales optimally.  

\subsection{Cleaning correlated anisotropies from a map}
\label{ss:cleaning}

Our estimator is optimal for \emph{statistically} estimating the level of (and, hence, cleaning) correlated anisotropies from angular cross-correlations between diffuse background/foreground maps and spectroscopic galaxies.  The fractional errors we quote on number are equivalent to the error with which anisotropies can be statistically removed. Thus, the survey optimizations for this application are equivalent to those discussed for $N^{(p)}_i$ estimates.  Our previous calculations suggest that correlating anisotropies can be cleaned statistically to the $1$ per cent level.  For wide field observations of diffuse redshifted 21cm emission, this factor of $100$ could be helpful if extragalactic sources are found to be a limiting factor.  
  For CMB analyses, cross-correlations could also be interesting for studying the redshift distribution and for expunging foregrounds.  For example, it could better enable the separation of the cosmic infrared background (CIB) from CMB anisotropies generated at higher redshift. (CIB contamination is currently the limiting factor in measurements of kinetic Sunyaev-Zeldovich effect, which conveniently does not correlate with the $s_i$; \citealt{reichardt12}).   \citet{kashlinsky07} investigated correlations on $\sim10'$ scales between diffuse anisotropies in Spitzer and HST deep fields.  Our results suggest the sensitivity to the clustering component would be increased with wider fields (perhaps using shallower ground based observations rather than HST, since we found that the extremely high number density in the HST fields is not useful). 

For diffuse anisotropies, gravitational lensing enters at second order because lensing preserves surface brightness.  Thus, at large scales its impact on correlating the anisotropies in a map with the spectroscopic sample is small.  If the ``spectroscopic'' sample is measured at sufficiently high redshifts that the magnification-magnification term becomes important, only then can magnification result in a linear order diffuse foreground--spectroscopic population cross-correlation signal.  Magnification also has the effect of correlating the $\widehat{s}_i$, which can bias the estimate.  However, both magnification effects are correctable as the $\alpha^{(s)}_i$ can be measured.

{\comment
Finally, the goal is sometimes to invert a measured 2D clustering signal to 3D
clustering of a population using knowledge of $dN/dz$.  In the cases where the
accuracy requirements are not stringent, knowledge of the mean redshift and
the redshift width suffices to make this conversion.  These quantities are
typically easier to constrain than the full $dN/dz$, and so far fewer spectra
are required for the cross-correlation.  Assuming a $z$-independent, power-law
power spectrum and $dN/dz$ that can be parameterized by a power of distance
times an exponential of a power of distance, we found knowing just the mean
and variance of $dN/dz$ sufficed to invert the 2D clustering to 3D at the
ten per cent level.
}

\section{Mock surveys}
\label{sec:mocks}

\begin{figure}
\begin{center}
{\epsfig{file=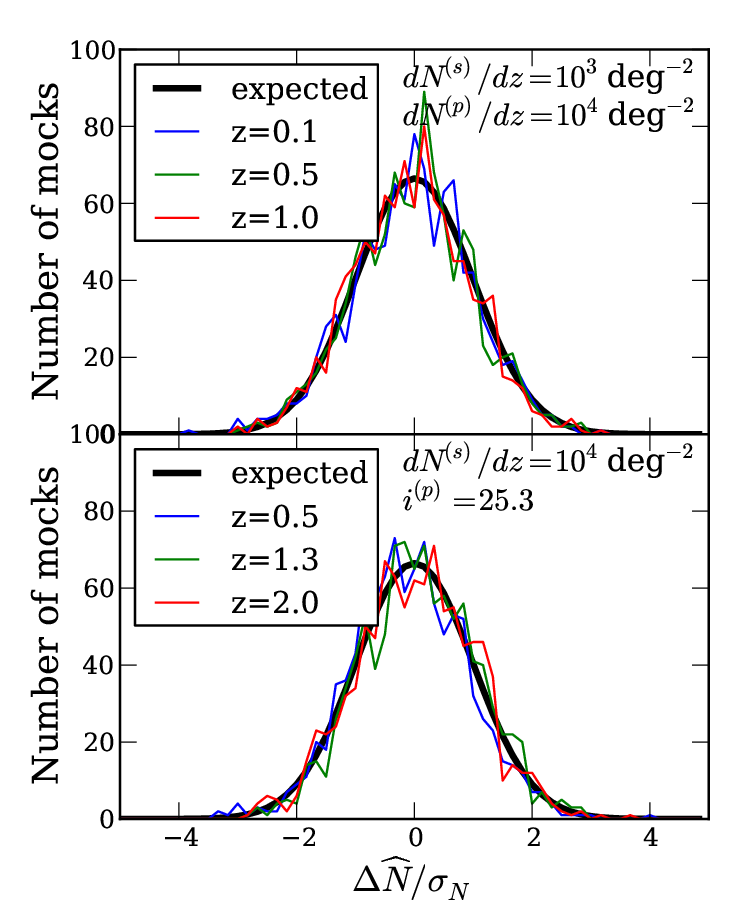, width=7.5cm}}
\end{center}
\caption{Test of our estimator's convergence, showing the
  distribution of the estimated value in units of the
  Fisher error for $1,000$ mocks.  The thick solid curve is the
  expected distribution of estimates.  For each mock, we start off
  with initial values for the $N^{(p)}_i$ that are each an order of
  magnitude smaller than their actual value.  The top panel shows the
  case of a $10\times10~$deg$^{2}$ field with the specified populations and $10$ bins spanning $0<z<1$ (resulting in $\sim 10$ per
  cent errors).  The bottom is a $30\times30~$deg$^{2}$ field with a
  photometric sample complete to $\rmi^{(p)} = 25.3$ and $50$ bins spanning
  $0<z<2.5$  (resulting in $\sim 1$ per cent errors). We find that the estimator robustly converges to its minimum, even when it starts far from it, and that in both cases there are zero outliers at $> 5~\sigma$ in the $1,000$ mocks.  
\label{fig:mocks}}
\end{figure}

\begin{figure}
\begin{center}
{\epsfig{file=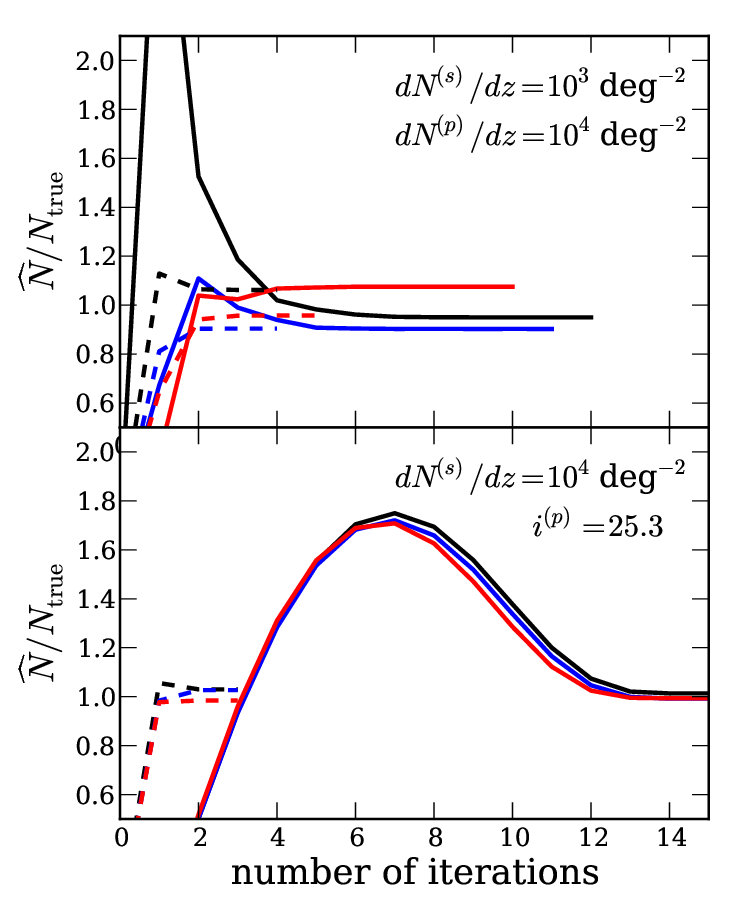, width=7.5cm}}
\end{center}
\caption{Walk of the estimated $N^{(p)}_i$ for $i= N_{\rm bin}/2$ as a function of iteration number for the two cross-correlation examples described in Fig.~\ref{fig:mocks} and the text.  The solid curves are the full minimum variance estimator, and the dashed curves are the Schur-Limber estimator (which converges more quickly).  The curves terminate after the last iteration changed the estimated $\widehat{N}^{(p)}_i$ by less than a part in $10^{5}$ when averaged over all $i$.  The initial guesses for the $\widehat{N}^{(p)}_i$ are taken to be an order of magnitude too small.  The asymptotic value of each $\widehat{N}^{(p)}_i$ shown in this figure is within $2 \, \sigma$ of the input ${N}^{(p)}_i$.
\label{fig:estwalk}}
\end{figure}

We are interested in understanding the robustness with which the proposed estimator converges to the input $N^{(p)}_i$.  To investigate its convergence, mock surveys are generated by decomposing the covariance matrix $\bfA$ into its eigenvectors $\mathbf{e}_\alpha$ and eigenvalues $\lambda_\alpha$ for $\alpha \in [0, N_{\rm bin}]$.  Then, a realization of the galaxy field that at multipole $\ell$ that has this covariance matrix is given by
\begin{equation}
g_\beta (\ell, m)= \sum_{\alpha=0}^{N_{\rm bin}} r_\alpha \lambda_\alpha(\ell)^{1/2} [\mathbf{e}_\alpha(\ell)]_\beta,
\end{equation}
where $r_\alpha$ is a Gaussian deviate with unit variance.  Here, $g_i$ corresponds to the overdensity in redshift bin $i$ of the spectroscopic survey, and $g_0$ is the overdensity in the photometric sample.   Our mocks assume that we are operating in a small enough patch such that there is a one-to-one mapping between wavevectors and spherical harmonics.    In addition, our mocks assume linear theory and the Limber approximation.  These approximations should not impact the conclusions per our previous results.\footnote{These mocks have one significant advantage over a real survey: they are periodic.  Hence, we do not have to worry about the survey window functions, and different modes on the lattice are truly independent.  We discussed how to deal with these real-world complications in Section \ref{ss:finitesky}.}

We generate $1,000$ mocks for two contrasting cases to illustrate the estimator's performance: 
\begin{itemize}
\item $10\times10~$deg$^{2}$ field with $dN^{(s)}/dz =
  10^3~$deg$^{-2}$, $dN^{(p)}/dz = 10^4~$deg$^{-2}$, and $10$
  redshift bins spanning $0<z<1$, each with $1,000^2$ angular pixels, specifications which result in $\sim 10$ per cent  errors on the $\widehat{N}^{(p)}_i$,
\item $30\times30~$deg$^{2}$ field with $dN^{(s)}/dz =
  10^4~$deg$^{-2}$ and photometry up to $\rmi^{(p)} = 25.3$, spanning
  $0<z<2.5$ with $50$ bins and $300^2$ angular pixels, which result in $\sim 1$ per cent errors on the $\widehat{N}^{(p)}_i$.
\end{itemize} 
The resolution of each mock is sufficient to resolve the scales that contain the bulk of the information (Section \ref{ss:finitesky}).  

Next, we apply the estimator to the harmonic space realization of these mocks.  (It would be equivalent to apply our estimator in real space using the results of Section \ref{sec:configspace}.)  Fig.~\ref{fig:mocks} demonstrates that the minimum variance quadratic estimator converges to the expected Gaussian distribution of errors.  This holds despite starting with initial estimates for the $\widehat{N}^{(p)}_i$ that are an order of magnitude smaller than the true values used in the mocks.  There are no outliers from the $5\,\sigma$ regions plotted in this figure for the cases shown.  Thus, our estimator does not tend to find local extrema.  We find that only when the Fisher errors become ${\cal O}(1)$ does the estimator no longer converge properly in all cases.  However, the Schur-Limber estimator in all cases we investigated successfully converged to the expected distribution of estimates.  This result is not surprising as the Schur-Limber estimator always minimizes $\sum_{\ell, m} v_i (A_{0i} - \widehat{p} \widehat{s}_i)$, with the $v_i$ being weak functions of the other $N^{(p)}_j$.  Thus, it is advisable to first use the Schur-Limber estimator ({\comment or a Markov chain to map the likelihood surface}) in cases where the $\widehat{N}^{(p)}_i$ are poorly constrained (and often in this limit the Schur-Limber estimator will in fact be optimal).  

Fig.~\ref{fig:estwalk} shows the walk of the
$\widehat{N}^{(p)}_{N_{\rm bin}/2}$ estimate as a function of
iteration number for the middle redshift bin in the two
cross-correlation cases. The solid curves are the full minimum
variance estimator, and the dashed curves are the Schur-Limber
estimator (which converges more quickly).  The curves terminate when
the next successive iteration changes the estimated
$\widehat{N}^{(p)}_i$ by less than a part in $10^{5}$ when averaged over all
$i$.  The Schur-Limber estimator converges rapidly in both examples
(after $3-4$ iterations).  This similar convergence rate is despite
the two cross-correlation cases being considerably different in terms
of their sensitivity, their $dN^{(p)}/dz$, and their $N_{\rm bin}$.  For the
minimum variance quadratic estimator, convergence requires additional
steps -- as many as $20$ iterations for the case in the bottom panel.

\section{Breaking the bias -- number degeneracy}
\label{sec:bias_number}

Much of our discussion has ignored that cross-correlations do not
constrain number alone but instead bias times number.  The bias often can be parametrized
as a smoothly and slowly varying function with redshift.  An exception is
samples with hard color cuts, where the underlying galaxy population, and
hence the large-scale bias, can change relatively quickly with $z$ at points
where spectral features transition in and out of filters.  In such cases,
knowledge of $b^{(p)}_i N^{(p)}_i$ is more difficult to translate into
knowledge about $N^{(p)}_i$.

For many applications, bias times number is in fact the quantity of interest,
including attempts to measure 3D correlations with angular
correlations or attempts to subtract correlated anisotropies from a
map of diffuse backgrounds.  However, knowing the bias is particularly
important to the application of calibrating the lens redshifts for
weak lensing surveys.  RSDs as well as lensing magnification formally provide
terms that break the bias--number degeneracy.  However, we argued that
breaking this degeneracy is unlikely with RSDs.  Cosmic magnification
is more promising:  We argued that surveys capable of percent-level $N^{(p)}_i$
determinations may be able to constrain the bias to $10$ per cent.

Other possibilities for breaking this degeneracy require using additional scales or constraints not included in our earlier estimates.   Such methods to break this degeneracy include
modeling of the one-halo term in $\langle p s_i \rangle$;
abundance matching or other modeling methods to map galaxy number to bias
\citep[e.g.][as $b^{(p)}_i$ is a weak function of mass for abundant halos]{conroy06};
galaxy-galaxy lensing with the photometric galaxies as both sources
and lenses (using the $b^{(p)}_i N^{(p)}_i$ from cross-correlation measurements -- the quantity needed for the lenses -- to constrain $dN^{(p)}/dz$ of the sources); {\comment breaking up the photometric sample into subsamples and using that the auto-correlation of each subsample provides an integral constraint on its bias}; and
measurements of the $2^{\rm nd}$ order bias,
either in the two point function or  higher order statistics.
While several of these avenues appear promising, we shall not pursue
them here.

\section{Conclusions}
\label{sec:conclusions}

Determining the redshift distribution of a particular population of astronomical
objects is often quite difficult.  However, since most cosmological objects are
clustered (i.e., they trace the same matter field on large scales),
objects that are close together on the sky are also likely to be close
together in redshift.  Thus, the redshift
distribution of a population of objects can be determined by cross-correlating it in angle with a
population whose redshift distribution is better known.
This paper presented a new, optimal estimator for the
redshift distribution of a given population in terms of cross-correlations.  
We found that this estimator (1) is quite intuitive in a number of limits, (2) is straightforward
to apply to observations, (3) robustly finds the posterior maximum, and (4) conveniently selects angular scales at which the
fluctuations are well approximated as independent between redshift bins and
at which linear theory applies.  In addition, we provided analytic formulae that can be used to quickly estimate the sensitivity of cross-correlations between overlapping surveys to $b \, dN/dz$ -- the linear bias times angular number density per redshift. We compared our estimator to others suggested in the literature, showing that
it produces considerably smaller errors than the familiar estimator of \citet{newman08}.


The optimal estimator's fractional error on the number of objects (times their bias) in a redshift bin is $\approx \sqrt{10^2 N_{\rm bin}' /{\cal N}^{(s)}}$ if the spectroscopic sample has a mean angular density of less than a few thousand and the unknown sample has a mean density larger than this value.  Here, ${\cal N}^{(s)}$ is the total number of spectra per unit redshift, and $N_{\rm bin}'$ is the number of redshift bins spanned by the bulk of the unknown population.\footnote{This formula is analogous to the sensitivity of direct spectroscopic followup to $dN/dz$, where the fractional error is the square root of the number of spectra in a redshift bin.  It indicates that cross correlations have an order of magnitude larger error at fixed number of spectra.  However, cross correlations have the significant advantage of not requiring the spectra to be of the same objects for which the redshift distribution is desired.}   Thus, it is not necessarily better to use a narrow, deep spectroscopic survey covering tens of degrees than a wide, shallow one.  Once the spectroscopic and unknown populations have $dN/dz \gg 10^4 \, b^{-2} ~$deg$^{-2}$, the sensitivity scales simply with the fraction of sky covered (again with an intuitive formula) and no longer depends on just the total number of spectra.  
 We found that upcoming spectroscopic surveys that aim for millions of spectra can potentially achieve percent-level constraints on the $b \, dN/dz$ of an unconstrained population.  Furthermore, we showed that our estimates for the constraints on $b \, dN/dz$ also apply to spectroscopically calibrating samples binned by their photometric redshift, and we also commented on the sensitivity of photometric self-calibration.

We investigated a number of approximations and how they bias the
estimator.  In the Limber approximation -- which we found to be
excellent for relevant redshift slice widths -- the covariance matrix for this problem can be analytically
inverted, allowing simple expressions for the estimator.  We showed that the
nearly optimal, Limber-approximation estimator can be expressed as an
iteration of
\begin{equation}
\widehat{N}_i = [\widehat{N}_i]_{\rm last} + \sum v_i \left( \widehat{p} \; \widehat{s}_i -  \langle \widehat{p} \; \widehat{s}_i \rangle  \right)/\sum v_i \frac{d\langle \widehat{p} \; \widehat{s}_i\rangle}{dN_i} ,
\end{equation}
where the $v_i$ are weights comprised of intuitive combinations of the
covariance matrix (Eq. \ref{eqn:vell_SchurLimber}) and $\widehat{p} \,
\widehat{s}_i$ is the cross-correlation between the unknown sample and the
spectroscopic sample in bin $z_i$.  The summations are either
evaluated over bins in angular separation or spherical harmonic
indices depending on whether $\widehat{p} \, \widehat{s}_i$  is measured
in configuration or harmonic space.  {\comment In many limits, this estimator has the same error as the maximum likelihood estimate for the cross-power amplitude.}  Furthermore, we found that the
bias from assuming the Limber approximation was minute and also argued
that the same holds for redshift space distortions.  We found that
cosmic magnification can be a significant source of estimator bias,
becoming important once surveys achieve $\lesssim 10$ per cent statistical errors (especially if the surveys extend to $z\gtrsim 2$ or if $dN/dz$ of the unknown sample falls off quickly).  We discussed strategies for correcting this bias.  

The techniques developed in this paper can be applied to a wide range of
existing and upcoming surveys from DES, GAMA and WISE, to LSST, Euclid
and the SKA.
We intend to apply this estimator to observational data in a future paper.\\

We thank Carlos Cunha, Gary Bernstein, Shirley Ho, Jeffrey Newman, and David Schlegel for helpful comments.  We especially thank Chris Blake and Andrew Johnson for pointing out an error previously in  equation~(\ref{eqn:minquad}). MM acknowledges support by the National Aeronautics and Space Administration
through Hubble Postdoctoral Fellowship awarded by the Space
Telescope Science Institute, which is operated by the
Association of Universities for Research in Astronomy,
Inc., for NASA, under contract NAS 5-26555.
MW is supported by NASA.

\bibliographystyle{mn2e}
\bibliography{cross_corr}

\appendix

\section{Estimator Details}

This appendix gives two generalizations of the minimum variance quadratic estimator (Appendix \ref{ap:fullestimator}), then shows how a prior would impact the estimator (Appendix \ref{ap:estwithprior}), and finally considers how the estimator and variance change with different basis choices to represent $dN^{(p)}/dz$ (Appendix \ref{ap:other_bases}).

\subsection{Full Estimator}
\label{ap:fullestimator}
Here we write two more complete expressions for the estimator than were given in the text.

First, the estimator given by Eq.~(\ref{eqn:MQest}) is biased by different cosmic realizations except in the limit in which a large number of modes are used with comparable weight.  
  The full, unbiased estimator replaces Eq.~(\ref{eqn:MQest}) with \citep[for more on derivation see ensuing appendix]{bond98}
\begin{eqnarray}
F_{i j}^{\rm full} &=& F_{ij} + \sum_{\ell, m} {\rm Tr} \Bigg[  \left \{ \left( \begin{array}{c}  \widehat{p} \\  \widehat{\bmath{s}}\end{array} \right) \left( \begin{array}{cc}  \widehat{p}  & \widehat{\bmath{s}}\end{array} \right) - \bfA \right \} \nonumber \\
 &\times &\left(\bfA^{-1}  \bfA_{, i} \,\bfA^{-1} \bfA_{, j} \bfA^{-1} -  \frac{1}{2}\bfA^{-1} \bfA_{,ij} \bfA^{-1}\right) \Bigg].
\label{eqn:fullest}
\end{eqnarray}
This expression shows that the estimator is biased by using $F_{ij}$ rather than $F_{i j}^{\rm full}$ at the level of $N_\ell^{-1/2}$, where $N_\ell$ is the number of modes that contribute.  There are $N_\ell = \ell_{-2}^2 \sim 10^6 f_{\rm sky}$ total modes that generally contribute to the estimator (at least when one sample is abundant).   Thus, this error will impact the estimator at the $10^{-3} f_{\rm sky}^{-1/2}$ level.  This additional sample variance noise should typically be below the statistical error.  We saw no evidence for this bias in the estimates from mock surveys in Section \ref{sec:mocks}.

All of our estimators can be written as sums over $\theta$ or $\ell$ and do not require keeping angular information.  This may come as a surprise because each individual $\ell, ~m$ mode contributes independent information and so it may seem suboptimal to combine them in annuli.  However, one can note that this is also a symmetry of the likelihood function as ${\cal L}$ can be written so that the argument in the exponent is proportional to $\sum_{\ell,\; m}{\rm Tr}[\widehat{\bfA}(\ell) \, \bfA^{-1}(\ell)]$, where $\widehat{\bfA}(\ell)$ is the estimated covariance matrix (e.g., $\widehat{A}_{00}(\ell) \equiv (2 \ell +1)^{-1} \sum_{m} |p(\ell, m)|^2$).

\subsection{Impact of Prior}
\label{ap:estwithprior}

The estimator given in Eq.s~(\ref{eqn:MQest}) and (\ref{eqn:NiLike2}) follows from using the multidimensional Newton's method to find the zeros of the derivative of the log of the data likelihood function, $\log { \cal L}$ \citep{bond98}:\footnote{Newton's method is applied to the log of the likelihood rather than the likelihood itself because Newton's method provides exact estimates for the extrema of a quadratic function.}
\begin{equation}
\widehat{N}_i = [\widehat{N}_i]_{\rm last} -([{\log {\cal L}}]_{,,})_{ij}^{-1} [\log {\cal L}]_{,j},
\label{eqn:NiLike}
\end{equation}
where $[\log {\cal L}]_{,,}$ is the Hessian of $\log { \cal L}$, which upon ensemble average is the negative of the Fisher matrix.  
  For a Gaussian likelihood with covariance matrix $\bfC$ and data vector $\Delta$, $[\log {\cal L}]_{,i} = \Delta^T \bfC^{-1} \bfC_{,i}  \bfC^{-1} \Delta/2$.

With this derivation in mind, it is straightforward to generalize
Eq.~\ref{eqn:NiLike} to include a prior:  
\begin{equation}
  \widehat{N}_i = [\widehat{N}_i]_{\rm last} -
  \left([\log {\cal L}]_{,,} + [\log {\cal L}_{\rm P}]_{,,} \right)_{ij}^{-1}
  \left( [\log {\cal L}]_{,j} + [\log {\cal L}_{\rm P}]_{,j} \right),
\label{eqn:NiLike2}
\end{equation}
where ${\cal L}_{\rm P}$ is the prior likelihood function.  The case of a Gaussian prior on the $N_i$ is given by Eq.~(\ref{eqn:Nprior}).



As an application of the above, let us consider the case of our $N^{(p)}_i$ estimator in which the $b^{(s)}_i$ are imperfectly known and instead are constrained by prior information. 
Remember that since the $N^{(p)}_i$ are estimated from large-scale cross-correlations,
they are degenerate (ignoring e.g. magnification) with $b^{(s)}_i$ and can only be
separated with a prior from the auto-correlation measurements.  In this case, the Fisher matrix of the parameters $N^{(p)}_i$ and $b^{(s)}_i$ plus a prior on $b^{(s)}_i$ yields the new error matrix:
\begin{equation}
\bfF^{bs} = \frac{F^S_{ii}}{[N^{(p)}_i]^2} \left( \begin{array}{cc}  [N^{(p)}_i]^2 & N^{(p)}_i  b^{(s)}_i \\  N^{(p)}_i  b^{(s)}_i & [b^{(s)}_i]^2   \end{array} \right) + \left( \begin{array}{cc}  0 & 0\\  0 & \sigma_{bs}^{-2}   \end{array} \right),
\end{equation}
where $\sigma_{bs}$ is the standard deviation of the Gaussian prior on
$b^{(s)}_i$ centered on $[b^{(s)}_i]_{\rm prior}$.  Our previous results
correspond to $\sigma_{bs}\to 0$.
(We are ignoring redshift-bin correlations in the prior for simplicity, but such
correlations can be easily incorporated.)  The fractional variance on a
measurement of $N^{(p)}_i$ is thus
\begin{equation}
  \left(\frac{\delta N^{(p)}_i}{N^{(p)}_i}\right)^2 \equiv
  [\bfF^{bs}]^{-1}_{ii}  =  [F_{ii}^S]^{-1} \,
  \left(1+\left[\frac{N^{(p)}_i\sigma_{bs}}{b_i^{(s)}}\right]^2 F^S_{ii}\right).
\end{equation}
Therefore, the fractional variance in the estimated $b_i^{(s)}$ is the limiting factor when it is larger than the fractional variance in the estimate of $N^{(p)}_i$ for the case that $b_i^{(s)}$ is held fixed.

The estimator in this limit is
\begin{eqnarray}
  \widehat{N}^{(p)}_{i} &=&   [\widehat{N}^{(p)}_{i}]_{\rm last} +
   \frac{1}{F_{i i}^S} \sum_{\ell, m} \, \frac{[A_{0i}]_{,i}}{A_{00} \, A_{ii}} 
  \left \{  \widehat{p} \; \widehat{s}_i  -  A_{0i} \right \} \nonumber \\
  &+& N^{(p)}_i \left( [b^{(s)}_i]_{\rm prior}/[\widehat{b}^{(s)}_i]_{\rm last} - 1 \right),
\label{est:Nhatbspec}
\end{eqnarray}
with the complementary estimator for the bias being trivially $\widehat{b}^{(s)}_i=[b^{(s)}_i]_{\rm prior}$.

For the case of SDSS or BOSS quasars (where ${\cal N}^{(s)} \sim 10^5$), the variance
in the measured bias is $\sigma_{bs} \sim 0.1$ \citep{ross09,white12},
which is comparable to the redshift error expected from cross-correlations
(Fig.~\ref{fig:Constn}).  However, for rare samples with fewer spectra than SDSS
quasars, the uncertainty in $b_i^{(s)}$ will dominate the error in
the $N^{(p)}_i$ that ignores the bias uncertainty.

\subsection{Estimator and constraints in other bases}
\label{ap:other_bases}

We have chosen a top hat basis set for convenience, which also leads to an estimator that converges robustly to the likelihood peak.  Other choices are clearly possible, and they may be preferred in some situations.  For example, instead of $N^{(p)}_i$ we could estimate the parameters of a particular functional form.  Or we could expand $dN^{(p)}/dz$ as a sum of overlapping Gaussians or (orthogonal) polynomials times basis functions (e.g.~a power law times an exponential).   While the quadratic estimator formalism is completely general,  it is not trivial to recast the estimator in terms of an arbitrary basis set as $\bfA$ needs to be recast in terms of the new parameter set.  In many cases, this is not analytically expressible (with an exception being the linear case discussed below).    However, it is trivial to translate our results for the error on a parameter into another basis set.  The new Fisher matrix is given by the chain rule:
\begin{equation}
 \bfF' = \bfW^T \, \bfF \; \bfW,
 \label{eqn:Fprime}
\end{equation}
where $\bfW$ is the Jacobian matrix between the $N^{(p)}_i$ and the new
parameter set $\lambda_i$.
We showed that the Fisher matrix is often well approximated as diagonal,
such as in the Schur-Limber limit.  In this case
\begin{equation}
 F'_{ij} \approx \sum_{k=1}^{N_{\rm bin}}
   \frac{1}{F^S_{kk}} \frac{dN^{(p)}_k}{d\lambda_i}
                   \, \frac{dN^{(p)}_k}{d\lambda_j}.
\end{equation}
Once the $N^{(p)}_i$ are estimated with our technique, they can be combined
to estimate the $\lambda_i$ with error given by $\bfF'$.

\begin{figure}
\begin{center}
\epsfig{file=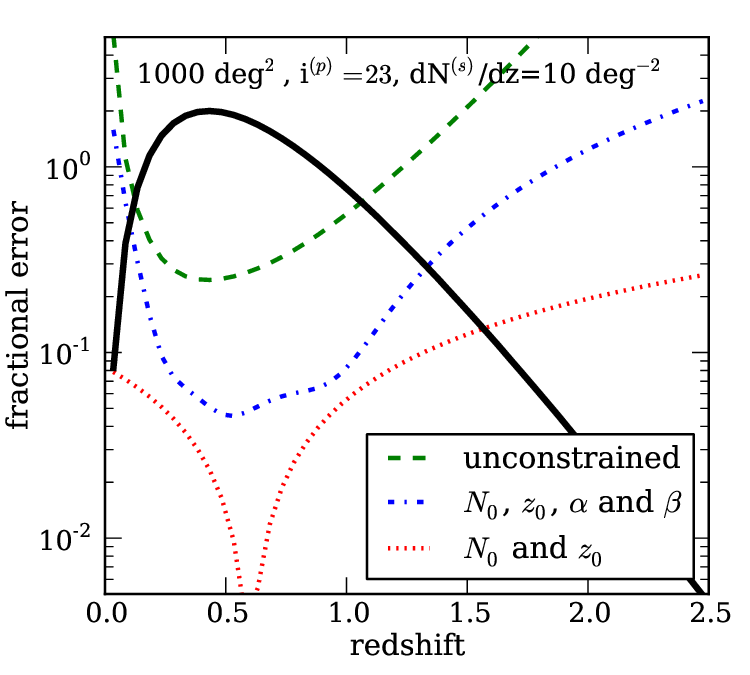, width=7.5cm}
\end{center}
\caption{Improvement in constraints from a constrained parametrization of $dN^{(p)}/dz$ rather than the case considered in the text in which the top-hat basis $N^{(p)}_i$ are free.  Shown are surveys with the parameters $\rmi^{(p)} = 23$, $dN^{(s)}/dz = 10~$deg$^{-2}
$, and $\Delta z = 0.05$ over $1,000\,$deg$^2$ and $0 < z < 2.5$.  The fractional errors for the unconstrained case -- the case investigated in the body of this paper -- are given by the green dashed curve, and the case where $b^{(p)} dN^{(p)}/dz$ is constrained by the functional form $N_0 \, (z/z_0)^\alpha \exp[-(z/z_0)^\beta]$, marginalizing over the parameters specified in the key, is given by the dot-dashed blue and dotted red curves.    This constraining functional is evaluated at the fiducial parameters given by Eq.~(\ref{eqn:pziband}) for these two cases.  The black solid curve shows $dN^{(p)}/dz$, arbitrarily normalized.
\label{fig:constrained}}
\end{figure}

Fig.~\ref{fig:constrained} shows an example using Eq.~\ref{eqn:Fprime} in which we changed basis to one in which $dN^{(p)}/dz$ is constrained to have the smooth functional form specified in the key (a generalization of our Eq.~(\ref{eqn:pziband}) for $P(z,i)$).  This figure investigates the case of a photometric population with $\rmi^{(p)} = 23$ and with a low density of spectroscopic objects given by $dN^{(s)}/dz = 10~$deg$^{-2}$, overlapping over a sky area of $1,000\,$deg$^2$ (although, the total number of spectra, here $10^4$, is the essential quantity). It shows that the constraints are substantially improved even if a fairly general functional form is assumed (varying two parameters for the dotted curves and four for the dot dashed).  One advantage of parametrizing $dN^{(p)}/dz$ with a smooth functional form is that the constraints do not depend on the choice of $\Delta z$.

Finally, we note that the formalism this paper developed for estimating the $N^{(p)}_i$ can be trivially recast for  models in which one instead aims to constrain some set of basis functions $\phi_i$ for which $dN^{(p)}/dz = \sum_i c_i \phi_i(z)$, where $c_i$ are a set of coefficients.  In this case, the  primarily difference is that for the $\alpha_\ell(k,z_i)$ that went into calculating $\bfC(\ell)$, the index $i$ no longer indices the redshift bin but rather the basis function. 

\section{Extended Limber approximation}
\label{ap:limber}
\label{ap:rsd}

The Limber approximation is most applicable on small angular scales,
where we may approximate the sky as flat and the spherical harmonic
transform as a Fourier transform \citep[e.g.][]{WCDH99, Papai08}.
With these approximations, the angular correlation function can be written as
\begin{eqnarray}
  w(\theta) &=& \int d\chi_1\,d\chi_2\ W(\chi_1)W(\chi_2) \nonumber \\
  &\times& \int\frac{d^3k}{(2\pi)^3} P(k)
  e^{i\mathbf{k}\cdot(\mathbf{x}_1-\mathbf{x}_2)},
  \label{eqn:extendedlimber} \\
  &\approx & \int\frac{d^3k}{(2\pi)^3} P(\mathbf{k}_\perp,k_\parallel)
  \int d\bar{\chi}\ W^2(\bar{\chi}) e^{i\mathbf{k}_\perp\cdot\mathbf{x}_\perp}
  \nonumber \\
  &\times& \int dZ\ e^{ik_\parallel Z}, \\
  &=& \int\frac{K_\perp\,dK_\perp}{2\pi} P(\mathbf{k}_\perp,k_\parallel=0)
  \nonumber \\
  &\times& \int d\bar{\chi}\ W^2(\bar{\chi}) J_0(k_\perp\bar{\chi}\theta), \label{eqn:Zchi}
\end{eqnarray}
where in the second line we have changed variables from $\chi_i$ to
center-of-mass and relative coordinates, $\bar{\chi}=(\chi_1+\chi_2)/2$ and
$Z=\chi_1-\chi_2$, and assumed that $W$ is so broad that
$W(\bar{\chi}\pm Z/2)\approx W(\bar{\chi})$ (which is not always the case for the $W$ considered in the text).
Writing $\ell=k_\perp\bar{\chi}$ and using
$J_0(\ell\theta)\simeq P_\ell(\cos\theta)$ for $\theta\ll 1$ and $\ell\gg 1$,
the angular power spectrum, $C_\ell$, is thus
\begin{equation}
  C_\ell = \int d\chi\ \frac{W^2(\chi)}{\chi^2}
  \ P(k_\perp=\ell/\chi,k_\parallel=0).
\label{eqn:cl_limber}
\end{equation}
The Limber approximation further results in correlations between non-overlapping redshift slices being zero.   

One can compare the Limber approximation to the analytic solution for certain cases to see when and how well these approximations work.
Let us assume $W(\chi)$ is a top-hat in $\chi$ in slices of width
$\Delta\chi$ (as in the main body of this paper).  Then, the cross-spectrum is
\begin{equation}
\ell^2 C_{ij} = k_\perp^2 \int \frac{dk_\parallel}{2\pi}
  e^{i k_\parallel(\chi_i-\chi_j)}
  {\rm sinc}\left[\frac{k_\parallel\Delta\chi}{2}\right]^2
  P(k_\perp,k_\parallel).
  \label{eqn:cl_limber_flatsky}
\end{equation}
Using the method of steepest descents (or approximating the power spectrum
as a power-law and using the asymptotic behavior of the resulting Bessel
functions), it can be shown that for $k_\perp |\chi_i-\chi_j|  \gg 1$
\begin{equation}
\ell^2 C_{ij} \rightarrow  \ell^2 C_{ij}^{\rm asymp} \equiv \frac{k_\perp}{\Delta\chi^2} P(k_\perp)
  e^{-k_\perp|\chi_i-\chi_j|},
\end{equation}
We can make further progress by assuming that $P(k)$ is a power-law.
In particular, if $P(k)$ is a power-law with index $-2$, roughly the index
on galaxy scales in our Universe, the integral in Eq.~\ref{eqn:cl_limber_flatsky} has simple poles that make the evaluation trivial:
\begin{equation}
\ell^2 C_{ij} = \ell^2C_{ij}^{\rm asymp}
  \begin{cases} k_\perp\Delta\chi+\exp[-k_\perp\Delta\chi]-1~~~i=j;\\
                {\rm cosh}[{k_\perp\Delta\chi}]-1 \qquad\qquad\quad i\ne j.
  \end{cases}
  \label{eqn:limberm2}
\end{equation}
Note that when $i=j$ and $k_\perp\Delta\chi\gg 1$ we recover the Limber
result $\ell^2C_{ii}\simeq (k_\perp^2/\Delta\chi)P(k_\perp)$.  In addition, at $k_\perp \Delta \chi = 2$ (the boundary of applicability used in Fig. 6), Eq.~\ref{eqn:limberm2} undershoots Limber by $40$ per cent with this percentage decreasing roughly linearly with increasing $k_\perp \Delta \chi$.  The errors from Limber will be smaller when $P(k)$ has a flatter power-law, as is the case at $k_\perp \Delta \chi \sim 1$ for the $\Delta \chi$ considered in the text.  That the Limber approximation works so well once $k_\perp \Delta \chi$ moderately exceeds unity helps explain why in the text we find it to be such a good approximation for our problem.  

Next, consider the impact of redshift-space distortions (RSDs) in the Limber approximation, which have been neglected in all of our prior discussion.  RSDs could be interesting for our purposes because they break the $b^{(x)}_i$--$N^{(x)}_i$ degeneracy.  On linear scales
the lowest-order correction owing to RSDs is to multiply the power spectrum by
$1+2\beta_i \, \mu^2$, where $\mu=k_\parallel/k$ and $\beta_i \simeq\Omega_m^{0.6}/b_i^{(x)}$, with the redefinition of $\chi$ and $\bfk$ to be the analogous redshift-space quantities \citep{kaiser87, hamilton92}.
In the Limber approximation, $|k_\parallel|\la \Delta\chi^{-1}$ and so we
expect $|\mu|\ll 1$ and the correction to be small.  However, how quickly this falls off depends on $W(\chi)$.  In the case of our top hat window function and with the replacement $P(k_\perp, k_\parallel) \rightarrow P(k_\perp)(1 + 2 \, \beta_i \,\mu^2)$ -- which is analogous to the Limber approximation --, Eq.~\ref{eqn:cl_limber_flatsky} can be integrated analytically yielding
\begin{equation}
\ell^2 C_{ii} = \frac{k_\perp^2}{\Delta\chi} P(k_\perp) \left(1 + \frac{2 \, \beta_i}{k_\perp \Delta\chi} \right),
\end{equation}
with the off-diagonals being zero.
Thus, the RSD correction falls off slowly as $(k_\perp \Delta\chi)^{-1}$ in the case of top hat $W$.  A curiosity is that if we had approximated $\mu$ as $k_\parallel/k_\perp$, the integral would have diverged.  Thus, in the case of a top hat $W$, the RSD term arises from modes with $\mu \sim 1$.

However, smoother $W(\chi)$ result in RSDs having a weaker scaling in the
Limber regime.  Consider the case in which $W(\chi)$ is a Gaussian with standard deviation $\sigma$.  The analogous equation to Eq.~\ref{eqn:cl_limber_flatsky} for this case is
\begin{equation}
\ell^2 C_{ij} = k_\perp^2 \int \frac{dk_\parallel}{2\pi}
  e^{i k_\parallel(\chi_i-\chi_j)}
  {\rm \exp}\left[-{k_\parallel^2 \, \sigma^2}\right]
  P(k_\perp,k_\parallel).
\end{equation}
For large $\sigma$ the integral is dominated by small $k_\parallel$, and
we can Taylor series expand about $k_\parallel=0$ as above.  In this case, the correction
due to redshift-space distortions enters at order $\mathcal{O}( [k_\perp \sigma]^{-2})$.  The RSD term is similar (merely increasing by a factor of $2$) if one of the two window functions were much narrower than $\sigma$.  In addition, exponential or triangle window functions also have RSDs entering at $\mathcal{O}( [k_\perp \sigma]^{-2})$.\footnote{This result that RSDs depend on the smoothness of $W(\chi)$ is analogous to the finding in \citet{nock10}.  There, the impact of RSDs on the correlation function measured in a top hat projection over $\sim 100~$Mpc was shown to be much more significant than when the effective window was smoothed with a pair-averaging scheme.}

It is important for our calculations if the RSDs in Limber --{\comment an approximation that we showed holds excellently at angles that contribute to the estimator} -- contribute at $\mathcal{O}( [k_\perp \sigma]^{-1})$ rather than $\mathcal{O}( [k_\perp \sigma]^{-2})$, where $\sigma$ is the width of our window function.  RSDs would be a promising signal to break the $b^{(x)}_i$--$N^{(x)}_i$ degeneracy if the former scaling holds, but are not in the case of the latter.  It may appear with the formalism in the text, which uses top hat $W_i$, that the $\mathcal{O}( [k_\perp \sigma]^{-1})$ scaling would apply.  However, for the case of interest where the $dN^{(p)}/dz$ is a smooth function that is not known, we posit that one is always in the regime where the RSD term falls off as $\mathcal{O}( [k_\perp \sigma]^{-2})$.  Basis functions can always be chosen that have smooth $W(\chi)$ and where the RSD terms contribute at $\mathcal{O}( [k_\perp \sigma]^{-2})$.  That they contribute at $\mathcal{O}( [k_\perp \sigma]^{-1})$ for top hat windows is a pathological result of our basis choice that implicitly assumes that the distribution of $dN^{(p)}/dz$ is a histogram with sharp breaks between redshift steps.  

 To include RSDs properly requires a smoother basis set for the $W_i$ than we take in the text.  Because of this added complication, we do not consider RSDs in our formulae in the text.  For the reasons espoused above and because the modes that contribute to our estimate are generally safely in the Limber regime, the bias from ignoring their impact on correlation functions with the photometric sample should be small. {\comment RSDs are a more important consideration for the spectroscopic--spectroscopic elements in $\bfA$.  (However, these elements do not impact our estimator in the Schur-Limber limit.)}

\section{Magnification bias}
\label{ap:magnification}

The spatial density of observed galaxies is modulated by an additional factor that we have ignored so far of $(1+ \delta_\mu)$ owing to lensing magnification \citep{turner84, fugmann88, narayan89, hui07, hui08}. In the weak lensing regime,
\begin{equation}
\delta_\mu(\nhat, z_i) \equiv 2\,(-\alpha_i^{(x)}- 1) \int_0^{\chi_i} d\chi \, \frac{\chi_i - \chi}{\chi_i} \chi \, \nabla_\perp^2 \phi(\chi, \nhat),
\end{equation} 
where $\nabla_\perp^2$ is the comoving Laplacian in the plane perpendicular to the radial direction and $\alpha^{(x)}_i$ is the power-law slope of the cumulative number of sources at the survey flux threshold and redshift $z_i$.  (Note that $\alpha^{(x)}_i$ is defined to be a negative number as long as the cumulative number decreases with increasing flux.)  Thus, magnification generates additional correlations 
such that
\begin{equation}
C_{ij} \rightarrow C_{ij} + C_{ij}^{\delta \mu} + C_{ji}^{\delta \mu},
\label{eqn:Cijremap}
\end{equation}
where $C_{ji}^{\delta \mu}$ is the cross-correlation function between the galaxy overdensity field in redshift slice $j$ and $\delta_{\mu, i}$, and we are dropping the smaller  $C_{ij}^{\mu \mu}$ term. 
In the Limber regime, the expression for the new terms in Eq.~\ref{eqn:Cijremap} is \citep[their Eq.~7.9]{bartelmann01}
\begin{equation}
C_{ij}^{\delta \mu} = -\left(\frac{\alpha_i^{(x)}+1}{b_i^{(x)}} \right) \frac{3 H_0^2 \Omega_0}{c^2} \int \frac{d\chi}{\chi a}  \,W_j(\chi) Y_i(\chi) D^2(\chi) P(\frac{\ell}{\chi}),
\end{equation}
for $i > j$.  Otherwise, $C_{ij}^{\delta \mu} = 0$ (we ignore the contribution of magnification to the $i=j$ elements), and we denote the source population in question by $x$ and lens by $y$ as it could be either the photometric or spectroscopic sample.  Here,
\begin{equation}
Y_i(\chi) =  \int_\chi^\infty d\chi' \, W_i(\chi')  \, \frac{\chi' - \chi}{\chi'}.
\end{equation}
Magnification depends only on the bias of the lens and not the source and so can break the degeneracy between bias and number.  (This dependence may be opaque in our notation as the $C_{ij}^{\delta \mu}$ enter $\bfA$ multiplied by factors of the bias.)

Noting that $c^2/(3H_0^2 \Omega_m) = 2\times 10^7\,\Mpc^{2}$, a back-of-the-envelope estimate for $C_{ij}^{\delta \mu}$ is
\begin{equation}
C_{ij}^{\delta \mu} \approx -\left( \frac{\alpha_i^{(x)}+1}{b_i^{(x)}} \right)  \frac{(1+z_j) \, D^2(z_j)\,P\left(\frac{\ell}{\chi_j} \right)}{(2\times 10^7 {\rm ~Mpc}^2)} \left(\frac{1}{\chi_j} - \frac{1}{\chi_i} \right)
\end{equation}
when $i>j$, and we have approximated $W_i$ and $W_j$ as sharply peaked around their respective redshifts.
This is similar to the $C_{jj}$ term without lensing (Eq.~\ref{eqn:limberapprox}), differing most importantly by the factor $[(1+z_j) \, \chi_j \, \Delta \chi_j]/ 2\times 10^7 {\rm ~Mpc}^2$.  This factor is ${\cal O}(10^{-2})$ for populations at $z\sim 1$ and $N_{\rm bin}\sim 50$, but can be larger for higher redshift populations.  Thus, magnification will add off-diagonal terms that are
${\cal O}(10^{-2})$ of the diagonal terms in $\bfC$ that were zero in much of our treatment in the text.  The new magnification terms have a larger impact on the components in $\bfA$ involving $p$, as these terms sum over $i$ and $j$ in $C_{ij}$

\subsection{Photo-$z$ calibration with magnification}
\label{ap:magnification_pz}

Here we discuss how magnification could potentially be corrected in the application of photo-$z$ calibration investigated in Section~\ref{ss:specphotoz} (and we use the same notation as introduced there).  We consider a simplified problem in which most of the $pm$ photo-$z$ sample is concentrated at redshift $z_m$.  Then, there is a significant bias if the error on $T^{(pm)}_i/T^{(pm)}_m$ is comparable to $C_{mi}^{\delta \mu}/C_{ii}$, which we just showed is ${\cal O} ([N_{\rm bin}]^{-1})$ for $z_i \sim 1$.

The minimum variance estimator with a prior on the $\alpha^{(x)}_i$ (which enters analogously to the number prior in Eq.~\ref{eqn:Nprior}) can also be written for this simplified problem:  First, the covariance matrix at some $\ell$ and in the Limber approximation is 
\begin{eqnarray}
D_{00} &\approx&
  [T^{(pm)}_m]^2 C_{mm} + w^{(pm)} + {\cal M} , \\
D_{01} &\approx& T^{(pm)}_j T^{(s)}_j C_{jj}
              +  T^{(pm)}_m T^{(s)}_j C_{mj}^{\delta \mu} + w^{(pms)}_j, \\
D_{11} &\approx& [T^{(s)}_j]^2 C_{jj}  + w^{(s)}_j,
\end{eqnarray}
where ${\cal M}$ encompasses the impact of photometric self-magnification, and we have dropped terms that do not contain $T^{(pm)}_m$ except the off-diagonal $T^{(pm)}_i$ terms for which the estimator's sensitivity to $T^{(pm)}_i$ derives.  For the specified $\bfD$ and a prior on $\alpha^{(x)}$ with variance $\sigma_\alpha$, the minimum variance quadratic estimator is
\begin{eqnarray}
\widehat{T^{(pm)}_i}  = [\widehat{T^{(pm)}_i}]_{\rm last} + [\bfF^{-1}]_{11} \sum_{\ell, m}  \frac{S' \, T^{(s)}_j C_{ij}}{D_{00} D_{11}} \, \left(\widehat{p_m}\, \widehat{s}_i - D_{01} \right),
\end{eqnarray}
where $S' = D_{00} D_{11} (D_{00} D_{11} + D_{01}^2)/\det[\bfD]^2$,
$\alpha^{(x)}$ is set by the prior, we have assumed that $T^{(pm)}_m$ is well constrained by other cross
(and auto) correlations (which is quite likely), and $\bfF$ also has a simple analytic representation.  This estimator is quite analogous to our previous estimator.  

It is instructive to look at the variance on a measurement of $T^{(pm)}_i$ in a single mode:
\begin{equation}
[\bfF^{-1}]_{11} = \frac{D_{00} D_{11} + S' [T^{(pm)}_m T^{(s)}_j C_{mj}^{\delta \mu}/(\alpha+1)]^2 \sigma_\alpha^2}{(S' T^{(s)}_j C_{ij})^2}.
\label{eqn:lens_var}
\end{equation}
This equation shows that error on the magnification bias times $S'$ (the latter term in the numerator) has to be comparable to the auto power terms (the former term) in order to change our previously quoted errors in Section~\ref{ss:specphotoz}.  It also suggests that it may be desirable to down weight large-angle modes where $S'$ is largest (that have the smallest noise) and, hence, where the fog from lensing is most disruptive.

\section{Recurrence relations for (and the evaluation of integrals over)
spherical Bessel functions}
\label{ap:jayell}

Our most general expressions for the auto and cross power spectra, Eqs.~\ref{Cij} and\ref{alphaij}, involved integrals over spherical Bessel functions.
Numerical methods for evaluating spherical Bessel functions and integrating
over them are well advanced, but do not seem to be widely known.  This appendix gives the
details of the algorithms used in this study.  Further details can be found in
\citep{Miller52,Corbato59,Gillman88,Poularikas00} or at
\url{http://www.utdallas.edu/~cantrell/ee6481/lectures/bessres1.pdf}.

First we address the evaluation of the $j_\ell$.
For small values of the argument, we use a series expansion of $j_\ell(x)$.
For larger values, we evaluate the $j_\ell$ using a downwardly stable recurrence
relation for $r_\ell\equiv j_\ell/j_{\ell-1}$.
Specifically we first initialize $r_L$ by setting $j_L(x)=0$ for $L$ much
larger than any $\ell$ of interest (and $x$).  Then the relation
\begin{equation}
  r_{\ell-1} = \frac{1}{(2\ell-1)/x-r_\ell}
\end{equation}
is downwardly stable and can be used to find $r_\ell$ for $0<\ell<L$.
The $j_\ell$ can then be evaluated by moving up the hierarchy after
initializing $j_0(x)=\sin(x)/x$.

Eqs.~(\ref{Cij}) and (\ref{alphaij}) are difficult integrals to evaluate
owing to the oscillatory nature of the $j_\ell$.
We experimented with using the scheme suggested in \citet{lucas95} of
decomposing the product of $j_\ell$ into a sum of functions that each
have a single oscillatory period at large arguments and then using the
transformations discussed therein on a series where the $n^{\rm th}$
member is our $k$-integral evaluated from $0$ out to the $n^{\rm th}$ zero.
This operation removes oscillatory behavior in this slowly converging series
so that it converges more quickly to the $n\to\infty$ limit, and the
integral converges for $n\sim 10$ \citep{lucas95}.
Experiments with some of the integral terms indicated that the \citet{lucas95}
method was much faster than a brute-force integration, but we were able to
find a simpler implementation which was sufficiently fast and accurate.  In particular, we ended up evaluating these integrals by brute force, integrating typically
out to the $1,000^{\rm th}$ zero of the $\alpha_\ell(k,z_i)$
(which were pre-computed and stored in a table).
A slight improvement in the convergence of the integral was obtained by
applying a Gaussian damping to the integrand -- based on the fact that
$k_\parallel\gg \ell/\chi$ should not contribute much to the integral.
The details of this damping did not affect our results.

\section{The power-law case}
\label{ap:powerlaw}

{\comment The main body of this paper used power-law approximations to the power-spectrum and correlation function to understand the mechanics of the Schur-Limber estimator.  To aid this discussion, here we work through expressions for the angular
power spectrum and correlation function (and their relation) under these approximations.}

Recall that within the Limber approximation (Section \ref{sec:limber})
\begin{equation}
  C_\ell = \int d\chi\,P(k)\,\frac{W^2(\chi)}{\chi^2},
\end{equation}
where $W(\chi)$ is the projection kernel that defines the 2D (projected) overdensity
in terms of the 3D, and it
integrates to unity against $d\,\chi$.
We shall assume that $W(\chi)$ is peaked at $\chi_0$ and of width
$\Delta\chi$ such that $k\chi_0\gg k\Delta\chi\gg 1$ for scales, $k$,
which contribute significantly.

Assuming a power-law power spectrum of the form
$\Delta^2(k) \equiv k^3 P(k)/2\pi^2=(k/k_\star)^{3+n}$, with $-2<n<-1$, the real-space
3D correlation function is
\begin{equation}
  \xi(r) = \left(\frac{r_0}{r}\right)^{\gamma}
         = \int \frac{dk}{k} \Delta^2(k)\ j_0(kr)
         = B_n\left(k_\star r\right)^{-3-n},
         \label{eqn:xipl}
\end{equation}
where $B_n\equiv-\sin(n\pi/2)\,\Gamma(2+n,0)$, which respectively equals $1.25$ and $1$ for $n=-3/2$ and $n=-1$ ($B_n$ diverges as $n\to -3^{+}$).
It follows from Eq.~\ref{eqn:xipl} that $\gamma=n+3$ and $r_0 = B_n^{1/\gamma}/k_\star$.

In the Limber approximation,
\begin{equation}
  C_\ell = \frac{2\pi^2}{k_\star^3\mathcal{V}}
           \left(\frac{\ell}{k_\star\chi_0}\right)^n,
\end{equation}
where $\mathcal{V}=\chi_0^2\,\Delta\chi$ is the volume per steradian.
Using analogous relations to Eq.~\ref{eqn:xipl}, the 2D or projected correlation function is
\begin{equation}
  w(\theta) = \left(\frac{\theta_\star}{\theta}\right)^{n+2}
            = \frac{\pi\,A_n}{k_\star^3\mathcal{V}}
              \, \left(k_\star\chi_0\right)^{-n}
              \ \theta^{-n-2},
\end{equation}
where 
$A_n\equiv 2^{n+1}\,\Gamma(1+n/2)/\Gamma(-n/2)\simeq 2.1$ and $1$
for $n=-3/2$ and $n=-1$ ($A_n$ diverges as $n\to -2^+$).

Particularly simple expressions hold in the case $n=-1$ for which
$A_n=B_n=1$, so $\Delta^2=(k/k_\star)^2$,
\begin{equation}
  \xi(r) = \left(\frac{r_0}{r}\right)^2 \quad {\rm where}
  \qquad \ r_0=k_\star^{-1},
\end{equation}
and 
\begin{equation}
  w(\theta) = \left(\frac{\theta_\star}{\theta}\right)
  = \pi\ \left(\frac{r_0}{\chi_0}\right)^2
    \left( \frac{\chi_0}{\Delta\chi} \right)
    \ \theta^{-1}.
\end{equation}

\end{document}